\documentclass[10pt]{article}
\usepackage{amsmath,mathrsfs}
\usepackage{lineno}
\usepackage{caption}
\usepackage{subcaption}
\usepackage{graphicx}
\usepackage{enumerate}
\usepackage{natbib}
\usepackage{bbm}
\usepackage{bm}
\usepackage{amssymb}
\usepackage[affil-it]{authblk}
\usepackage{verbatim}
\usepackage{extarrows}
\usepackage{framed}
\usepackage[colorlinks=true,citecolor=blue]{hyperref}
\usepackage{algorithm}
\usepackage{algorithmic}
\usepackage{url} 
\usepackage{float}
\usepackage{geometry}
\usepackage{lscape} 
\usepackage{multirow}
\usepackage{booktabs}
\newtheorem{assumption}{Assumption}

\newtheorem{proposition}{Proposition}

\newtheorem{remark}{Remark}
\newtheorem{lemma}{Lemma}

\begin{document}
\def\spacingset#1{\renewcommand{\baselinestretch}%
{#1}\small\normalsize} \spacingset{1}
\title{False Discovery Rate Control via Data Splitting}
\author{Chenguang Dai\footnote{The first two authors contributed equally to this work.}}
\author{Buyu Lin}
\author{Xin Xing}
\author{Jun S. Liu}
\affil{Department of Statistics, Harvard University}
\maketitle
\begin{abstract}
Selecting relevant features associated with a given response variable is an important issue in many scientific fields. Quantifying quality and uncertainty of a selection result via false discovery rate (FDR) control has been of recent interest. This paper introduces a  way of using data-splitting strategies to asymptotically control the FDR while maintaining a high power.
For each feature, the method constructs a test statistic by estimating two independent regression coefficients via data splitting. FDR control is achieved  by taking advantage of the statistic's property
that, for any null feature, its sampling distribution 
is symmetric about zero. Furthermore, 
we propose  Multiple Data Splitting (MDS) to stabilize the selection result and boost the power. Interestingly and surprisingly, with the FDR still under control, MDS not only helps overcome the power loss caused by sample splitting, but also results in a lower variance of the false discovery proportion (FDP) compared with all other methods in consideration.
We prove that the proposed data-splitting methods can asymptotically control the FDR at any designated level for 
linear and Gaussian graphical models in both low and high dimensions. Through intensive simulation studies and a real-data application,
we show that the proposed methods are robust to the unknown distribution of features, easy to implement and computationally efficient, and are often the most powerful ones amongst competitors 
especially when the signals are weak and the correlations or partial correlations are high among features. 
\end{abstract}

\section{Introduction}
\subsection{Background for  FDR control in regression models}
\label{subsec:motivation}
Scientific researchers in the current big data era often have the privilege of collecting or accessing a large number of explanatory features targeting a specific response variable.  For instance, population geneticists often need to  profile thousands of single nucleotide polymorphisms (SNPs) in  genome-wide association studies. A ubiquitous belief is that the response variable depends on  only a small fraction of the collected features. Therefore, researchers are highly interested in identifying these relevant features so that the computability of the downstream analysis, the reproducibility of the reported results, and the interpretability of the scientific findings can be greatly enhanced. 
Throughout the paper, we denote the explanatory features as $\{X_1, \cdots, X_p\}$, with $p$ being potentially large, and denote the response variable as $y$.
Although the methodological developments presented here are in the context of feature selection for regression models, they can also be adapted to solve general multiple testing problems. 

Many advances in  feature selection methods for regression analyses have been made in the past few decades, such as stepwise regressions \citep{efroymson1960multiple}, Lasso regression \citep{tibshirani1996regression}, and  Bayesian variable selection methods \citep{o2009review}. A desired property of a selection procedure is its capability of controlling the number of false positives, which can be mathematically calibrated by the false discovery rate (FDR) \citep{benjamini1995controlling} defined as follows:
\begin{equation}\nonumber
\text{FDR} = \mathbbm{E}[\text{FDP}],\ \ \ \text{FDP} = \frac{\#\{j: j\in S_0,\ j \in \widehat{S}\}}{\#\{j \in \widehat{S}\}\vee 1},
\end{equation}
where $S_0$ denotes the index set of the null features (irrelevant features), $\widehat{S}$ denotes the index set of the selected features, and FDP stands for ``false discovery proportion''. The expectation is taken with respect to  the randomness in both the data and the selection procedure if it is stochastic. 

One popular class of  FDR control methods is based on the Benjamin-Hochberg (BHq) procedure \citep{benjamini1995controlling}.
BHq requires p-values and guarantees exact FDR control when all the p-values are independent.
\citet{benjamini2001control}  generalized BHq to handle dependent p-values. They proved that BHq achieves  FDR control under positive dependence, and is also valid under any arbitrary dependence structure if a shrinkage of the control level by $\sum_{j=1}^p1/j$ is applied.
Further discussions on  generalizing BHq 
can be found in  \citet{2002Sanat} for general stepwise multiple testing procedures with positive dependence, \citet{storey2004strong} for weak dependence, 
\citet{2008WeibiaoWu} and \citet{clarke2009robustness} for Markov models and linear processes.

Another class of methods is based on the  ``knockoff filtering'' idea, which does not require p-values for individual features, and achieves  FDR control by creating  ``knockoff" features in a similar spirit as adding spike-in controls in biological experiments. 
\citet{barber2015controlling}   first proposed the fixed-design knockoff filter, which achieves  exact FDR control for  low-dimensional Gaussian linear models regardless of the dependency structure among features.
The model-X knockoff filter \citep{candes2018panning} further extends the applicability of knockoff filtering to high-dimensional problems, and can be applied without having to know the underlying true relationship between the response and features. However, it requires the exact knowledge of the joint distribution of features. If this distribution is unknown,  \citet{barber2020} showed that the inflation of the FDR is  proportional to the estimation error in the conditional distribution of $X_j$ given $\bm X_{-j}$.\footnote{$\bm X_{-j} = \{X_1, \cdots, X_p\}\backslash \{X_j\}$.}
For details on how to generate good knockoff features, see
\citet{Romano_2019}, \citet{Jordon2019KnockoffGANGK} (using deep generative models) and \citet{2020wenshuo} (using sequential MCMC algorithms). 
\cite{huang2020relaxing} 
generalized the model-X knockoff filter using conditioning to allow features to follow an exponential family distribution with unknown parameters.
Further developments include  the multilayer knockoff filter \citep{katsevich2019multilayer}, which achieves FDR control at both group and individual levels, and DeepPINK \citep{lu2018deeppink}, which models the relationship between the response and features by a neural network.
Successful applications of the knockoff filter in genetics have been reported \citep{2018genehunting,2020Sesiamulti}. 

In this paper, we propose an FDR control framework based on data splitting. Historically, data splitting has been used for evaluating statistical predictions (e.g., cross validation)
\citep{stone1974cross} and selecting efficient test statistics \citep{moran1973dividing, cox1975note}. 
Later, data splitting has been employed to overcome difficulties in statistical inference in high dimensions. For example,
\citet{wasserman2009high} proposed to split the data into three parts to implement a three-stage regression method.
Specifically, the user first fits a suite of candidate models to the first part of the data. The second part of the data is then used to select one of those models based on cross validations. Finally, the null features are eliminated based on hypothesis
testing using the third part of the data. Other practices of data splitting in feature selection/multiple hypotheses testing can be found in 
\citet{rubin2006method} (estimating the optimal cutoff for test statistics) 
and \citet{ignatiadis2016data} (determining  proper weights for individual hypotheses).
More recently,
\citet{barber2019knockoff} extended the applicability of the fixed-design knockoff filter to high-dimensional linear models via data splitting, in which the first part of the data is used to screen out enough null features so that the fixed-design knockoff filter can be applied to the selected features using the second part of the data. 

FDR control originally introduced by \citet{benjamini1995controlling} is formulated as a sampling property of the procedure, and 
all the aforementioned methods, including our proposed ones, take this Frequentist point of view. 
Bayesian views of FDR control have also been studied in the literature, such as  the ``local'' FDR  control method \citep{2005localfdr}, which has been successfully applied to analyze microarray data \citep{2001localfdr}.
The local FDR control framework is more delicate in the sense that it attaches each hypothesis/feature a probabilistic quantification of being null. 
However, it requires accurately estimating the densities of the test statistics, which can be challenging in practice. 
It is worth noting that there is also a Bayesian interpretation of the positive FDR,\footnote{The positive FDR is defined as $\mathbbm{E}[\text{FDP}\mid |\widehat{S}| > 0]$.} as pointed out by  \citet{2003Storeyqvalue}.

\subsection{Motivations and main contributions}
\label{subsec:main-contribution}
In high-dimensional regressions, it can be challenging to either construct  valid p-values (even asymptotically) or estimate accurately the joint distribution of features, thus limiting the applicability of both BHq and the model-X knockoff filter. The data-splitting framework proposed here appears to fill in this gap. Throughout, we use 
DS and MDS to denote the proposed single data-splitting procedure and its refinement, the multiple data-splitting procedure, respectively. Main contributions of this work are summarized as follows.
\begin{itemize}
\item We propose a general FDR control framework based on data splitting, which is both conceptually simple and computationally  efficient. 
Compared to BHq and the model-X knockoff filter, our methods require neither the p-values nor the joint distribution of features. 
\item We propose a general strategy to aggregate the selection results obtained from multiple independent data splits, which stabilizes the selection result and  improves the  power of a single data split.
\item The general FDR control theories of DS and MDS are established in a model-free setting under certain assumptions, which can be further verified for the tasks of feature selection and graph estimation in linear and  Gaussian graphical models under standard conditions. 
\item We empirically demonstrate that DS and MDS  control the FDR at the designated level in all cases we have tested, and MDS achieve the best or nearly the best power  in a wide range of simulation scenarios and real data applications.
\end{itemize}

DS starts by splitting the data into two halves, and then applies two potentially different statistical learning procedures to each part of the data. 
As mentioned in Section \ref{subsec:motivation}, the idea of using data splitting to make valid statistical inferences has been around for some time.
While the main motivation of most existing methods is to handle the high-dimensionality (e.g.,
to obtain valid p-values or apply the fixed-design knockoff filter), we aim at obtaining two independent measurements of the importance of each feature via data splitting.
FDR control is achieved by constructing a proper test statistic for each feature based on these two measurements.

\begin{figure}[h]
\centering
\includegraphics[width=0.8\columnwidth]{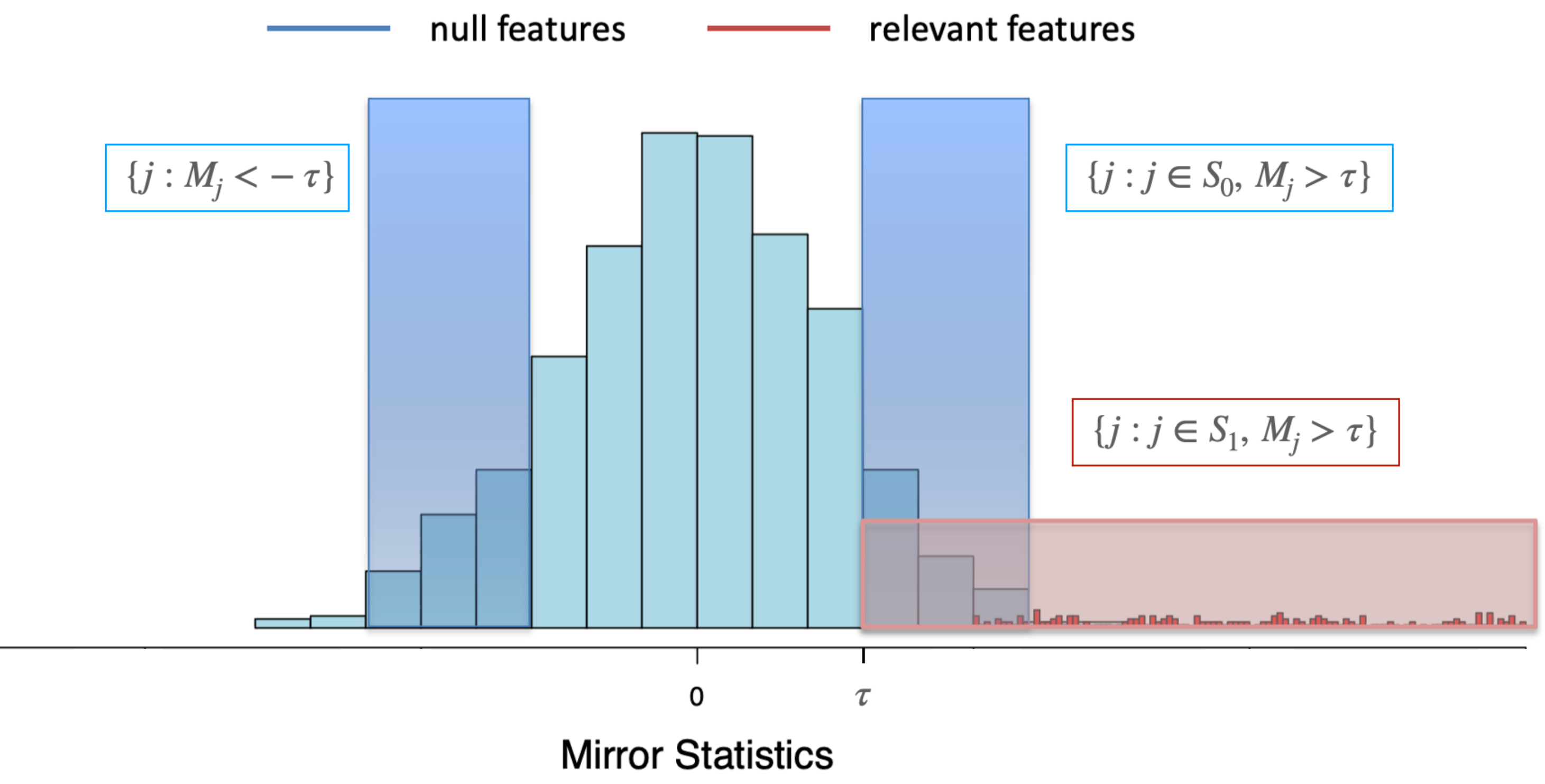}
\caption{A cartoon illustration of the mirror statistic. $M_j$ denotes the mirror statistic of feature $X_j$. $S_0$ and $S_1$ denote the index set of the null features and the relevant features, respectively. Features with mirror statistics larger than the cutoff $\tau$ are selected.}
\label{fig:mirror-statistic-demo}
\end{figure}

Without resorting to p-values, we follow a similar strategy as in the knockoff filter to estimate the number of false positives. The main  idea is to construct
a test statistic $M_j$ for each feature $X_j$,
referred to as the ``mirror statistic"  in \citet{xin2019GM},
which has the following two key properties as illustrated by  Figure \ref{fig:mirror-statistic-demo}.
\begin{enumerate}[({A}1)]
\item A feature with a larger mirror statistic is more likely to be a relevant feature.
\vspace{-0.2cm}
\item The sampling distribution of the mirror statistic of any null feature is symmetric about 0.
\end{enumerate}

Property (A1) suggests that we can rank the importance of each feature by its mirror statistic, and select those features with mirror statistics larger than a cutoff ($\tau$ in Figure \ref{fig:mirror-statistic-demo}). Property (A2) implies that we can estimate (conservatively) the number of false positives, i.e., $\#\{j: j\in S_0,\ M_j > \tau\}$, by $\#\{j: M_j < -\tau\}$, if the mirror statistics of the null features are not too correlated.
As we will see, in our FDR control framework, Property (A1) will be naturally satisfied if the mirror statistic is properly constructed, thus our main concern is Property (A2).

MDS is built upon multiple independent replications of DS, aiming at reducing the variability of the selection result.
Instead of ranking features by their mirror statistics, MDS ranks features by their 
{\it inclusion rates}, which are selection frequencies adjusted by selection sizes, among multiple DS replications.
Empirically, we observe that MDS simultaneously reduces the FDR and boosts the power in most cases, suggesting that MDS yields better rankings of features than DS. 
We provide some useful insights on MDS by analyzing the simple  Normal means model, in which MDS can be shown to achieve nearly the optimal detection power (see Section \ref{subsec:mds-normal-mean}). MDS is conceptually most similar to the stability selection method \citep{meinshausen2010stability}, and a more detailed discussion about them is deferred to Section \ref{subsec:multiple-data-split}. 

We specialize the applications of DS and MDS to linear and Gaussian graphical models, and show that both DS and MDS achieve FDR control under standard conditions including sparsity conditions, regularity conditions on the design matrix, and signal strength conditions.
For high-dimensional linear models, we propose a Lasso + ordinary least squares (OLS) procedure. More precisely, we first screen out some null features by Lasso
using one part of the data, and then run OLS on the selected features using the other part of the data.
Property (A2) is satisfied if all the relevant features are selected in the first step (the so-called sure screening property).
The methods designed for linear models are also applicable to  Gaussian graphical models because of the linear representation of the conditional dependence structure \citep{lauritzengraphical}. Given a nominal level $q$, we apply DS or MDS to each nodewise regression targeting at an FDR control level $q/2$, and then combine the nodewise selection results using the OR\footnote{OR stands for ``or'', i.e., the edge $(i,j)$ is selected in the final graph if vertices $X_i$ and $X_j$ are identified as conditional dependent in either of the nodewise regressions, $X_i$ on $\bm{X}_{-i}$ or $X_j$ on $\bm{X}_{-j}$. See Section \ref{subsec:gaussian-graphical-model} for more details.} rule \citep{meinshausen2006high}.
Numerical experiments show that DS and MDS performed significantly better than two existing methods 
including BHq based on the partial correlation test and the GFC\footnote{GFC stands for Gaussian graphical model estimation with false discovery rate control.} 
method proposed in \citet{liu2013gaussian} specifically for handling graphical models.

The rest of the paper is structured as follows. Section \ref{subsec:single-data-split} introduces DS, with a detailed discussion on the construction of the mirror statistics.
Sections \ref{subsec:multiple-data-split} and \ref{subsec:mds-normal-mean} focus on MDS, in which we show that,  with high probability, the inclusion rate is a monotone decreasing function of the p-value for the Normal means model.
The desired FDR control properties for DS and MDS are also established in Section \ref{sec:DS-MDS} in a model-free setting under certain conditions.
Section \ref{sec: statistical-model} discusses the applications of DS and MDS to linear and Gaussian graphical models. Sections \ref{subsec:sim-linear} and \ref{subsec:sim-GGM} demonstrate the competitive performances of DS and MDS through simulation studies.
Section \ref{sec:HIV} applies DS and MDS to the task of identifying mutations associated with  drug resistance in a HIV-1 data set. Section \ref{sec:conclusion} concludes with a few final remarks. We give proofs and more details on the simulation studies in Supplementary Materials.

\section{Data Splitting for FDR Control}
\label{sec:DS-MDS}
\subsection{Single data splitting}
\label{subsec:single-data-split}
Suppose a set of random features $(X_1, \ldots, X_p)$ follow a $p$-dimensional distribution.
Denote the $n$ independent observations of these features as $\bm{X}_{n \times p} = (\bm{X}_1, \ldots, \bm{X}_p)$, also known as the {\it design matrix}, where $\bm{X}_j = (X_{1j}, \ldots, X_{nj})^{\intercal}$ denotes the vector containing $n$ independent realizations of feature $X_j$. 
We assume that each feature except the intercept with all 1's has been normalized to have zero mean and unit variance. For each set of the observed features $(X_{i1}, \ldots, X_{ip})$, there is an associated response variable $y_i$ for $i \in \{1,\ldots, n\}$. Let $\bm{y} = (y_1, \ldots, y_n)^\intercal$ be the vector of $n$ independent responses.
We assume that the response variable $y$ depends on  only a subset of features $X_{S_1} = \{X_j: j \in S_1\}$, and the task of feature selection is to identify the set $S_1$. 
Let $S_0 = \{1, \ldots, p\}\backslash S_1$ be the index set of the null features.
Let $p_0 = |S_0|$ and $p_1 = |S_1|$ be the number of the null and the relevant features, respectively.

Feature selection  commonly relies on a set of coefficients $\widehat{\bm{\beta}} = (\widehat{\beta}_1, \cdots, \widehat{\beta}_p)^\intercal$
to measure the importance of each feature.
The larger $|\widehat{\beta}_j|$ is, the more likely feature $X_j$ is useful in predicting $y$ (since features have been normalized).
For example, in linear regressions, $\widehat{\bm{\beta}}$ can be the vector of coefficients estimated via OLS or some shrinkage methods.
In contrast to those commonly used approaches that select features based on a single set of coefficients, we  construct two independent sets of coefficients, $\bm{\widehat{\beta}}^{(1)}$ and $\bm{\widehat{\beta}}^{(2)}$, potentially with two different statistical procedures, in order to set up an FDR control framework. 
The independence between $\bm{\widehat{\beta}}^{(1)}$ and $\bm{\widehat{\beta}}^{(2)}$ can be ensured by employing a data-splitting strategy.
More precisely, we split the $n$ observations into two groups, denoted as $(\bm{y}^{(1)}, \bm{X}^{(1)})$ and $(\bm{y}^{(2)}, \bm{X}^{(2)})$. We then estimate $\bm{\widehat{\beta}}^{(1)}$ based on $(\bm{y}^{(1)}, \bm{X}^{(1)})$ and  $\bm{\widehat{\beta}}^{(2)}$ based on $(\bm{y}^{(2)}, \bm{X}^{(2)})$. 
The splitting procedure is flexible, as long as the split is independent of the response vector $\bm{y}$.
The sample sizes for the two groups can also be potentially different. Empirically, we find that the half-half sample splitting leads to the highest power.
To achieve FDR control under our data-splitting framework,  the two sets of coefficients shall satisfy the following assumption besides being independent.
\begin{assumption} 
\label{assump:symmetric}
(Symmetry) For each null feature index $j \in S_0$, the sampling distribution of either $\widehat{\beta}^{(1)}_j$ or $\widehat{\beta}^{(2)}_j$ is symmetric about 0.
\end{assumption}

Note that the symmetry assumption is only required for the null features and can be further relaxed to asymptotic symmetry. Furthermore, for $j\in S_0$, it is sufficient that only one of $\widehat{\beta}^{(1)}_j$ and $\widehat{\beta}^{(2)}_j$ is symmetric about 0.
In Section \ref{sec: statistical-model}, we propose a Lasso + OLS procedure for linear and Gaussian graphical models so that the symmetry assumption can be satisfied with probability approaching 1 under certain conditions. 

Our FDR control framework  starts with the construction of a mirror statistic that satisfies Properties (A1) and (A2) as discussed in Section \ref{subsec:main-contribution}. 
A general form of the mirror statistic $M_j$ is
\begin{equation}
\label{eq:mirror-statistic}
M_j = \text{sign}\big(\widehat{\beta}_j^{(1)} \widehat{\beta}_j^{(2)}\big)f\big(|\widehat{\beta}_j^{(1)}|, |\widehat{\beta}_j^{(2)}|\big),
\end{equation}
where  function $f(u,v)$ is non-negative, symmetric about $u$ and $v$, and monotonically increasing in both $u$ and $v$. For a relevant feature, the two coefficients tend to be large (in the absolute value) and have the same sign if the estimation procedures are reasonably efficient. Since $f(u, v)$ is monotonically increasing in both $u$ and $v$, the corresponding mirror statistic is likely to be positive and relatively large, which implies Property (A1). In addition, the independence between the two coefficients, together with the symmetry assumption, imply Property (A2) as shown by Lemma \ref{lemma:mirror-statistic-symmetry}. 

\begin{lemma}
\label{lemma:mirror-statistic-symmetry}
Under Assumption \ref{assump:symmetric}, regardless of the data-splitting procedure, the sampling distribution of $M_j$ is symmetric about 0 for $j \in S_0$.
\end{lemma}

The proof is elementary and thus omitted.
Three convenient choices of $f(u,v)$ are:
\begin{equation}\label{eq:contrast_choice}
f(u, v) = 2 \min(u,v) ,\ \ \ f(u,v) = uv,\ \ \ f(u, v) = u + v.
\end{equation}
The first choice equals to the mirror statistic 
proposed in \citet{xin2019GM}, and the third choice corresponds to the ``sign-maximum" between $\big|\widehat{\beta}_j^{(1)}+ \widehat{\beta}_j^{(2)}\big|$ and $\big|\widehat{\beta}_j^{(1)}- \widehat{\beta}_j^{(2)}\big|$, and
is optimal in a simplified setting as described in Proposition \ref{prop:optimality-mirror-statistics}.  The optimally of the sign-max mirror statistic  has also been empirically observed by \citet{barber2015controlling}, and been recently proved by \cite{ke2020power} based on a more delicate analysis under the weak-and-rare signal setting.

\begin{proposition}
\label{prop:optimality-mirror-statistics}
Suppose the set of coefficients $(\widehat{\beta}_j^{(1)}, \widehat{\beta}_j^{(2)})$ are independent over the feature index $j\in\{1,\ldots,p\}$. Furthermore, suppose (a) for $j\in S_0$, the two coefficients $\widehat{\beta}_j^{(1)}$ and $\widehat{\beta}_j^{(2)}$  follow $N(0, 1)$ independently; (b) for $j\in S_1$, the two coefficients  follow $N(\omega, 1)$ independently; and (c)  
$p_1/p_0 {\rightarrow} r$ as $p\to \infty$. Then, $f(u, v) = u + v$ is the optimal choice that yields the highest power.
\end{proposition}

Proposition \ref{prop:optimality-mirror-statistics} still holds if the set of coefficients $(\widehat{\beta}_j^{(1)}, \widehat{\beta}_j^{(2)})$ are only weakly correlated over the feature index $j$, as long as the following law of large numbers is satisfied:
\begin{equation}\nonumber
\lim_{p\to\infty}\frac{\#\{j: j \in S_0, j \in \widehat{S}\}}{\# \{j: j \in \widehat{S}\}} = \frac{\mathbbm{P}(j \in \widehat{S}\mid  j\in S_0)}{\mathbbm{P}(j \in \widehat{S} \mid  j\in S_0) + r\mathbbm{P}(j \in \widehat{S} \mid j\in S_1)}.
\end{equation}
The proof of Proposition \ref{prop:optimality-mirror-statistics} (see Supplementary Materials) might be of general interest. We  rephrase the FDR control problem under the hypothesis testing framework, and prove the optimality using the Neyman-Pearson lemma.
The form $f(u, v) = u + v$ is derived based on the rejection rule of the corresponding likelihood ratio test. 
For linear models in more realistic settings, we empirically compare the performances of the three choices of $f(u, v)$ listed in  \eqref{eq:contrast_choice} in Section \ref{subsec:sim-linear}.

The symmetric property of the mirror statistics for the null features gives us an upper bound of the number of false positives:
\begin{equation}
\label{eq:estimate-false-positives}
\# \{j \in S_0: M_j > t\} \approx \# \{j \in S_0: M_j < -t\} \leq \# \{j: M_j < -t\},\ \ \ \forall \ t > 0.
\end{equation}
The $\text{FDP}(t)$  of the selection $\widehat{S}_t = \{j: M_j >t\}$, as well as an ``over estimate" of it, referred to as $\widehat{\text{FDP}}(t)$ in the following, are thus given by
\begin{equation}\nonumber
\text{FDP}(t) = \frac{ \# \{j: M_j > t,\ j\in S_0\}}{ \# \{j: M_j > t\} \vee 1},\ \ \ \ \ \ \widehat{\text{FDP}}(t) = \frac{ \# \{j: M_j < -t\}}{ \# \{j: M_j > t\} \vee 1}.
\end{equation}
For any designated FDR control level $q\in(0, 1)$, we can choose the data-driven cutoff $\tau_q$ as follows:
\begin{equation}\nonumber
\tau_q = \min\{t > 0: \widehat{\text{FDP}}(t) \leq q\},
\end{equation}
and the final selection is $\widehat{S}_{\tau_q} = \{j: M_j > \tau_q\}$. The proposed FDR control procedure is summarized in Algorithm \ref{alg:FDR-data-splitting}.

\begin{algorithm*}
\caption{False discovery rate control via a single data split}
\label{alg:FDR-data-splitting}
\begin{enumerate}
\item Split the data into two groups $(\bm{y}^{(1)}, \bm{X}^{(1)})$ and $(\bm{y}^{(2)}, \bm{X}^{(2)})$, independent to the response vector $\bm{y}$.
\item Estimate the ``impact'' coefficients $\bm{\widehat{\beta}}^{(1)}$ and $\bm{\widehat{\beta}}^{(2)}$ on each part of the data. The two estimation procedures can be potentially different.
\item Calculate the mirror statistics following \eqref{eq:mirror-statistic}.
\item Given a designated FDR level $q \in (0, 1)$, calculate the cutoff $\tau_q$ as:
\begin{equation}
\label{eq:select_cutoff}
\tau_q = \min\left\{t > 0: \widehat{\text{FDP}}(t) = \frac{ \# \{j: M_j < -t\}}{ \# \{j: M_j > t\} \vee 1} \leq q\right\}.
\end{equation}
\item Select the features $\{j: M_j > \tau_q\}$.
\end{enumerate}
\end{algorithm*}

In order to obtain a good estimate of the number of false positives using  \eqref{eq:estimate-false-positives}, the mirror statistics of the null features cannot be too correlated. Mathematically, we require the following weak dependence assumption.

\begin{assumption}
\label{assump:weak-dependency}
(Weak dependence among the null features) 
The mirror statistics $M_j's$ are continuous random variables, and there exist  constants $c > 0$ and $\alpha \in (0, 2)$ such that 
\begin{equation}\nonumber
\textnormal{Var}\bigg(\sum_{j\in S_0}\mathbbm{1}(M_j > t)\bigg) \leq cp_0^\alpha,\ \ \forall\ t \in \mathbbm{R}, \ \ \mbox{where } p_0=|S_0| .
\end{equation}
\end{assumption}

Assumption \ref{assump:weak-dependency} only restricts  the correlations among the null features, regardless of the correlations associated with the relevant features. We note that if the  mirror statistics of the null features have constant pairwise correlations, or can be clustered into a fixed number of groups (say, 2) so that their within-group correlation is a constant, $\alpha$ has to be 2 and Assumption \ref{assump:weak-dependency} does not hold. Except for these extreme cases, we believe that Assumption \ref{assump:weak-dependency}  holds in  fairly broad settings.
For example, in Section \ref{subsec:linear-model}, we show that for linear models, the weak dependence assumption holds as long as the covariance matrix of the null features satisfies some regularity condition (e.g., the eigenvalues are doubly bounded). Empirically, we observed that even in the case where the null features have constant pairwise correlations, our methods still perform very well, often outperforming BHq and the knockoff filters (see Figure \ref{fig:normal-design-constant-correlation} in Supplementary Materials).  

Recall that $\text{FDP}(t)$ refers to the FDP of the selection $\widehat{S}_t = \{j: M_j >t\}$. 
We assume that the variances of the mirror statistics do not diverge to infinity and are also bounded away from 0.
The proposition below shows that for any nominal level $q\in(0, 1)$, $\text{FDP}(\tau_q)$ and the corresponding $\text{FDR}(\tau_q)$ are under control, in which $\tau_q$ is the data-dependent cutoff chosen following \eqref{eq:select_cutoff}. 

\begin{proposition}
\label{prop:FDR}
For any designated FDR control level $q \in (0,1)$, assume that there exists a constant $t_q > 0$ such that
$\mathbbm{P}(\text{FDP} (t_q)\leq q)\to 1$ as $p\to\infty$.
Then, under Assumptions \ref{assump:symmetric} and  \ref{assump:weak-dependency}, 
the procedure in Algorithm~\ref{alg:FDR-data-splitting} satisfies
\begin{equation}\nonumber
\text{FDP}(\tau_q) \leq q + o_p(1)\ \ \ \text{and}\ \ \ \limsup_{p\to\infty}\text{FDR}(\tau_q) \leq q.
\end{equation}
\end{proposition}

We note that the existence of $t_q > 0$ such that $\mathbbm{P}(\text{FDP} (t_q)\leq q)\to 1$ 
essentially guarantees the asymptotic feasibility of  FDR control based upon  the rankings of features by their mirror statistics.
Specifically, it implies that the data-dependent cutoff $\tau_q$ is bounded
with probability approaching 1, thus does not diverge to infinity.
It is  a technical assumption for handling the most general setting without specifying a parametric model between the response and features.
When we work with specific models such as linear or Gaussian graphical models, this technical assumption is no longer required (see Section \ref{subsec:linear-model}).
Similar assumptions also appear in \citet{storey2004strong} and \citet{2008WeibiaoWu} in order to achieve a high level of generality.

In Assumption \ref{assump:symmetric}, the exact symmetry can be relaxed to asymptotic symmetry. 
Suppose for $j \in S_0$, the sampling distribution of either $\widehat{\beta}^{(1)}_j$ or $\widehat{\beta}^{(2)}_j$ is asymptotically symmetric about 0. In addition, the asymptotic symmetry is uniform over $j\in S_0$ in the sense that the resulting mirror statistics satisfy the following condition,
\begin{equation}\nonumber
\max_{j\in S_0}\left|\mathbbm{P}(M_j > t) - \mathbbm{P}(M_j < -t)\right| \to 0, \ \ \ \forall\ t.
\end{equation}
Then Proposition \ref{prop:FDR} still holds. 
As for high-dimensional generalized linear models, one way to construct the mirror statistic is to use the debiased Lasso estimator \citep{van2014asymptotically, zhang2014confidence, javanmard2014confidence}, which is asymptotically normal, and therefore symmetric. Furthermore, under certain conditions, the bias in the debiased Lasso estimator vanishes uniformly over the features. 
Thus, the proposed methodologies are applicable, and we refer the readers to \citet{dai2020scale} for more details.

Before concluding this section, we remark that DS is inspired by the recently proposed Gaussian mirror method \citep{xin2019GM}, which
perturbs the features one by one and examines the corresponding impact. 
Compared to the Gaussian mirror method, DS  is easier to implement and computationally more efficient especially for large $n$ and $p$. For linear models, the Gaussian mirror method  requires $p$ linear fittings. In contrast, DS perturbs all the features simultaneously by randomly splitting the data into two halves, 
thus requiring only two linear fittings.
The gain of the computational efficiency can be more significant for Gaussian graphical models (see Section \ref{subsec:gaussian-graphical-model}). DS requires $2p$ nodewise linear fittings, whereas
the Gaussian mirror method would require $p^2$ nodewise linear fittings, which is generally unacceptable when $p$ is large.
In addition, since DS is conceptually simpler, it is more convenient to adapt DS to other statistical models.

\subsection{Multiple data splitting}
\label{subsec:multiple-data-split}
There are two major concerns about DS. First, splitting the data into two halves inflates the variances of the estimated regression coefficients, thus DS can potentially suffer  a power loss in comparison with competing methods  that properly use the full data. Second, the selection result of DS may not be stable and can vary substantially across different sample splits. 

To remedy these issues, we propose a multiple data-splitting (MDS) procedure to aggregate the selection results obtained from independent replications of DS.
For linear and Gaussian graphical models, we prove that MDS achieves FDR control under certain conditions.
Simulation results in Section \ref{sec:simulation} confirm FDR control of MDS and demonstrate a fairly universal power improvement of MDS over DS.
Going beyond these two models, we empirically found that MDS can work competitively for a much wider class of models.
MDS is also generally applicable without requiring p-values or any knowledge regarding the joint distribution of features.

Given $(\bm{X}, \bm{y})$, suppose we independently repeat DS $m$ times with random sample splits. Each time the set of selected features is denoted as $\widehat{S}^{(k)}$ for $k=1,\ldots, m$. For each feature $X_j$, we define the associated  {\it inclusion rate} $I_j$ and its estimate $\widehat{I}_j$ as 
\begin{equation} \label{eq:inclustion-rate}
I_j = \mathbbm{E}\left[\frac{\mathbbm{1}(j \in \widehat{S})}{|\widehat{S}|\vee 1} \biggm| \bm{X}, \bm{y}\right], \ \ \ \widehat{I}_j = \frac{1}{m}\sum_{k = 1}^m\frac{\mathbbm{1}(j \in \widehat{S}^{(k)})}{|\widehat{S}^{(k)}|\vee 1},
\end{equation}
in which the expectation is taken with respect to the randomness in data splitting. Note that this rate is not an estimate of the selection probability, but rather an importance measurement of each feature relative to the DS selection procedure. For example, in the case where feature $X_j$ is always selected by DS and DS always selects 20 features across $m$ random sample splits, the inclusion rate $I_j$ equals to 1/20.  MDS is most useful if the following informal statement is approximately true: if a feature is selected less frequently in the repeated sample splitting, it is less likely to be a relevant feature. In other words, the rankings of features by the inclusion rates should roughly reflect the importance of features.
If this holds, we can choose a proper inclusion-rate cutoff to control the FDR, and select those features with inclusion rates larger than the cutoff. 

The way of choosing a proper inclusion-rate cutoff is detailed in Algorithm \ref{alg:multiple-splits} and briefly discussed here. 
Let the sorted inclusion rate estimates  be $0 \leq \widehat{I}_{(1)} \leq \widehat{I}_{(2)} \leq \cdots \leq \widehat{I}_{(p)}$. 
Proposition \ref{prop:FDR} suggests a backtracking approach to select the cutoff based on the following argument: if we had $m$ independent sets of data $(\bm{X}, \bm{y})$ and applied DS to all of them for feature selection, the average FDP would be (asymptotically) no larger than the designated FDR control level $q$. 
Although it is not possible to generate new data, we can consider $\{\widehat{S}^{(k)},\ k=1,\ldots,m\}$ as an approximation to $m$ independent selection results obtained via data regeneration.
We thus find the largest cutoff such that, if we assume that the features with inclusion rates larger/smaller than the cutoff are ``true'' relevant/null features, respectively, the average FDP among $\{\widehat{S}^{(k)},\ k = 1,\ldots, m\}$ is no larger than $q$. Empirically, we find that MDS often results in a  lower FDR than the nominal level but still enjoys a competitive power. The proposition below gives some intuitions regarding how MDS is guaranteed to control the FDR properly. 

\begin{algorithm*}
\caption{Aggregating selection results from multiple data splits.}
\label{alg:multiple-splits}
\begin{enumerate}
\item Sort the estimated inclusion rates (see (\ref{eq:inclustion-rate})):  $0 \leq \widehat{I}_{(1)} \leq \widehat{I}_{(2)} \leq \cdots \leq \widehat{I}_{(p)}$.
\item Find the largest $\ell \in \{1,\ldots, p\}$ such that
$\widehat{I}_{(1)} + \cdots + \widehat{I}_{(\ell)} \leq q$.
\item Select the features $\widehat{S} = \{j: \widehat{I}_j > \widehat{I}_{(\ell)}\}$.
\end{enumerate}
\end{algorithm*}

\begin{proposition}
\label{prop:MDS-FDR} 
Suppose we can asymptotically control the FDP of DS for any designated level $q\in(0, 1)$.\footnote{In the sparse regime, we shall asymptotically control the FDP of DS at some level $q'<q$ in order to bypass the technical difficulties. Since $q'$ can be arbitrarily close to $q$, this modification is purely for the technical purpose and has almost no practical implications.}
Furthermore, we assume that with probability approaching 1, the power of DS is  bounded below by some $\kappa > 0$.
We consider the following two regimes with $n, p\to\infty$ at a proper rate.
\begin{enumerate}
\item[(a)] In the non-sparse regime where
$\liminf p_1 / p > 0$,  we assume that the mirror statistics are consistent at ranking features, i.e., $\sup_{i\in S_1, j\in S_0 }\mathbbm{P}(I_i<I_j)\to 0$.
\item[(b)] In the sparse regime where $\limsup p_1/p=0$,  we assume that the mirror statistics are  strongly consistent at ranking features, i.e., 
$\sup_{i\in S_1}\mathbbm{P}(I_i<\max_{j\in S_0} I_j) \to 0$. 
\end{enumerate}
Then, for MDS (see Algorithm \ref{alg:multiple-splits}) in both the non-sparse and the sparse regimes, we have
\begin{equation}\nonumber
\text{FDP} \leq q + o_p(1)\ \ \ \text{and}\ \ \ \limsup_{n,p\to\infty}\text{FDR} \leq q.
\end{equation}
\end{proposition}

Although some of the conditions in Proposition \ref{prop:MDS-FDR} are not explicit enough or directly verifiable without imposing a specific model and the associated assumptions, the proposition points out a key factor for MDS to achieve FDR control: the ranking consistency of the baseline algorithm. 
More precisely, assuming that the mirror statistics are consistent at ranking features (see Proposition \ref{prop:MDS-FDR}(a)), we can show that the number of false positives is in the order of $o_p(p_0)$. In the non-sparse regime, this leads to the desired FDR control property for MDS as $p_1$ and $p_0$ are in the same order. 
In the sparse regime, since $p_0\gg p_1$, we require a stronger ranking consistency condition
(see Proposition \ref{prop:MDS-FDR}(b)), under which we can show that the number of false positives is in the order of $o_p(p_1)$.
In Section \ref{sec: statistical-model}, we show that the ranking consistency condition holds for linear and Gaussian graphical models under more explicit conditions.

The idea of using data perturbation and replicating the procedure multiple times to stabilize the selection results is not new. For example, \citet{meinshausen2010stability} proposed a stability selection method, which   perturbs the data via subsampling and runs a feature selection algorithm multiple times across a set of regularization parameters. The final selection set only contains  ``stable" features, of which the selection frequencies under different magnitudes of regularization are above some user-defined threshold. 
Compared with the stability selection method, the motivation of MDS is  different. The stability selection method aims at overcoming the difficulty of finding a proper regularization parameter in  high-dimensional regression, whereas MDS is designed to stabilize DS and compensate for its power loss due to sample splitting.
Theoretically, under certain conditions, the stability selection method provides some finite-sample bound on the  number of false positives, whereas MDS asymptotically controls the perhaps more delicate FDR. Indeed, MDS requires a careful selection of the inclusion-rate cutoff in order to achieve FDR control. In contrast, for the stability selection method, the corresponding selection-probability cutoff can be much less stringent.
Furthermore, the two methods perturb the data in different ways. For each regularization parameter, 
the stability selection method obtains a collection of selection sets using different sub-samples of the data, whereas MDS always uses the full data, but replicates DS with independent sample splits so as to obtain multiple selection results.

There are also some relevant 
works on p-value aggregation in high-dimensional settings.
For example, 
\citet{meinshausen2009p} proposed to obtain a collection of p-values via repeated sample splitting. The multiple p-values of each feature are then aggregated  by choosing a proper quantile among them. After that, BHq can be applied to control the FDR. 
Empirically, we found that the resulting procedure is often too conservative, with a near-zero FDR but also a sub-optimal power (see Section \ref{subsec:sim-linear}).
For other related works, we refer the readers to \citet{van2009testing} and \citet{romano2019multiple}.

\subsection{A theoretical study of MDS for the Normal means model}
\label{subsec:mds-normal-mean}
We consider the simple Normal means model to gain some insights on how MDS compensates for the power loss of DS due to sample splitting.
For $i=1,\ldots, n$ and $j=1,\ldots,p$, we assume that $X_{ij}$ follows $N(\mu_j, 1)$ independently.
To test whether $\mu_j$ is 0, the p-value is given by $p_j = 2\Phi(-|\sqrt{n}\bar{\bm{X}_j}|)$, where $\bar{\bm{X}}_j = \sum_{i = 1}^n X_{ij}/n$, and $\Phi$ is the CDF of the standard Normal distribution. 

For DS, we construct the mirror statistic $M_j$ using the sample means of $\bm{X}_j^{(1)}$ and $\bm{X}_j^{(2)}$, and select $\widehat{S}$, i.e., reject the null hypotheses that $\mu_j$'s are 0, following Algorithm \ref{alg:FDR-data-splitting}. For MDS, we replicate DS $m$ times and estimate the inclusion rates following \eqref{eq:inclustion-rate}.
Proposition \ref{prop:rank-DS-MDS} below holds for any designated FDR control level $q \in (0, 1)$, and for all three choices of $f$ detailed in \eqref{eq:contrast_choice} for constructing the mirror statistics. For simplicity, we only prove the case for 
\begin{equation}
\label{eq:mirror-statistics-normal-means}
M_j = |\bar{\bm{X}}_j^{(1)}+\bar{\bm{X}}_j^{(2)}|-|\bar{\bm{X}}_j^{(1)}-\bar{\bm{X}}_j^{(2)}|.
\end{equation}

\begin{proposition}
\label{prop:rank-DS-MDS}
For any pair $(i, j)$ and two arbitrary  constants $0<c < c^\prime$, as $n \to \infty$, we have 
\begin{equation}\nonumber
\mathbbm{P}\left(M_i < M_j \mid c\leq \sqrt{n}(|\bar{\bm{X}}_i| - |\bar{\bm{X}}_j|\right)\leq c^\prime) \geq \gamma\ \ \mbox{and} \ \ \mathbbm{P}(I_i < I_j \mid c\leq \sqrt{n}(|\bar{\bm{X}}_i| - |\bar{\bm{X}}_j|)\leq c^\prime)  = o_p(1),
\end{equation}
in which $\gamma$ is a strictly positive constant depending on $c$ and $c^\prime$.
\end{proposition}

\begin{remark}
Note that the p-value $p_i$ is a monotone decreasing function of the sufficient statistic $|\bar{\bm{X}}_i|$.
Proposition \ref{prop:rank-DS-MDS} shows that for any pair $\mu_i$ and $\mu_j$ that have a fairly close separation between their p-values $p_i$ and $p_j$, i.e., $|\bar{\bm{X}}_i| - |\bar{\bm{X}}_j|=O_p(1/\sqrt{n})$, DS ranks $\mu_i$ and $\mu_j$ differently from their p-values with a non-vanishing probability, whereas MDS ranks them  consistently with their p-values with probability approaching 1.
Imagine a  perfect knockoff  procedure for this Normal means problem, which
ranks $\mu_j$'s using the knockoff statistic $|\bar{\bm{X}}_j| - |\bar{\bm{X}}^{\prime}_j|$ with $\bar{\bm{X}}^{\prime}_j$ being the  mean of $n$ independent samples from $N(0, 1)$.
Based on the same argument, we can show that the knockoff statistics also rank $\mu_i$ and $\mu_j$  differently from  their p-values with a non-vanishing probability if $|\bar{\bm{X}}_i| - |\bar{\bm{X}}_j|=O_p(1/\sqrt{n})$.
\end{remark}

\begin{figure*}
\centering
\includegraphics[width=0.4\columnwidth]{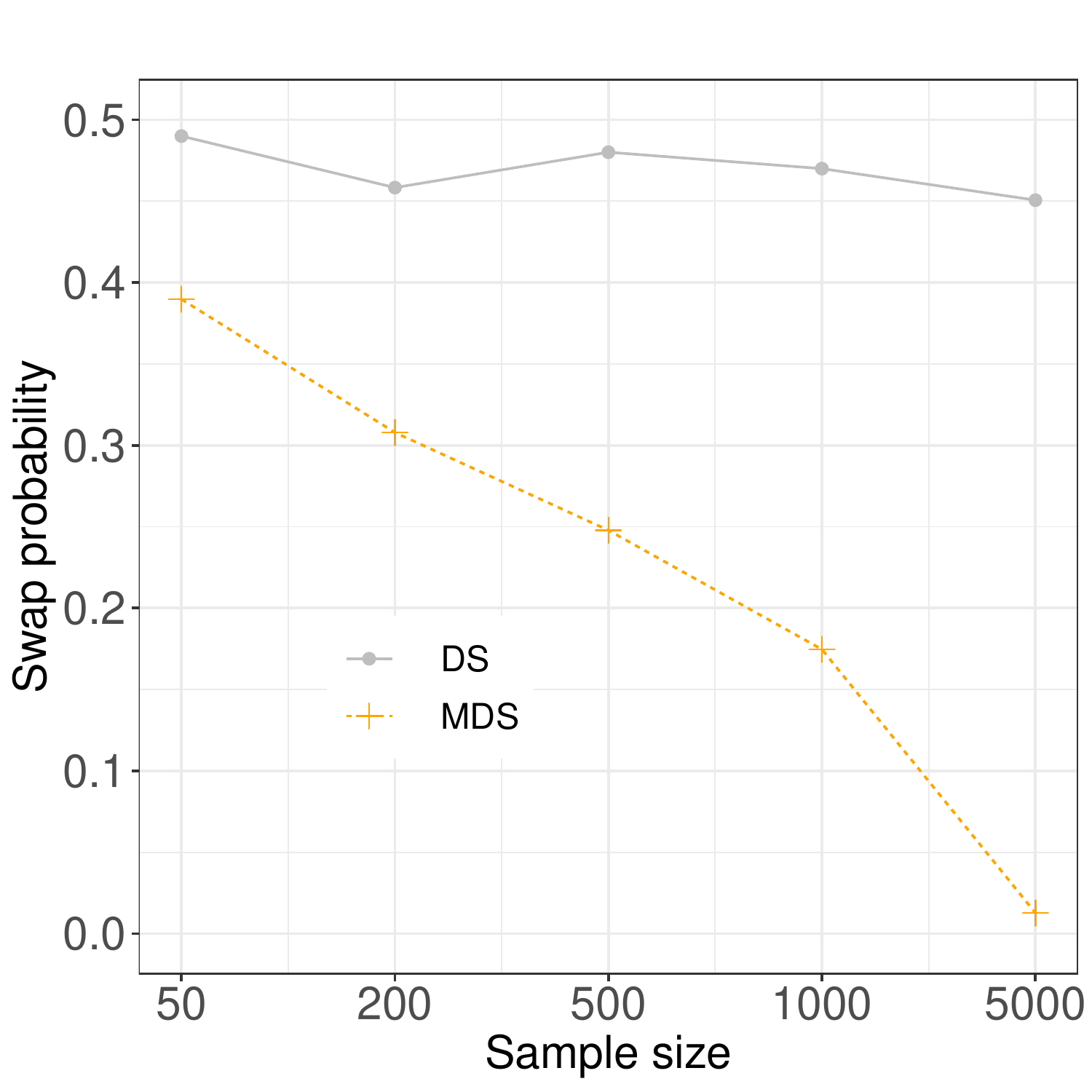}
\hspace{0.3cm}
\includegraphics[width=0.445\columnwidth]{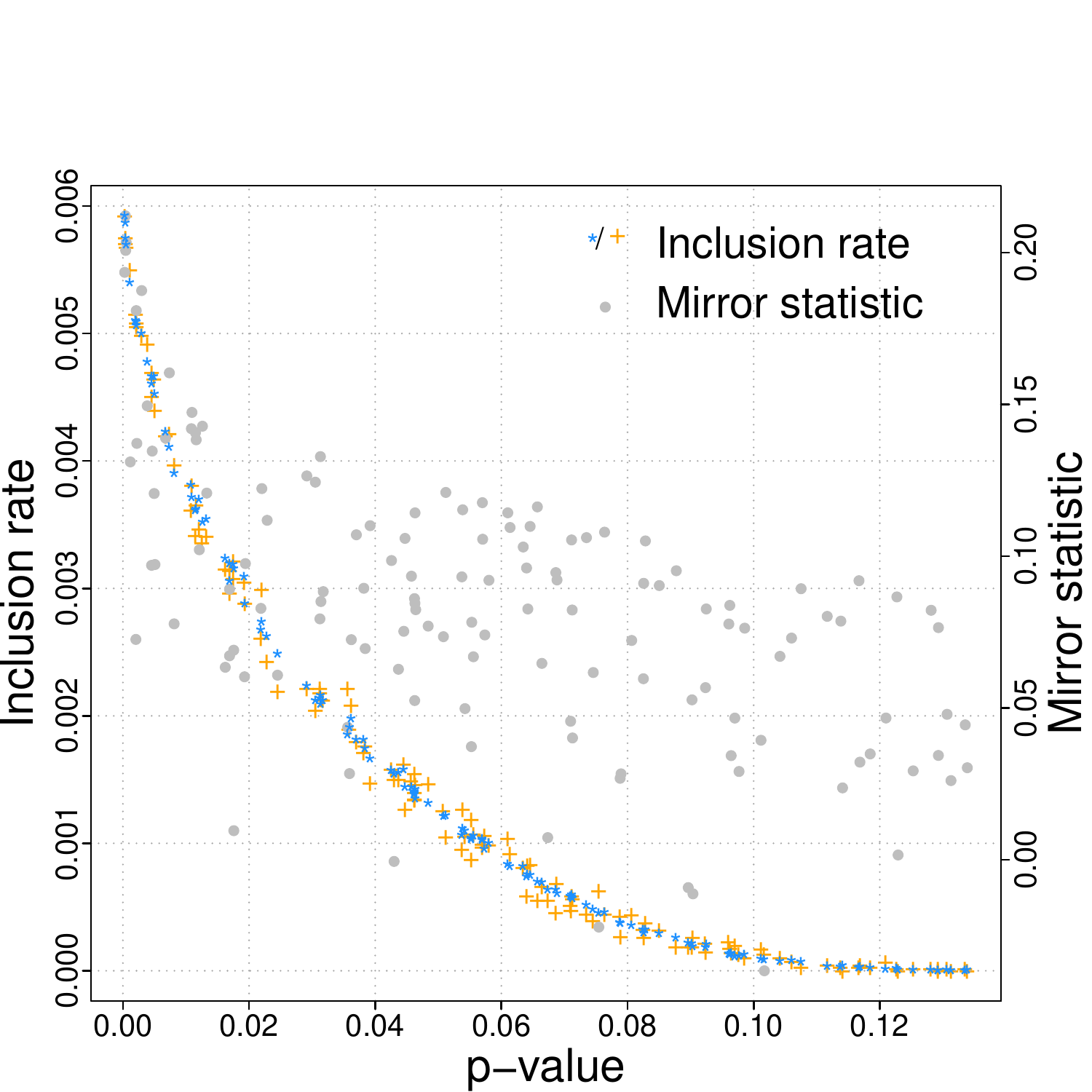}
\caption{Comparison of DS and MDS on the Normal means model. Given the two p-values $p_1, p_2$ with $p_1 < p_2$, the left panel plots the estimated swap probability, i.e., $\mathbbm{P}(M_1 < M_2)$ for DS and $\mathbbm{P}(I_1 < I_2)$ for MDS, against the sample size $n$. The right panel plots the inclusion rates of MDS and the mirror statistics of DS against the p-values. The detailed simulation setting can be found in Section \ref{subsec:mds-normal-mean}.}
\label{fig:MDS-normal-mean}
\end{figure*}

To illustrate Proposition \ref{prop:rank-DS-MDS}, we fix 
$p = 800$, and consider a weak separation between $p_1$ and $p_2$ by setting $p_1 < p_2$ with $p_1 = 0.020$ and $p_2 = 0.021$. That is, we sample $X_{i1}$'s conditioning on $\bar{\bm{X}}_1 = |\Phi(0.01)/\sqrt{n}|$ and 
$X_{i2}$'s conditioning on $\bar{\bm{X}}_2 = \bar{\bm{X}}_1 - 0.02/\sqrt{n}$.
For $j \geq 3$, we set $20\%$ of $\mu_j$'s to be nonzero, and sample them independently from $N(0, 0.5^2)$. 
We vary the sample size $n \in \{50, 200, 500, 1000, 5000\}$, and estimate the swap probability $\mathbbm{P}(M_1 < M_2)$ for DS and $\mathbbm{P}(I_1 < I_2)$ for MDS over 500 independent runs. For DS, we construct the mirror statistics following \eqref{eq:mirror-statistics-normal-means}. For MDS, we set the number of DS replications to be $m = 10n$.
The results in Figure \ref{fig:MDS-normal-mean} empirically validate Proposition \ref{prop:rank-DS-MDS}. The left panel shows that for MDS, the swap probability $\mathbbm{P}(I_1 < I_2)$ gets very close to 0 when the sample size is large enough (say, $n \geq 5000$). However, for DS, the swap probability $\mathbbm{P}(M_1 < M_2)$  approximately remains as  a constant (slightly below 0.5) as the sample size increases. 

Proposition \ref{prop:rank-DS-MDS} also implies that for this simple model, when the sample size is reasonably large, the inclusion rates and the p-values nearly yield the same rankings of $\mu_j$'s with high probability.  
To illustrate this, we consider a similar simulation setting as above with the sample size $n = 1000$, but without fixing $\bar{\bm{X}}_1$ or $\bar{\bm{X}}_2$.
In the right panel of Figure \ref{fig:MDS-normal-mean}, we plot the inclusion rates (blue ``$\ast$" and orange ``+") of MDS and the mirror statistics of DS (grey ``$\cdot$") against the p-values. 
For MDS, the blue ``$\ast$" and the orange ``+" refer to the estimated inclusion rates based upon $m = 10,000$ and $m = 400$ DS replications, respectively.
We see that the rankings of $\mu_j$'s given by the inclusion rates are significantly less noisy compared to the rankings by the mirror statistics, 
and the inclusion rate is approximately a monotone decreasing function of the p-value.  Thus, for this simple model, MDS almost recovers the power loss of DS due to sample splitting since
the p-values, which are calculated using the full data,  summarize all the information related to the testing task. 
Figures \ref{fig:normal-mean-power-splits} and \ref{fig:normal-mean-ROC} in Supplementary Materials provide more empirical comparisons between DS, MDS and BHq across various signal strengths.

Ideally, we would like to conduct as many sample splits as possible for MDS
so as to estimate the inclusion rates accurately. In practice, however, we find that the power of MDS no longer improves much after  a relatively small number of DS replications (say, $m=$ 100 or 200).
See Figure \ref{fig:normal-mean-power-splits} in Supplementary Materials and Figure \ref{fig:num_splits} in Section \ref{subsec:sim-linear} for some empirical evidences on the Normal means model and linear models, respectively.
For the Normal means model, Figure \ref{fig:MDS-normal-mean} shows that increasing $m$ from 400 to 10,000 leads to slightly less noisy rankings of $\mu_j$'s by the inclusion rates.

\section{Specializations for Different Statistical Models}
\label{sec: statistical-model}
In this section, we discuss how we construct the coefficients $\bm{\widehat{\beta}}$ for linear and Gaussian graphical models. Our main concerns are that (1) the coefficients shall satisfy the symmetry assumption (Assumption \ref{assump:symmetric}); (2) the mirror statistics of the null features are weakly correlated (Assumption \ref{assump:weak-dependency}). Throughout this section, we split the data into two parts of equal size.

\subsection{Linear models}
\label{subsec:linear-model}
Assuming that the true data generating process is $\bm{y} = \bm{X}\bm{\beta}^\star + \bm{\epsilon}$, where $\bm{\epsilon}\sim N(\bm{0}, \sigma^2 I_n)$, we consider the random-design scenario, in which each row of the design matrix $\bm{X}$ independently follows a $p$-dimensional distribution with a covariance matrix $\Sigma$.
In the context of feature selection, the true coefficient $\bm{\beta}^\star$ is often assumed to be sparse, and the goal is to identify the set $S_1 = \{j, \ \beta^\star_j \neq 0\}$. 

We   consider a Lasso + OLS procedure described as follows.
On the first half of the data $(\bm{y}^{(1)}, \bm{X}^{(1)})$, we run Lasso for dimension reduction. 
Let $\widehat{\bm{\beta}}^{(1)}$ be the estimated regression coefficients and denote $\widehat{S}^{(1)} = \{j,\ \widehat{\beta}^{(1)}_j \neq 0\}$.
Being restricted to the subset of features $\widehat{S}^{(1)}$ selected by Lasso, we run OLS using the second half of the data $(\bm{y}^{(2)}, \bm{X}^{(2)})$ to obtain the estimated coefficients $\widehat{\bm{\beta}}^{(2)}$. We then construct the mirror statistics by (\ref{eq:mirror-statistic}) using $\bm{\widehat{\beta}}^{(1)}$ and $\bm{\widehat{\beta}}^{(2)}$.

The symmetry assumption is satisfied if the sure screening property holds for Lasso, that is, all the relevant features are selected by Lasso in the first step. If this is the case, for any  selected null feature $j\in S_0\cap \widehat{S}^{(1)}$, its second coefficient
$\widehat{\beta}_j^{(2)}$ follows a centered Normal distribution conditioning on $\bm{X}^{(2)}$, thus is symmetric about 0. For Lasso, sufficient conditions for the sure screening property have been well established in the literature (e.g., see Remark \ref{remark:Lasso-sure-screening}). More generally, we can substitute Lasso by any other dimension reduction method as long as the sure screening property holds with probability approaching 1. 
For the weak dependence assumption, we calibrate the correlation structure among the mirror statistics using Mehler's identity \citep{kotz2000bivariate}, and show that the weak dependence assumption holds with probability approaching 1 under the regularity condition and the tail condition in Assumption \ref{assump:linear}.
Proposition \ref{prop:linear-FDR} establishes the desired FDR control property for both DS and MDS.

\begin{assumption}
\label{assump:linear}
\text{}
\begin{enumerate}
\item (Signal strength condition) $\min_{j\in S_1}|\beta^\star_j|\gg \sqrt{p_1\log p/n}$.
\item (Regularity condition) $1/c < \lambda_{\min}(\Sigma) \leq \lambda_{\max}(\Sigma) < c$ for some $c > 0$.
\item (Tail condition) $\bm X\Sigma^{-1/2}$ has independent sub-Gaussian rows.
\item (Sparsity condition) $p_1 = o(n/\log p)$.
\end{enumerate}
\end{assumption}

\begin{proposition}
\label{prop:linear-FDR}
Consider both DS and MDS, of which the two regression coefficients $\widehat{\bm{\beta}}^{(1)}$ and $\widehat{\bm{\beta}}^{(2)}$ are constructed using the Lasso + OLS procedure. For any designated FDR control level $q \in (0,1)$, 
under Assumption \ref{assump:linear}, we have
\begin{equation}\nonumber
\limsup_{n, p \to\infty}\text{FDR} \leq q\ \ \ \text{and}\ \ \ \liminf_{n, p \to\infty}\text{Power} = 1
\end{equation}
in the asymptotic regime where $\log p = o(n^\xi)$ for some $\xi\in(0, 1)$.
\end{proposition}

\begin{remark}
\label{remark:Lasso-sure-screening}
The sure screening property is implied by the signal strength condition and the compatibility condition \citep{vandegeer2009}.
The compatibility condition means that the sample covariance matrix $\widehat\Sigma$  of features satisfies $\phi(\widehat\Sigma, S_1)\geq\phi_0$ for some $\phi_0>0$, in which $\phi(\widehat\Sigma, S)$ is defined for any subset $S\subseteq \{1,\ldots,p\}$ as
\begin{equation}\nonumber
    \phi^2(\widehat \Sigma, S) = \min_{\bm\theta\in \mathbbm{R}^p}\left\{\frac{|S|\bm\theta^\intercal\widehat\Sigma\bm\theta}{||\bm\theta_S||_1^2}:\ \bm\theta\in\mathbbm{R}^p,\ ||\bm\theta_{S^c}||_1\leq 3||\bm\theta_S||_1 \right\}.
\end{equation}
By Theorem 2.4 in \citet{javanmard2014confidence}, if the regularity condition and the tail condition in Assumption \ref{assump:linear} hold, the compatibility condition holds with high probability for $n\geq cp_1\log(p/p_1)$. Furthermore, assuming that the compatibility condition holds, with a properly chosen regularization parameter, the Lasso regression coefficients $\widehat{\bm\beta}$ satisfy
\begin{equation}\nonumber
    ||\widehat{\bm\beta}-{\bm\beta}^\star||_2 = o_p(\sqrt{p_1\log p/n}).
\end{equation}
Together with the signal strength condition, we see that the sure screening property holds with probability approaching 1. 
The sure screening property also appears crucial in many other related methods, e.g., see \citet{barber2019knockoff} and \citet{rank2020fan} for the knockoff filter. 
\end{remark}

Besides Proposition \ref{prop:linear-FDR}, more detailed power analyses of DS and MDS are still unknown and await future investigations.
In comparison, some theoretical studies on the power of knockoff filters begin to appear recently.
For example, \citet{rank2020fan} showed that, under a similar signal strength condition, i.e., $\min_{j\in S_1}|\beta_j^\star|\gg \sqrt{\log p/n}$, 
and when features  follow a multivariate Normal distribution with known covariance matrix, the model-X knockoff filter has asymptotic power one. 
Moving beyond this ideal scenario, i.e, when the covariance matrix of features is unknown, they proposed a modified knockoff procedure based on data splitting and show that the power of the modified  procedure converges to one asymptotically if the sure screening property holds.
In a different asymptotic regime where both $n/p$ and $p_1/p$ converge to some fixed constants, 
the power analysis has also been rigorously carried out in the setting with i.i.d. Gaussian features (e.g., see \citet{weinstein2017power} for the ``counting"-knockoffs, and see \citet{weinstein2020power} and \citet{WW-LJ:2020} for the model-X knockoff filter and also the conditional randomization test).
For correlated designs, \citet{liu2019power} provided some explicit conditions under which the knockoff filter enjoys FDR zero and power one asymptotically.   Under the weak-and-rare signal setting, \cite{ke2020power} analyzed both the knockoff filter and the Gaussian mirror method for some special covariance structures, identifying key components that can influence the power of these methods.

Compared to BHq, one main advantage of DS and MDS is that they do not require p-values, which are generally difficult to obtain in high-dimensional linear models. 
Notable theoretical contributions on constructing valid p-values include the post-selection inference and the debiased Lasso procedure. 
Conditioning on the selected model, the post-selection inference derives the exact distribution of the regression coefficients.
Details have been worked out for several popular selection procedures including Lasso \citep{lee2016exact}, the forward stepwise regression, and the least angle regression \citep{tibshirani2016exact}. However, this type of theory is mostly developed case by case, and cannot be easily generalized to other selection procedures. 
The debiased Lasso procedure removes the biases in the Lasso regression coefficients so that they enjoy the asymptotic Normality under certain conditions \citep{van2014asymptotically, zhang2014confidence, javanmard2014confidence}. We refer the readers to \citet{javanmard2019false} for an FDR control procedure that applies BHq to the p-values obtained via the debiased Lasso procedure.

BHq may still perform poorly using the p-values obtained from the aforementioned methods. 
For the post-selection inference, the transformation that converts the regression coefficients to p-values may seriously dilute the true signals.
For the debiased-Lasso procedure, the asymptotic null p-values may appear highly non-uniform in finite-sample cases (e.g., see some empirical evidences in \citet{dezeure2015high} and \citet{candes2018panning}).
To avoid using p-values, several authors suggested selecting a proper penalty in penalized regressions based upon the p-value cutoff in order to achieve  FDR control. We refer the readers to  \citet{2009Benjaminiforward} and  \citet{2015Bogdan} for more details.

We conclude this section on linear models by briefly commenting on how to use the proposed methods in the low-dimensional setting with $n/p\to\infty$.
For the first half of the data $(\bm{y}^{(1)}, \bm{X}^{(1)})$, on a case-by-case basis, we can choose any sensible method (e.g., OLS, Lasso, ridge, or other regularization methods) to obtain the coefficients $\bm{\widehat{\beta}}^{(1)}$. For $(\bm{y}^{(2)}, \bm{X}^{(2)})$, we run OLS using all features to obtain the coefficients $\bm{\widehat{\beta}}^{(2)}$.  
The symmetry assumption is automatically satisfied since the model in the OLS step is well specified.
The weak dependence assumption still holds under the regularity condition and the tail condition in Assumption \ref{assump:linear}.
Therefore, similar to Proposition \ref{prop:linear-FDR}, we can show that DS asymptotically controls the FDR without requiring the signal strength  and the sparsity conditions.
We note that in the low-dimensional setting, the Lasso + OLS procedure is also applicable as long as Assumption \ref{assump:linear} is satisfied, and can be still favorable if both $n$, $p$ are large and the relevant features are sparse. In particular, in the asymptotic regime where $p/n\to c\in(0, 1/2)$, it can be problematic to directly run OLS on $(\bm{y}^{(2)}, \bm{X}^{(2)})$ with all features since the resulting covariance matrix of  $\bm{\widehat{\beta}}^{(2)}$ may be ill-conditioned,
thus the weak dependence assumption may not hold.

\subsection{Gaussian graphical models}
\label{subsec:gaussian-graphical-model}
Suppose $\bm{X} = (X_1, \ldots, X_p)$ follows a $p$-dimensional multivariate Normal distribution $N(\bm{\mu}, \Sigma)$.
Let $\Lambda = \Sigma^{-1} = (\lambda_{ij})$ be the precision matrix.
Without loss of generality, we assume $\bm{\mu} = \bm{0}$. One can define a corresponding Gaussian graphical model $(V, E)$, in which the set of vertices is $V = (X_1, \ldots, X_p)$, and there is an edge between two different vertices $X_i$ and $X_j$ if $X_i$ and $X_j$ are conditionally dependent given $\{X_k,\ k \neq i, j\}$. The graph estimation can be recast as a nodewise regression problem. To see this, for each vertex $X_j$, we can write 
\begin{equation}\nonumber
X_j = \bm{X}_{-j}^\intercal \bm{\beta}^j + \epsilon_j\ \ \ \text{with}\ \ \bm{\beta}^j = -\lambda_{jj}^{-1}\Lambda_{-j,j},
\end{equation}
where $\epsilon_j$, independent of $\bm{X}_{-j}$, follows a centered Normal distribution.
Thus, $\lambda_{ij} = 0$ implies that $X_i$ and $X_j$ are conditionally independent. We denote the neighborhood of vertex $X_j$ as $ne_j = \{k: \ k\neq j,\ \beta^{j}_{k} \neq 0\}$.
Given i.i.d. samples $\bm{X}_1, \ldots, \bm{X}_n$ from $N(\bm{\mu}, \Sigma)$, it is natural to consider first recovering the support of each $\bm{\beta}^j$ using a feature selection method such as Lasso \citep{meinshausen2006high}, and then combining all the nodewise selection results properly to estimate the graph. In view of this,
for a designated level $q\in(0,1)$, we propose an FDR control procedure as summarized in Algorithm \ref{alg:GFC-FDR}.

\begin{algorithm}
\caption{False discovery rate control in Gaussian graphical models via a single data split.}
\label{alg:GFC-FDR}
\begin{enumerate}
\item Targeting at the level $q/2$, apply the Lasso + OLS procedure (see Section \ref{subsec:linear-model}) to each nodewise regression. Denote the nodewise selection results as $\widehat{ne}_j = \{k:\  k\neq j,\ \widehat{\beta}^j_{k} \neq 0\}$ for $j\in\{1,\ldots,p\}$.
\item Combine the nodewise selection results using the OR rule to estimate the graph:
\begin{equation}\nonumber
\begin{aligned}
\widehat{E}_{\text{OR}} & = \{(i, j):\ i\in \widehat{ne}_j\ \text{or}\ j \in \widehat{ne}_i\}.
\end{aligned}
\end{equation}
\end{enumerate}
\end{algorithm}

A heuristic justification of the proposed method is given below:
\begin{equation}
\label{eq:graphical-intuition}
\begin{aligned}
\text{FDP} & = \frac{\#\{(i, j)\in \widehat{E}_{\text{OR}}, (i, j)\notin E\}}{|\widehat{E}_{\text{OR}}| \vee 1} \leq \frac{\sum_{j = 1}^p \#\{i\notin ne_j, i \in \widehat{ne}_j\}}{\frac{1}{2}\sum_{j = 1}^p \#\{i \in \widehat{ne}_j\} \vee 1}
 = \frac{\sum_{j = 1}^p \#\{i\notin ne_j, M_{ji} > \tau_{q/2}^j\}}{\frac{1}{2}\sum_{j = 1}^p \#\{M_{ji} > \tau_{q/2}^j\}\vee 1} \\ 
 & \approx \frac{\sum_{j = 1}^p \#\{i\notin ne_j, M_{ji} < -\tau_{q/2}^j\}}{\frac{1}{2}\sum_{j = 1}^p \#\{M_{ji} > \tau_{q/2}^j\}\vee 1}
\leq 2\max_{1\leq j\leq p}\frac{\#\{i\notin ne_j, M_{ji} < -\tau_{q/2}^j\}}{\#\{M_{ji} > \tau_{q/2}^j\}\vee 1} \leq q.
\end{aligned}    
\end{equation}
For the $j$-th nodewise regression, $M_{ji}$ is the mirror statistic of $X_i$, $i \neq j$, and $\tau_{q/2}^j$ is the selection cutoff of the mirror statistics. The first inequality in Equation \eqref{eq:graphical-intuition} is based on the fact that each edge can be selected at most twice. The approximation in the middle utilizes the symmetric property of the mirror statistics. The second to last inequality follows from the elementary inequality that $\left(\sum_n a_n\right)/\left(\sum_n b_n \right)\leq \max_n a_n/b_n$ for $a_n \geq 0,\ b_n > 0$.

We note that there are potentially two strategies to implement MDS for Gaussian graphical models: (i) We can apply MDS in each nodewise regression (Step 1 in Algorithm \ref{alg:GFC-FDR}) and then aggregate all the selection results using the OR rule; (ii) We can replicate the whole procedure in Algorithm \ref{alg:GFC-FDR} (both Steps 1 and 2) multiple times, and use MDS to aggregate all the selections results by
Algorithm \ref{alg:multiple-splits}. Empirically we found that both strategies achieve  FDR control, and the first one tends to have a higher power. Throughout the following theoretical justification and simulation studies, we focus on the first strategy for MDS.

Let $s = \max_{j\in \{1, \ldots, p\}} |ne_j|$. To theoretically justify our methods, we first show that with probability approaching 1, the symmetry assumption is simultaneously satisfied in all nodewise regressions under the following assumptions.

\begin{assumption}
\label{assump:graphical-model}
\text{}
\begin{enumerate}
\item (Regularity condition) $c \leq \lambda_{\min}(\Sigma)\leq \lambda_{\max}(\Sigma) \leq 1/c$ for some $c > 0$.
\item (Sparsity condition) 
$s = o(n/\log p)$.
\item (Signal strength condition)
$\min\{|\lambda_{ij}|: \lambda_{ij} \neq 0\} \gg  \sqrt{s\log p/n}$.
\end{enumerate}
\end{assumption}

Assumption \ref{assump:graphical-model} serves the same purpose  as Assumption \ref{assump:linear} for  linear models (e.g., ensure that the sure screening property holds simultaneously  in all nodewise regressions; see Remark \ref{remark:Lasso-sure-screening}). Similar assumptions also appear in \citet{liu2013gaussian} and \citet{meinshausen2006high}. Under Assumption \ref{assump:graphical-model}, we have the following proposition.

\begin{proposition}
\label{prop:gaussian-graphical-model}
Under Assumption \ref{assump:graphical-model}, 
as $n,p\to\infty$ satisfying $\log p = o(n)$, 
the symmetry assumption (Assumption \ref{assump:symmetric}) is simultaneously satisfied in all nodewise regressions with probability approaching 1. 
\end{proposition}

Similar to  linear models, the weak dependence assumption is implied by the regularity condition in Assumption \ref{assump:graphical-model}. The following proposition shows that both DS and MDS asymptotically control the FDR.
\begin{proposition}
\label{prop:graphical-FDR}
Assume that  Assumption \ref{assump:graphical-model} holds and that $\min_{j\in \{1, \ldots, p\}}|ne_j|/\log p \to \infty$. 
For any designated FDR control level $q\in(0,1)$, both DS (see Algorithm \ref{alg:GFC-FDR}) and the corresponding MDS procedure achieve
\begin{equation}\nonumber
\limsup_{n, p \to\infty}\text{FDR} \leq q\ \ \ \text{and}\ \ \ \liminf_{n, p \to\infty}\text{Power} = 1
\end{equation}
in the asymptotic regime where $\log p = o(n^\xi)$ for some $\xi\in(0,1)$.
\end{proposition}

The assumption $\min_{j\in \{1, \ldots, p\}}|ne_j|/\log p \to \infty$ is mainly for the technical purpose so that a union bound can be applied for all nodewise regressions. Empirically, we find that the data-splitting methods and the GFC method proposed in \citet{liu2013gaussian} are effective in quite different scenarios. GFC tends to work well if the underlying true graph is ultra-sparse, i.e., the nodewise sparsity is in the order of $o(\sqrt{n}/(\log p)^{3/2})$. In contrast, DS and MDS are capable of handling cases where the graph is not too sparse, but may suffer from the ultra-sparsity. 
A similar issue also exists in general knockoff-based methods, and we refer the readers to \citet{li2019nodewise} for relevant discussions.

\section{Numerical Illustrations}
\label{sec:simulation}
\subsection{Linear model}
\label{subsec:sim-linear}
We simulate the response vector $\bm{y}$ from the linear model $\bm{y}_{n\times 1} = \bm{X}_{n\times p}\bm{\beta}^\star_{p\times 1} + \bm{\epsilon}_{n\times 1}$ with $\bm{\epsilon}\sim N(\bm{0}, I_n)$, and
 randomly locate the signal index set $S_1$. For $j\in S_1$, we sample $\beta^\star_j$ from $N(0, \delta\sqrt{\log p/n})$, and refer to $\delta$ as the signal strength.
Throughout, the designated FDR control level is set to be $q = 0.1$.
The penalization parameter of Lasso is selected based on 10-fold cross-validation.

\begin{figure}
\begin{center}
\includegraphics[width=0.45\columnwidth]{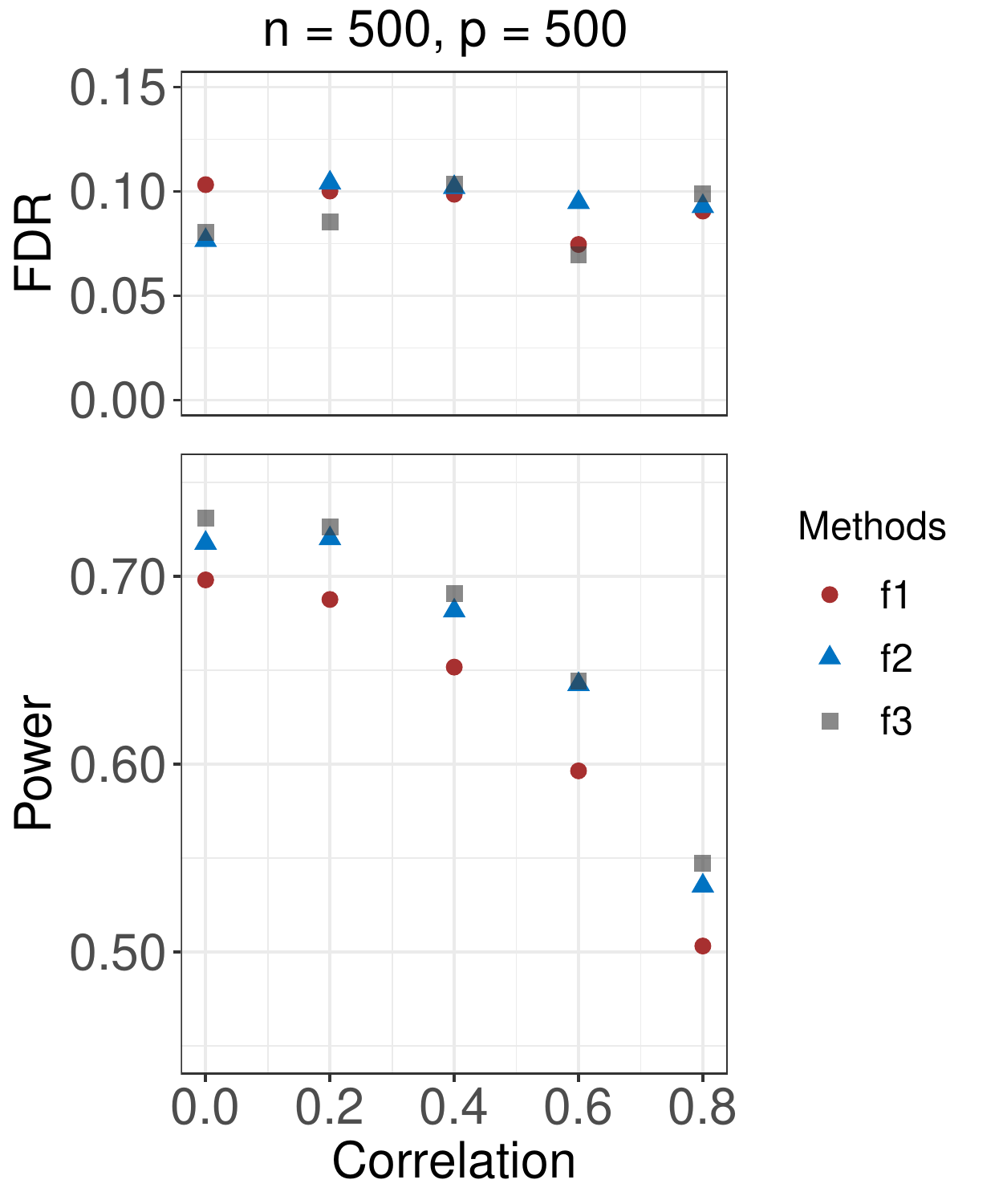}
\includegraphics[width=0.45\columnwidth]{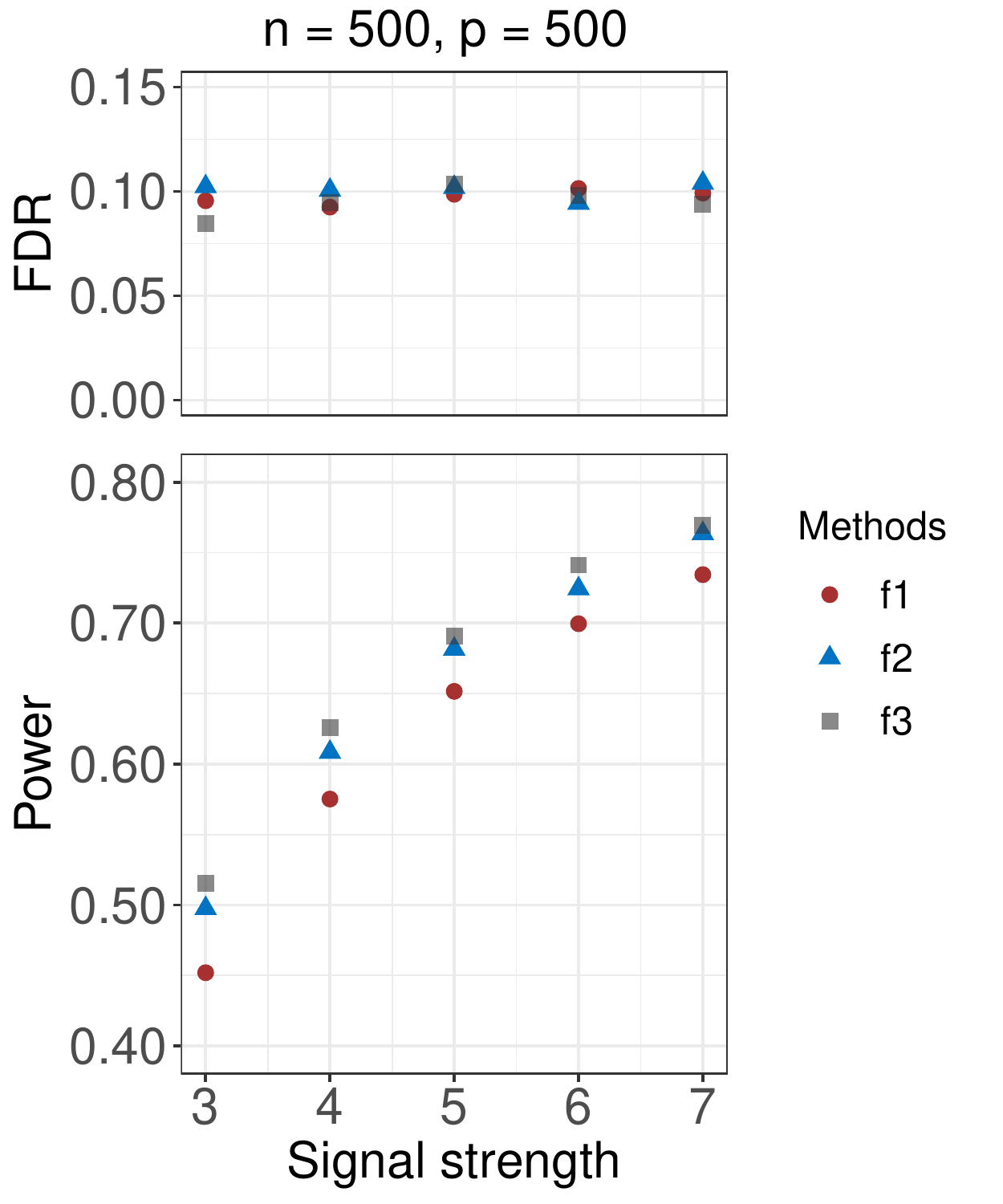}
\end{center}
\caption{Empirical FDRs and powers of DS using three different mirror statistics constructed with $f_1, f_2, f_3$ specified in \eqref{eq:contrast_choice}. Features are independently drawn from $N(0, \Sigma)$ with $\Sigma$ being a Toeplitz covariance matrix.
In the left panel, we fix the signal strength at $\delta = 5$ and vary the correlation $\rho$.
In the right panel, we fix the correlation at $\rho = 0.4$ and vary the signal strength $\delta$.
The number of relevant features is $p_1 = 50$ across all settings, and the designated FDR control level is $q = 0.1$.
Each dot in the figure represents the average from 50 independent runs.}
\label{fig:mirror-statistic-optimality-DS}
\end{figure}

We first investigate the performance of DS/MDS using different mirror statistics constructed with $f_1, f_2, f_3$ specified in \eqref{eq:contrast_choice}. We set the sample size $n = 500$, the number of features $p = 500$, and the number of relevant features $p_1 = 50$. Each row of the design matrix is independently drawn from $N(0, \Sigma)$. We consider a similar setup as in \citet{ma2020global}, where $\Sigma$ is a blockwise diagonal matrix of 10 Toeplitz submatrices whose
off-diagonal entries linearly descend from $\rho$ to 0. The detailed formula of $\Sigma$ is given in \eqref{eq:blockwise-diagonal-formula} in Supplementary Materials, and we refer to it as the Toeplitz covariance matrix throughout. We vary the correlation $\rho$ and the signal strength $\delta$, and the results are summarized in Figure \ref{fig:mirror-statistic-optimality-DS}. We see that across different settings, all three choices of mirror statistics achieve FDR control, and $f_3$ yields the highest power. Proposition \ref{prop:optimality-mirror-statistics} shows that $f_3$ is optimal for orthogonal designs, and the empirical results suggest that $f_3$ might also be a good choice in more realistic settings. Among all the simulation studies described below, we construct the mirror statistics with $f_3$. It is worth noting that the performance of MDS appears to be more robust to the choice of mirror statistics compared to DS (see Figure \ref{fig:mirror-statistics-optimality-MDS} in Supplementary Materials).

\begin{figure}
\begin{center}
\includegraphics[width=0.4\columnwidth]{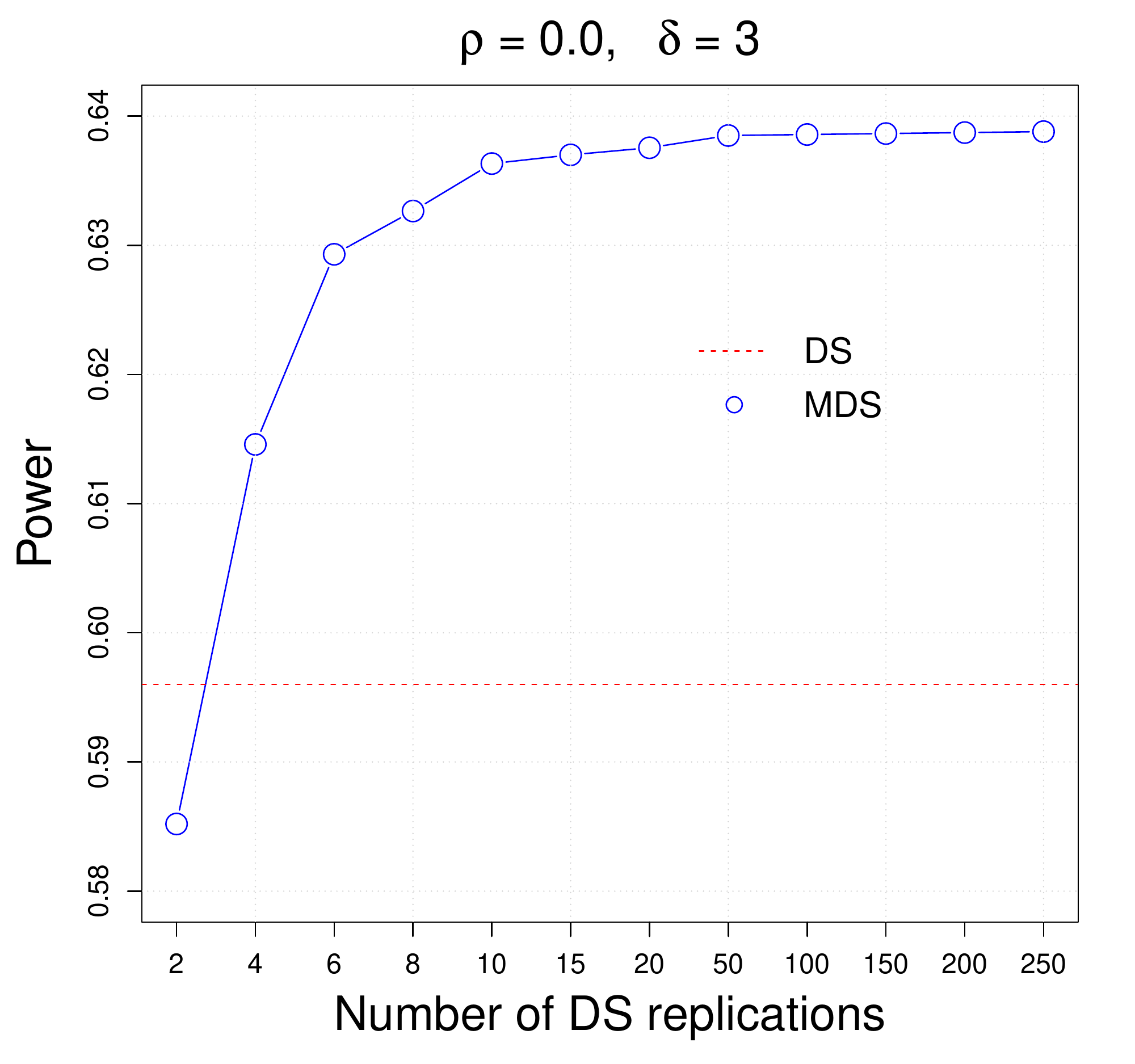}
\includegraphics[width=0.4\columnwidth]{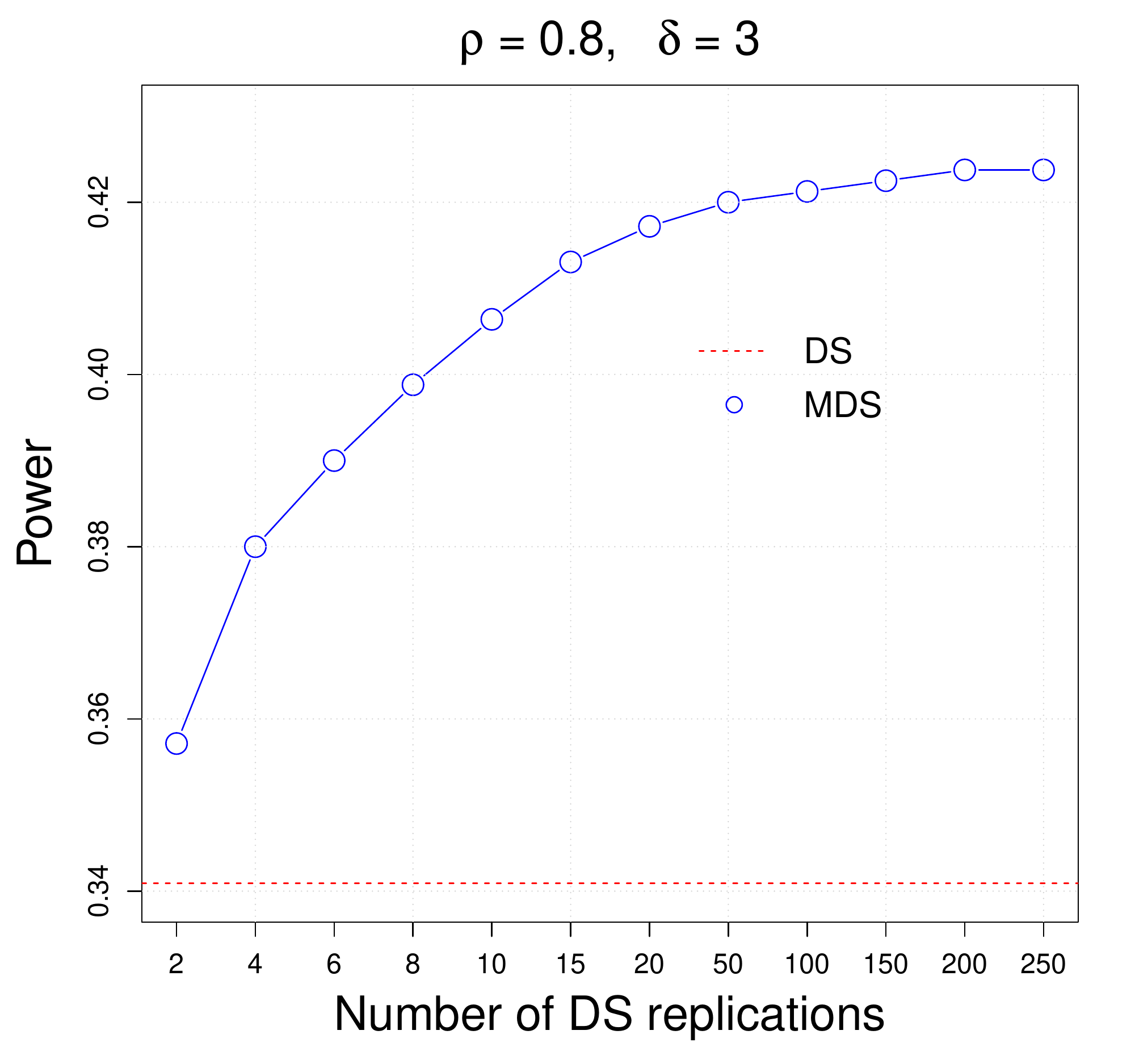}
\end{center}
\caption{Empirical powers of MDS with different number of DS replications. Each row of the design matrix is independently drawn from $N(0, \Sigma)$ with $\Sigma$ being a Toeplitz covariance matrix.
The sample size is $n =500$, the number of features is $p = 500$, and the number of relevant features is $p_1 = 50$ in both settings.
Over 50 independent runs, the blue dots and the red lines represent the average powers of MDS and DS, respectively.}
\label{fig:num_splits}
\end{figure}

We then examine the effect of the number of DS replications $m$ on the power of MDS. With $n = 500$, $p = 500$ and $p_1 = 50$, we generate features independently from $N(0, \Sigma)$ with $\Sigma$ being a Toeplitz covariance matrix. We set the signal strength $\delta = 3$ and test out two scenarios with the correlation $\rho = 0.0$ and $\rho = 0.8$. Figure \ref{fig:num_splits} shows that the power of MDS monotonically increases with the number of DS replications $m$, and becomes relatively stable after $m \geq 50$. 
Empirical evidence suggests that it only requires a small number of DS replications to realize the full power of MDS. Thus, MDS is  computationally more feasible for large data sets compared to other methods such as the knockoff filter and the Gaussian mirror method.
In the following examples, we set $m = 50$ for MDS.

We proceed to compare DS/MDS with two popular methods in high-dimensional settings under various design matrices:  MBHq \citep{meinshausen2009p} and  the model-X knockoff filter \citep{candes2018panning}. 
For their comparisons in low-dimenisonal settings, we refer the readers to Figures \ref{fig:low-d-toeplitz} and \ref{fig:low-d-toeplitz-std} in Supplementary Materials.
For MBHq, we obtain 50 p-values for each feature via repeated sample splitting. More precisely, we
run Lasso for feature screening on one half of the data,
and calculate the p-values for the selected features by running OLS on the other half of the data. We then combine the p-values across different sample splits using the R package \textit{hdi}.\footnote{https://cran.r-project.org/web/packages/hdi/hdi.pdf}
For the knockoff filter, we use the equi-correlated knockoffs, in which the covariance matrix of features is estimated using the R package \textit{knockoff}.\footnote{https://cran.r-project.org/web/packages/knockoff/index.html}
For all the simulation settings in Section \ref{subsec:sim-linear}, we empirically found that the equi-correlated knockoffs yields a more powerful knockoff filter compared to the default \textit{asdp} construction.

\begin{enumerate}
\item \textbf{Normal design matrices.} With $n = 800$, $p \in \{1000, 2000\}$ and $p_1 = 50$, we generate features independently from $N(0, \Sigma)$ with $\Sigma$ being a Toeplitz covariance matrix. We compare the performances of the competing methods under different correlations $\rho$ and signal strengths $\delta$. The results for $p = 2000$ are summarized in Figure \ref{fig:normal-design-toeplitz-correlation-P2000}, and the results for $p = 1000$ are summarized in Figure \ref{fig:normal-design-toeplitz-correlation-P1000} in Supplementary Materials. The FDRs of all the four methods are under control across different settings.
In terms of power, the knockoff filter and MDS are the two leading methods. MDS appears more powerful when features are more correlated, or when the signal strength is relatively weak, whereas the knockoff filter enjoys a higher power in the opposite regimes. 
We observed that MDS is more robust to highly correlated design matrices compared to the knockoff filter. Figure \ref{fig:normal-design-constant-correlation} in Supplementary Materials report the performances of the competing methods in the case where $\Sigma$ has constant pairwise correlation $\rho$. We see that the knockoff filter appears significantly less powerful than MDS when $\rho \geq 0.4$. The simulation results also suggest that MDS yields better rankings of features compared to DS, thus enjoys simultaneously  a lower FDR and a higher power.

\begin{figure}
\begin{center}
\includegraphics[width=0.45\columnwidth]{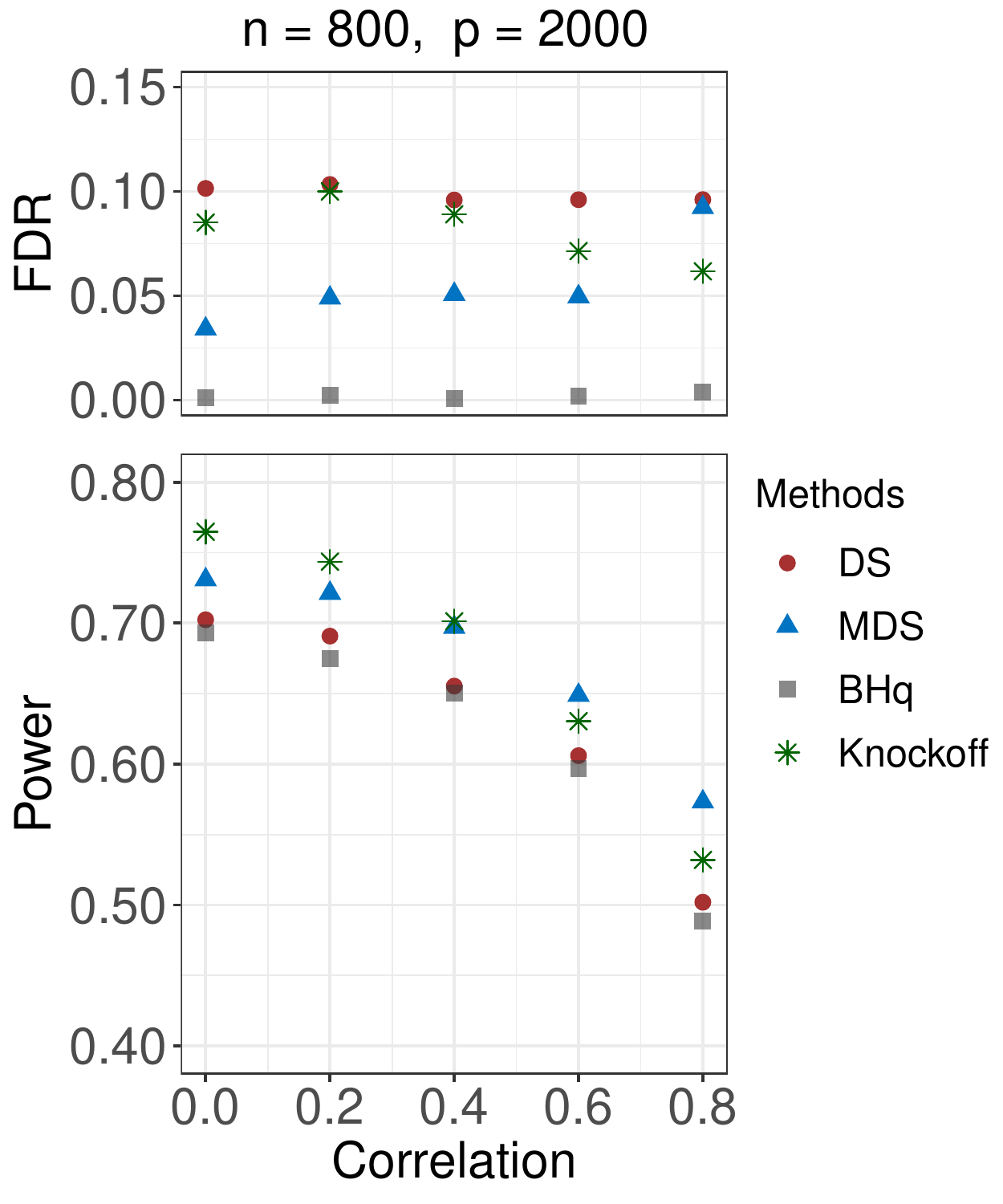}
\includegraphics[width=0.45\columnwidth]{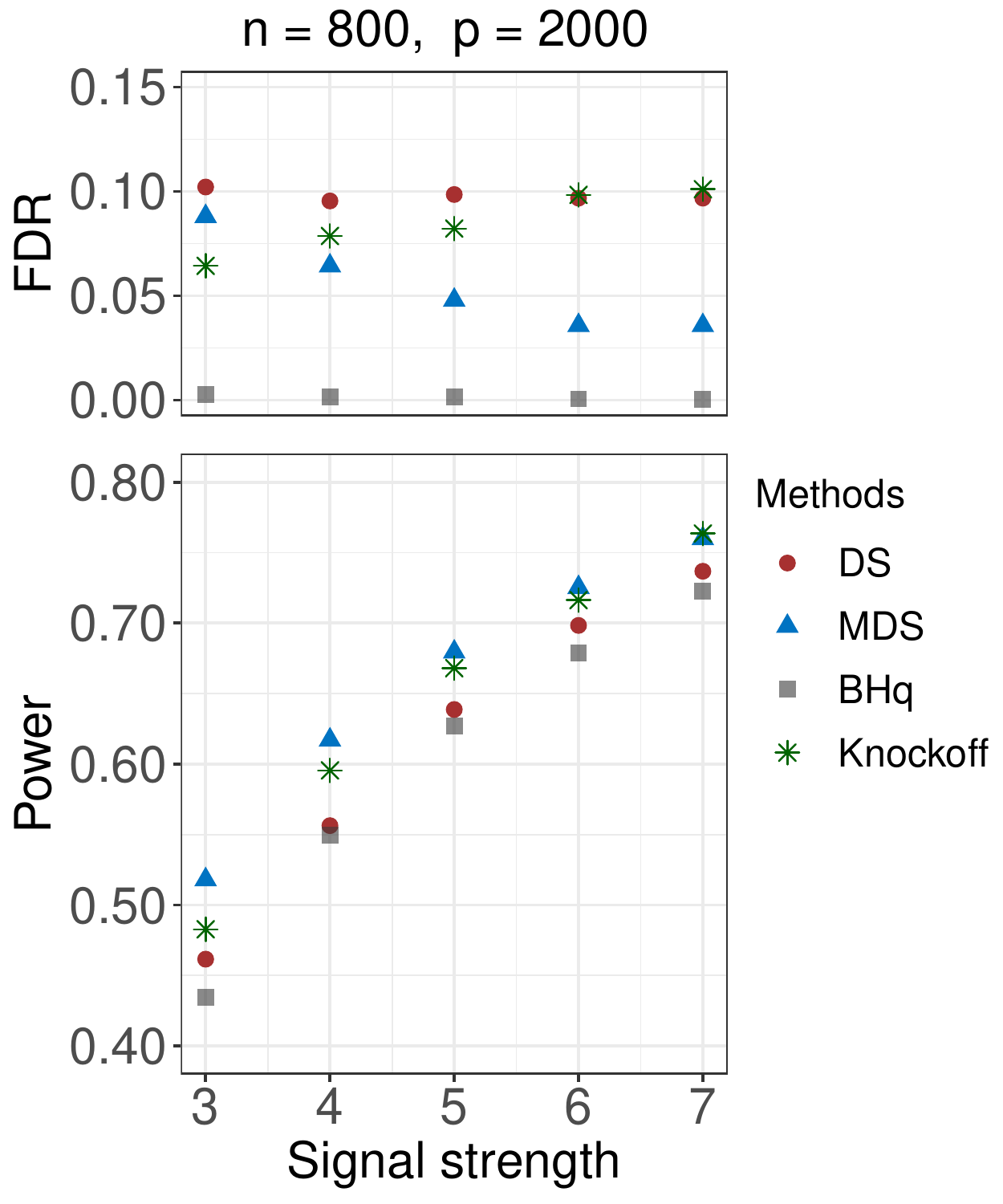}
\end{center}
\caption{Empirical FDRs and powers for linear models with Normal design matrices. Features are independently drawn from $N(0, \Sigma)$ with $\Sigma$ being a Toeplitz covariance matrix.
In the left panel, we fix the signal strength at $\delta = 5$ and vary the correlation $\rho$. In the right panel, we fix the correlation at $\rho = 0.5$ and vary the signal strength $\delta$.
The  designated FDR control level is $q = 0.1$, and the number of relevant features is $p_1= 50$ across all settings.
Each dot in the figure represents the average from 50 independent runs.}
\label{fig:normal-design-toeplitz-correlation-P2000}
\end{figure}

\item \textbf{Non-Normal design matrices.} When the joint distribution of features is unknown and non-Normal, the performance of the knockoff filter is not guaranteed if the knockoffs are generated based upon a naive fit of the multivariate Normal distribution using the design matrix. We here illustrate the robustness of DS/MDS with respect to the non-Normality by considering the following two design matrices: (1) a two-component Gaussian mixture  distribution centered at $0.5\times \mathbbm{1}_p$ and  $-0.5\times \mathbbm{1}_p$; (2) a centered multivariate $t$-distribution with 3 degrees of freedom. Throughout, the covariance matrix $\Sigma$ is set to be a Toeplitz matrix. Note that in both scenarios, the marginal distribution of each feature is still unimodal, and does not differ much from the Normal distribution in appearance. We fix $n = 800$, $p = 2000$, $p_1 = 70$, and test out different correlations $\rho$ and signal strengths $\delta$. The results are summarized in Figure \ref{fig:t-design-toeplitz-correlation}. Because of the model misspecification in the knockoff construction, the knockoff filter appears over conservative when features follow a Gaussian mixture distribution, and loses FDR control when features follow a $t$-distribution. The latter is perhaps a more concerning issue in the context of controlled feature selection, 
although the performance of the knockoff filter can be potentially improved by carefully modeling the joint distribution of features based on some structural assumptions (e.g., see \citet{2018genehunting}).
In comparison, MDS maintains FDR control and enjoys a reasonably high power in both scenarios. We also note that, except being overly conservative, MBHq performs quite competitively in all settings. 

\begin{figure}
\begin{center}
\includegraphics[width=0.45\columnwidth]{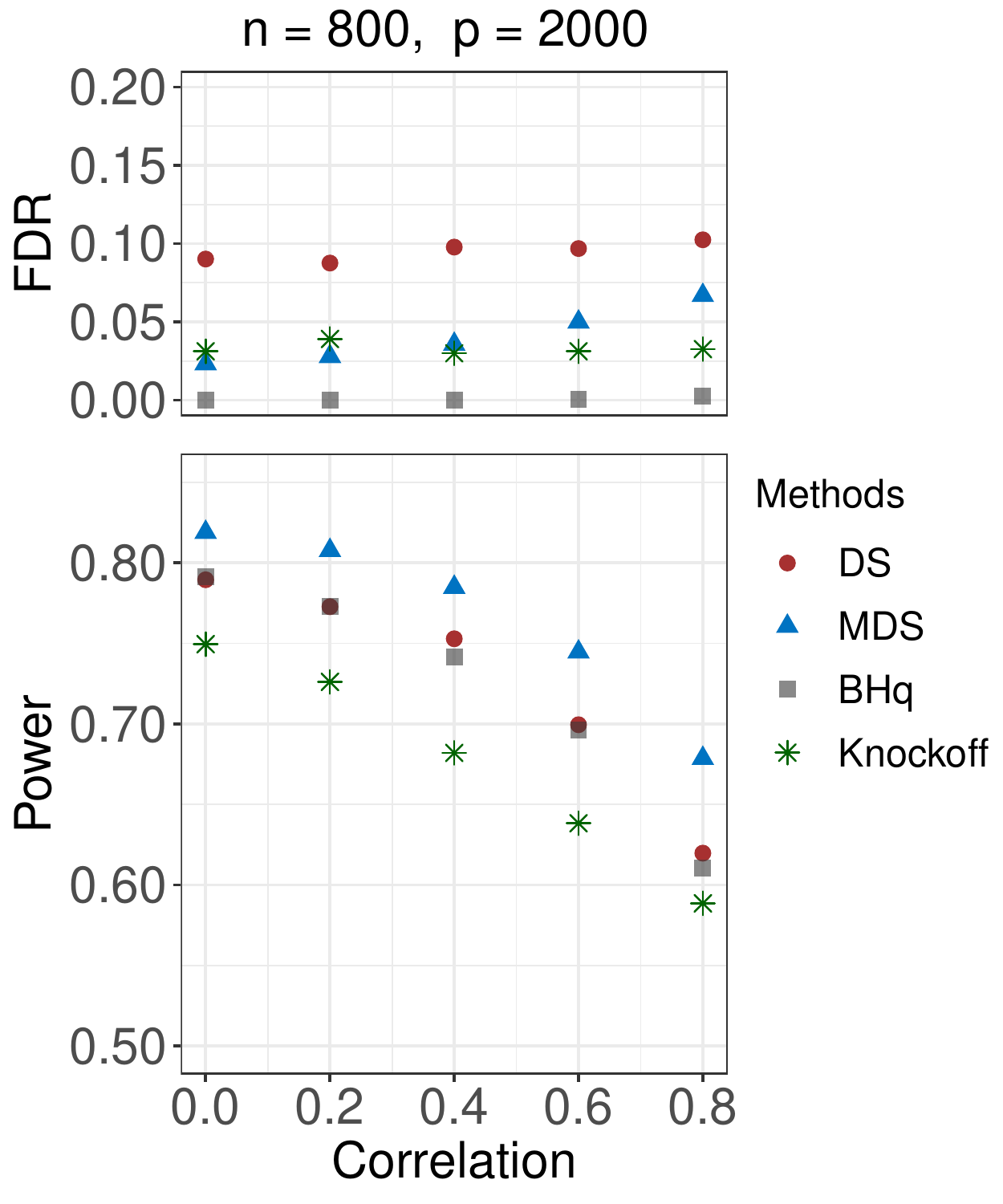}
\includegraphics[width=0.45\columnwidth]{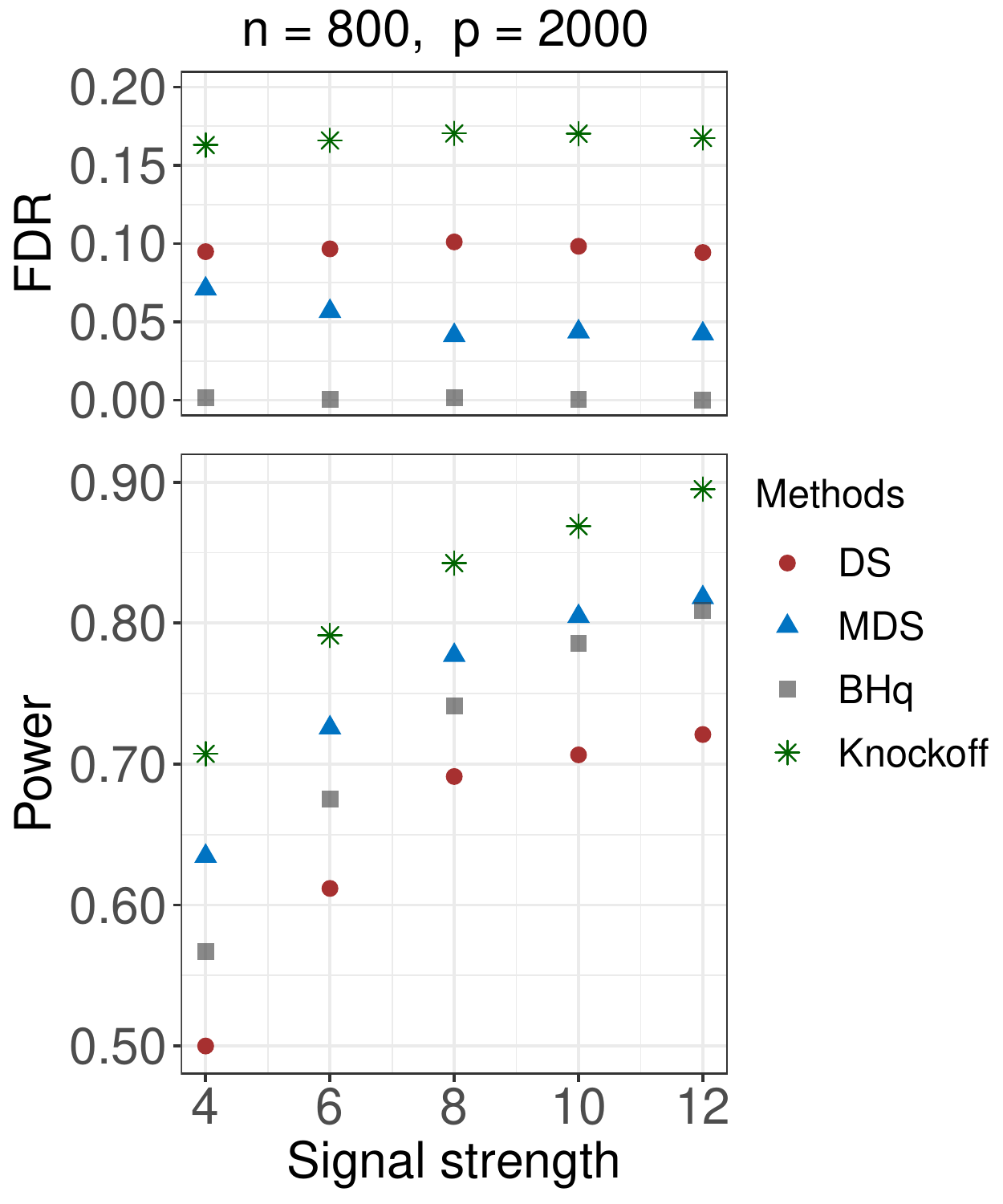}
\end{center}
\caption{
Empirical FDRs and powers for linear models with non-Normal design matrices. 
In the left panel, features are independently drawn from a two-component mixture Normal distribution centered at $0.5\times \mathbbm{1}_p$ and  $-0.5\times \mathbbm{1}_p$.
In the right panel, features are independently drawn from a centered multivariate $t$-distribution with 3 degrees of freedom.
The covariance matrix $\Sigma$ in both panels is set to be a Toeplitz matrix.
In the left panel, we fix the signal strength at $\delta = 8$ and vary the correlation $\rho$.
In the right panel, we fix the correlation at $\rho = 0.5$ and vary the signal strength $\delta$.
The number of relevant features is $p_1 = 70$ across all settings, and
the designated FDR control level is $q = 0.1$.
Each dot in the figure represents the average from 50 independent runs.}
\label{fig:t-design-toeplitz-correlation}
\end{figure}

\item \textbf{Real-data design matrices.} We consider using the scRNAseq data in \citet{hoffman2020single} as the design matrix. 
A total of 400 T47D A1–2 human breast cancer cells were treated with 100 nM synthetic glucocorticoid dexamethasone (Dex).
An scRNASeq experiment was performed after 18h of the Dex treatment, leading to a total of 400 samples of gene expressions for the treatment group. 
For the control group, there are 400 vehicle-treated control cells. An scRNAseq experiment was performed at the 18h timepoint to obtain the corresponding profile of gene expressions. After proper normalization, the final scRNAseq data\footnote{The data is available at \url{https://www.ncbi.nlm.nih.gov/geo/query/acc.cgi?acc=GSE141834}.} contains 800 samples, each with 32,049 gene expressions. To further reduce the dimensionality, we first screen out the genes detected in fewer than 10\% of cells, and then pick up the top $p$ most variable genes following \citet{hoffman2020single}.
We fix $p_1 = 70$, and simulate the response vector $\bm{y}$ with various $p$ and signal strengths. The results are summarized in Figure \ref{fig:GWAS-design-toeplitz-correlation}. We see that all the methods achieve FDR control, among which MDS enjoys the highest power. The knockoff filter appears to be conservative (with its FDR significantly below the nominal level 0.1), likely due to the fact that the joint distribution of gene expressions is non-Normal, resulting in a misspecified construction of the knockoffs.

\begin{figure}
\begin{center}
\includegraphics[width=0.45\columnwidth]{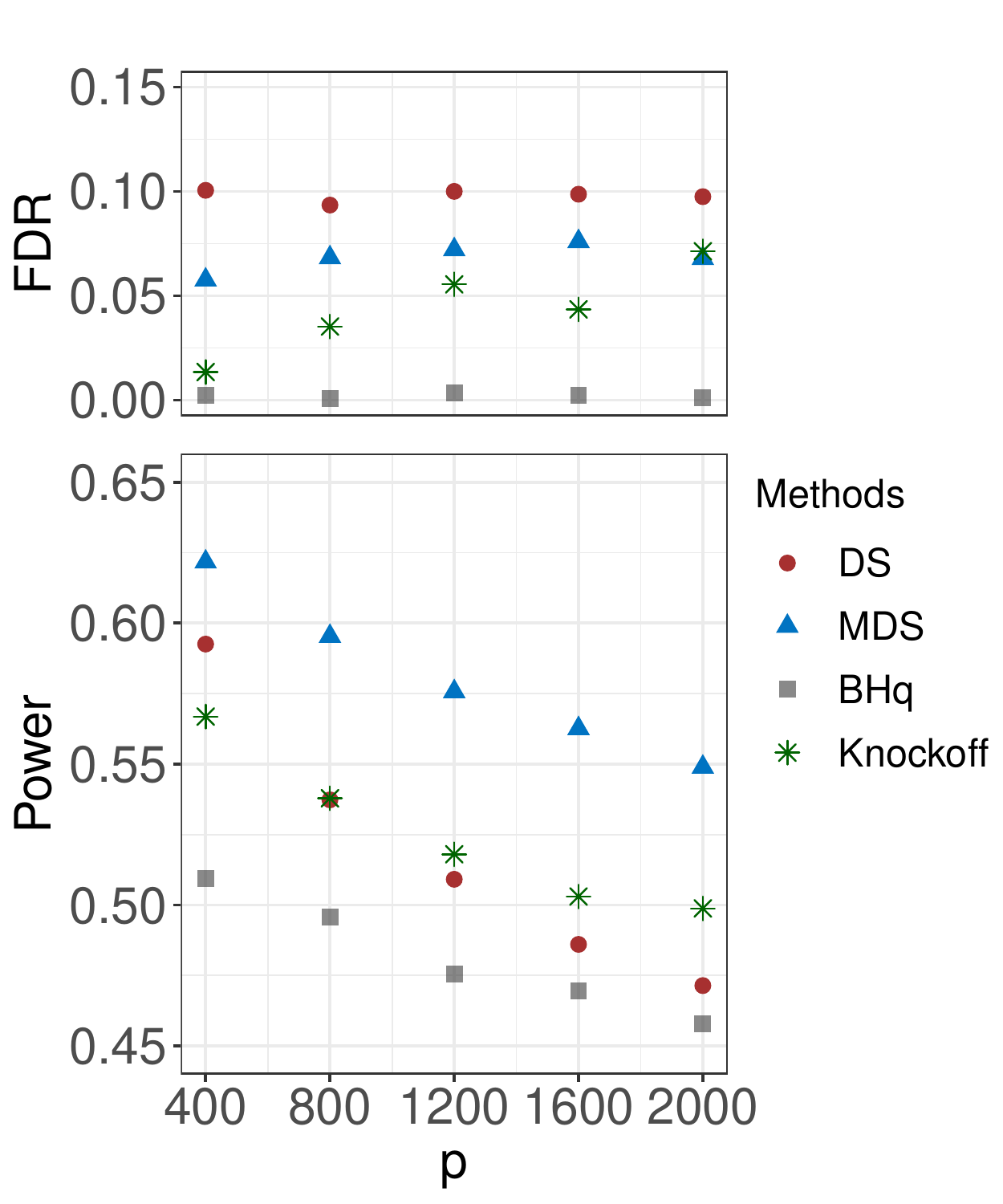}
\includegraphics[width=0.45\columnwidth]{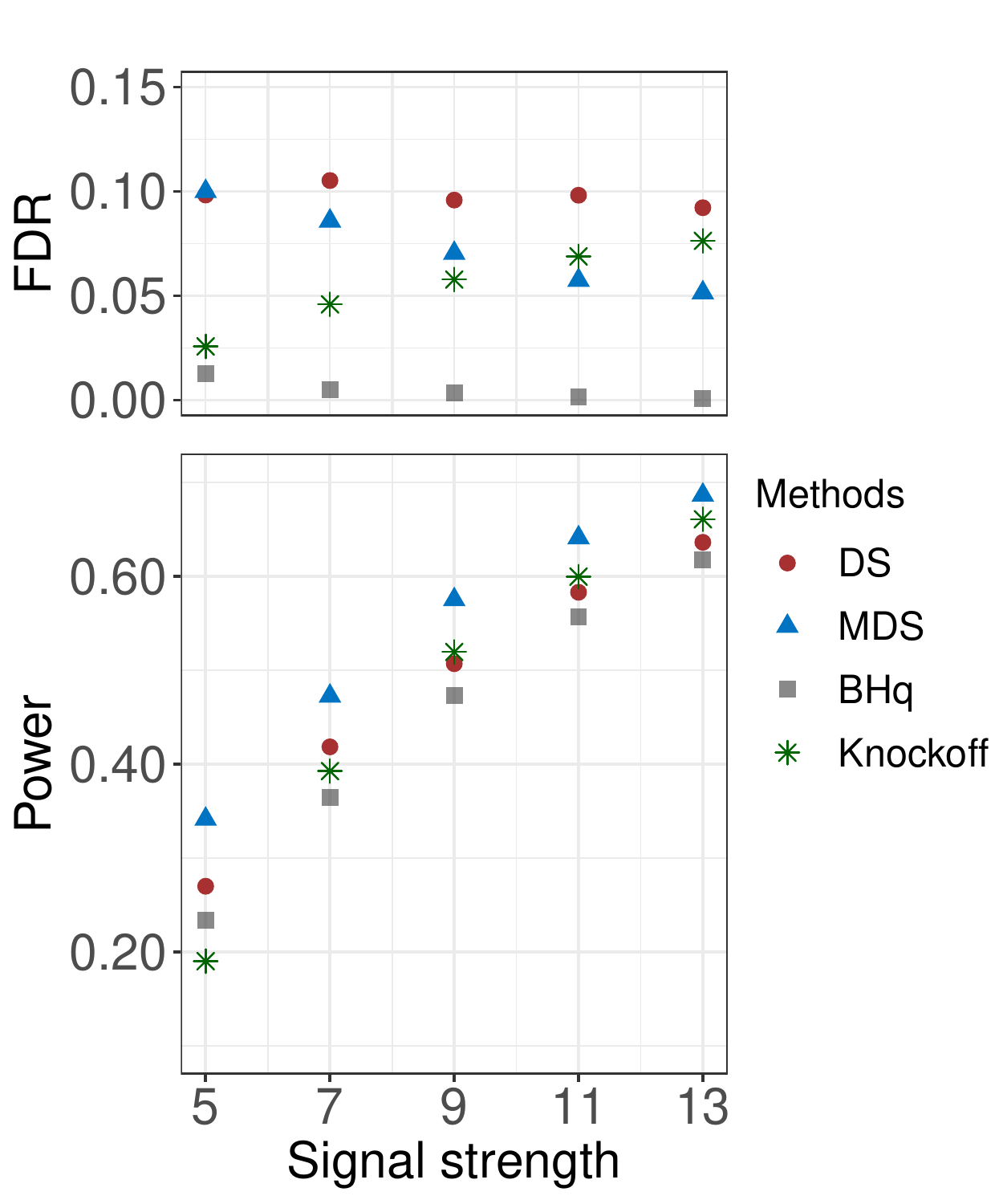}
\end{center}
\caption{
Empirical FDRs and powers for linear models with a GWAS design matrix.
The sample size is $n = 800$.
The signal strength along the $x$-axis in the right panel shows multiples of
$1/\sqrt{n}$.
In the left panel, we fix the signal strength at 9 and vary the dimension $p$.
In the right panel, we fix the dimension at $p = 1200$ and vary the signal strength.
The number of relevant features is $p_1 = 70$ across all settings, and
the designated FDR control level is $q = 0.1$.
Each dot in the figure represents the average from 50 independent runs.}
\label{fig:GWAS-design-toeplitz-correlation}
\end{figure}

\end{enumerate}

\begin{figure}
\begin{center}
\includegraphics[width=0.49\columnwidth]{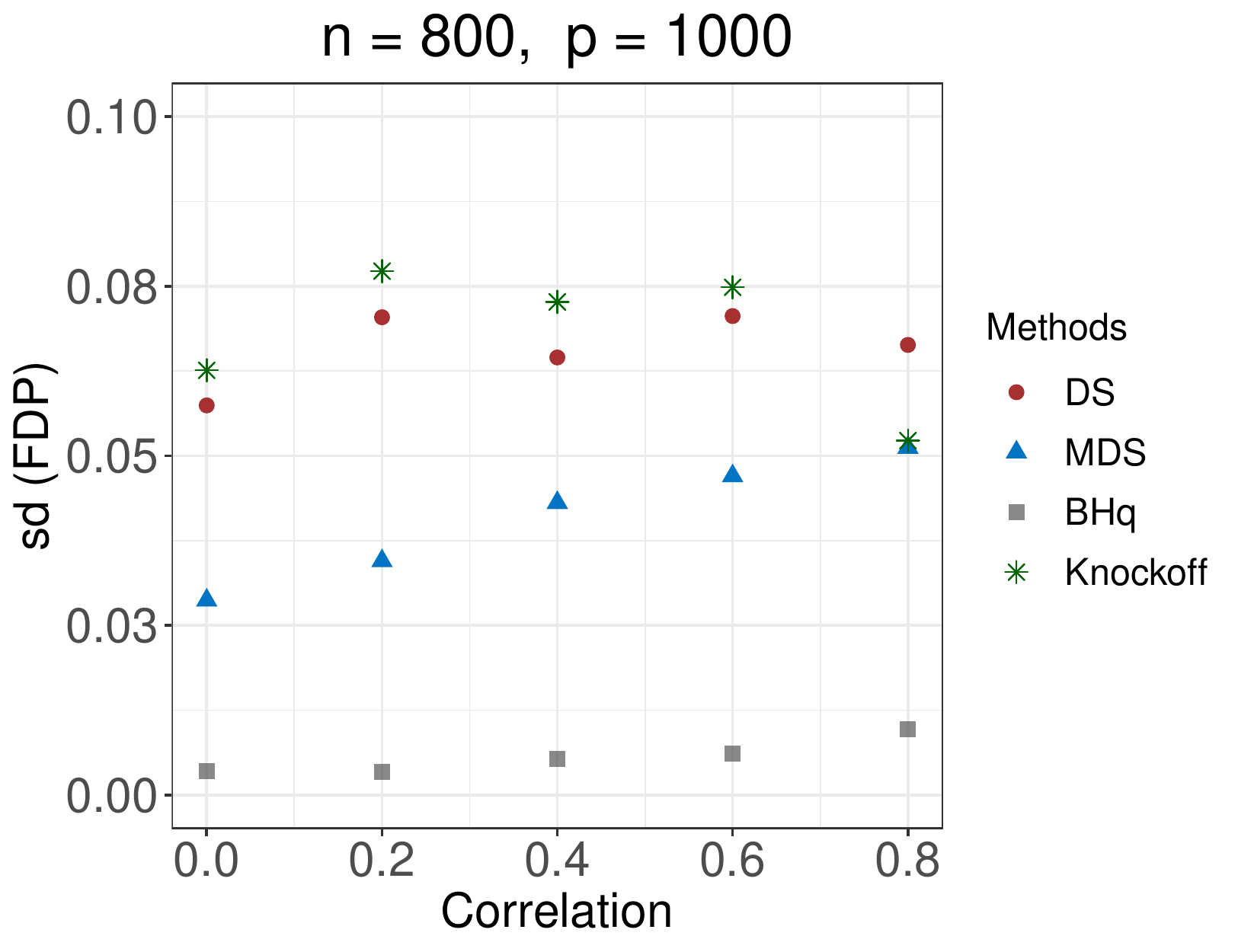}
\includegraphics[width=0.49\columnwidth]{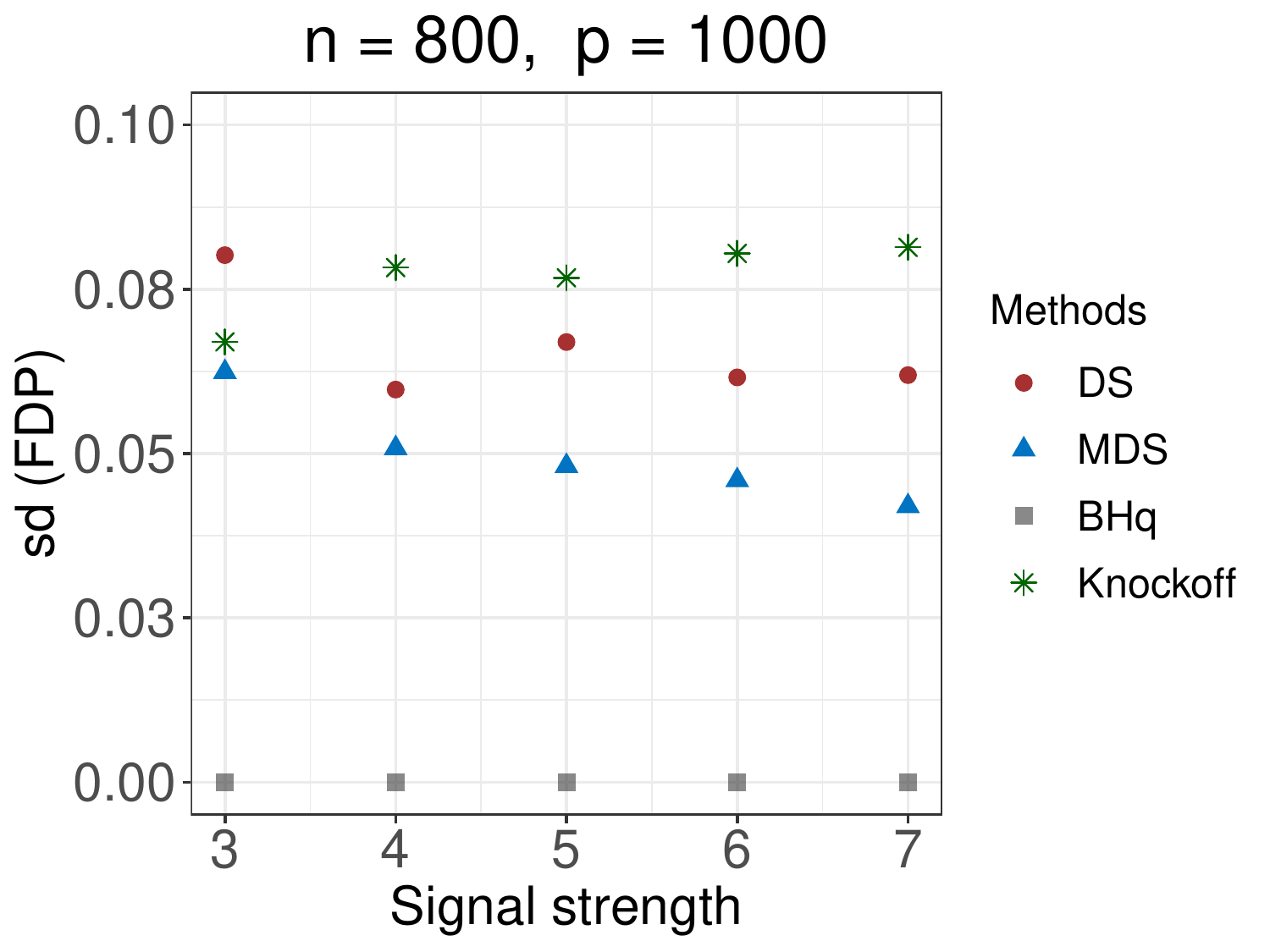}
\includegraphics[width=0.49\columnwidth]{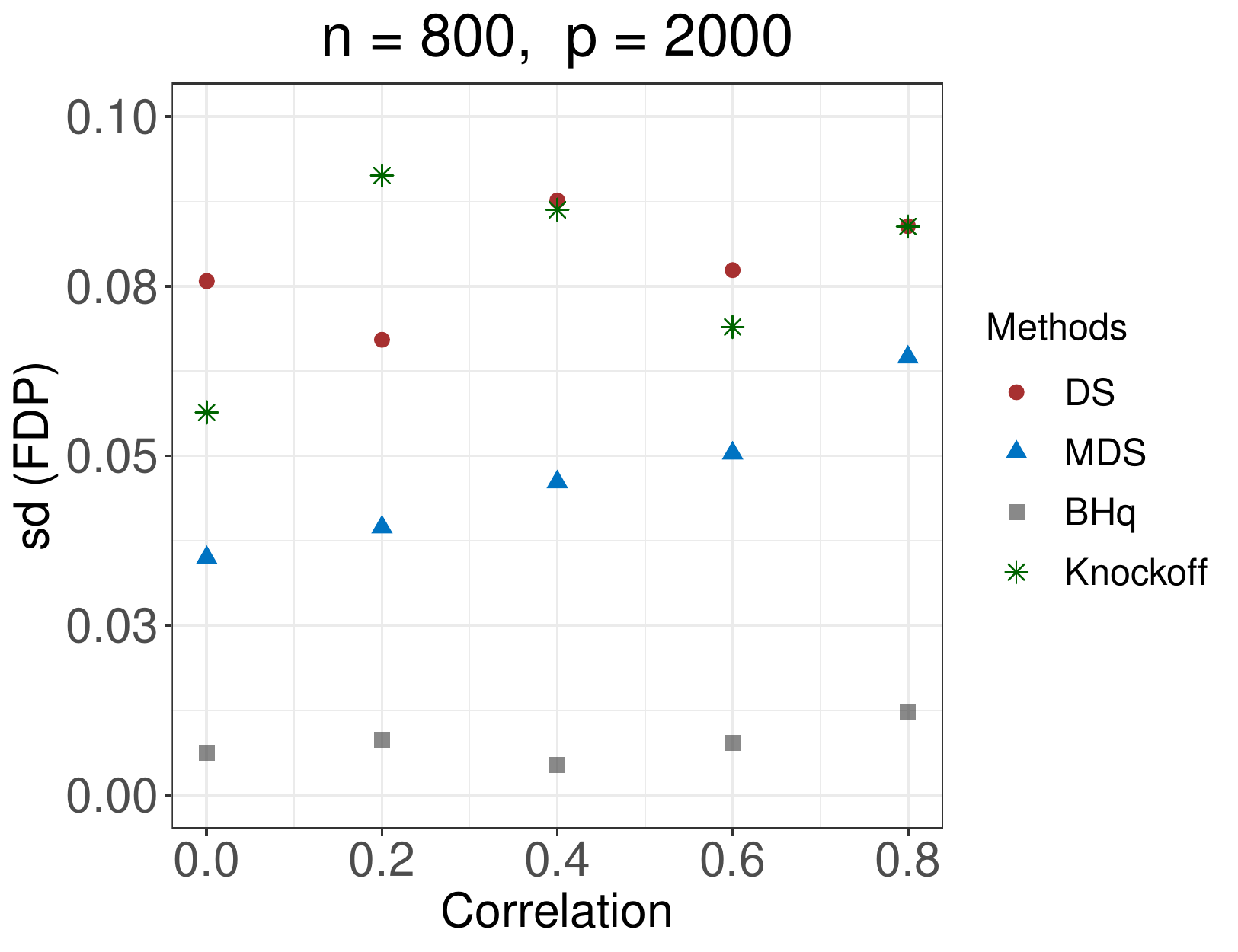}
\includegraphics[width=0.49\columnwidth]{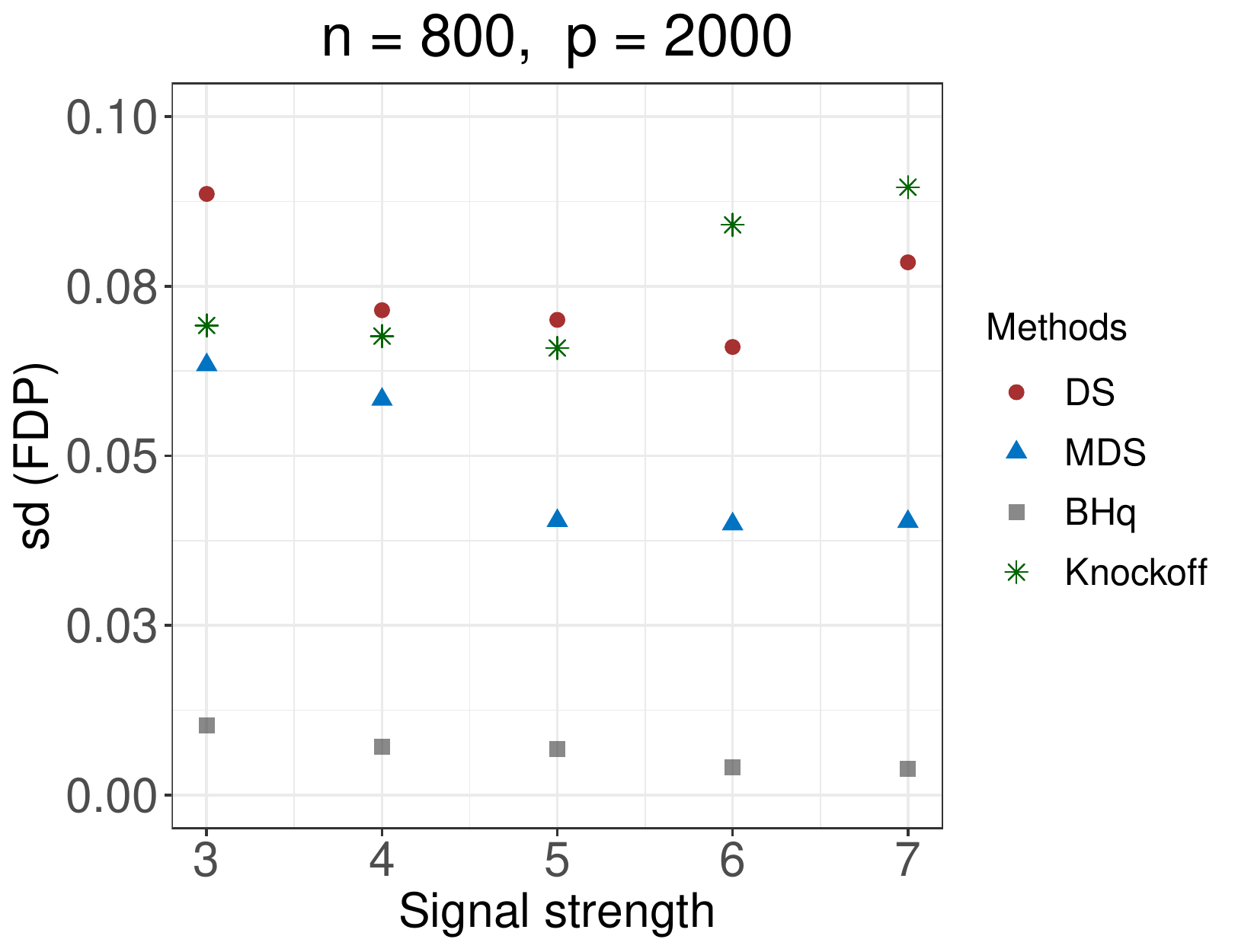}
\end{center}
\caption{Standard deviations of the FDP for linear models with a Toeplitz correlation structure. The algorithmic settings are as per Figure \ref{fig:normal-design-toeplitz-correlation-P2000}.}
\label{fig:normal-design-toeplitz-correlation-FDP-std}
\end{figure}

We conclude this section with some remarks on the variance of the FDP. 
Note that DS and the knockoff filter rank features using the mirror statistics $M_j$'s and the statistics $W_j$'s (see Section 3.2 in \citet{candes2018panning}), respectively. 
The statistic $W_j$ enjoys a \textit{flip-sign property}, that is, the signs of $W_j$'s for $j\in S_0$ are independent,  thus the FDP of the knockoff filter fluctuates and concentrates around the FDR.
For DS, the signs of the mirror statistics $M_j$'s for $j\in S_0$ are dependent so the variance of the FDP can be a more concerning issue for DS. FDR control becomes less meaningful if the variance is unacceptably large.
We empirically check the variances of the FDP for the four competing methods across the aforementioned simulation settings. The results are summarized in Figures \ref{fig:normal-design-toeplitz-correlation-FDP-std}, \ref{fig:normal-design-constant-correlation-FDP-std}, \ref{fig:t-design-constant-correlation-FDP-std}, and \ref{fig:GWAS-design-constant-correlation-FDP-std} in Supplementary Materials.
We observed that except for the cases where the knockoff filter appears overly conservative (e.g., see Figure \ref{fig:normal-design-constant-correlation-FDP-std}), the variances of the FDP are comparable for DS and the knockoff filter. More interestingly, perhaps due to its de-randomized nature,  MDS achieves a lower variance of the FDP than the knockoff filter in a majority of simulation settings.

\subsection{Gaussian graphical model}
\label{subsec:sim-GGM}
We set the designated FDR control level at $q = 0.2$ and consider two types of graphs:
\begin{enumerate}
\item \textbf{Banded graph.} The precision matrix $\Lambda$ is constructed such that $\lambda_{jj} = 1$, $\lambda_{ij} = \text{sign}(a)\cdot|a|^{|i - j|/c}$ if $0 < |i - j| \leq s$, and $\lambda_{ij} = 0$ if $|i - j| > s$. Throughout, we set $c = 1.5$ following \citet{li2019nodewise}. Other parameters including the sample size $n$, the dimension $p$, the partial correlation (signal strength) $a$, and the nodewise sparsity $s$ will be specified case by case. 
\item \textbf{Blockwise diagonal graph.} The precision matrix $\Lambda$ is blockwise diagonal with equally sized squared blocks generated in the same fashion. Throughout, we fix the block size to be 25 $\times$ 25. In each block, all the diagonal elements are set to be 1, and the off-diagonal elements are independently drawn from the uniform distribution $\text{Unif}((-0.8, -0.4)\cup(0.4, 0.8))$. 
\end{enumerate}
For both types of graphs, the precision matrix $\Lambda$ generated from the aforementioned processes may not be positive definite. If $\lambda_{\min}(\Lambda) < 0$, we reset $\Lambda \leftarrow \Lambda + (\lambda_{\min}(\Lambda) + 0.005)I_p$ following \citet{liu2013gaussian}. 
Three classes of competing methods are tested out, including (1) DS and MDS; (2) BHq; (3) GFC \citep{liu2013gaussian}. 
For MDS, nodewisely, we replicate DS 50 times and aggregate the selection results using Algorithm \ref{alg:multiple-splits}.
For BHq, the p-values are calculated based on the pairwise partial correlation test using the R package \textit{ppcor} \citep{kim2015ppcor}. 
For GFC, we use the R package \textit{SILGGM} \citep{zhang2018silggm} to implement it.

For the banded graph, we test out the following four scenarios:
\begin{enumerate}[\hspace{0.3cm}(a)]
\item fix $p = 100$, $s = 8$, $a = -0.6$, and vary the sample size $n \in \{500, 1000, 1500, 2000, 2500\}$;
\item fix $n = 1000$, $s = 8$, $a = -0.6$, and vary the dimension $p \in \{50, 100, 150, 200, 250\}$;
\item fix $n = 1000$, $p = 100$, $a = -0.6$, and vary the nodewise sparsity $s \in \{4, 6, 8, 10, 12\}$; 
\item fix $n = 1000$, $p = 100$, $s = 8$, and vary the signal strength $a \in \{-0.5, -0.6, -0.7, -0.8, -0.9\}$. 
\end{enumerate}
For the blockwise diagonal graph, we test out the following two scenarios:
\begin{enumerate}[\hspace{0.3cm}(a)]
\item fix $p = 100$, and vary the sample size $n \in \{200, 300, 400, 500, 600\}$; 
\item fix $n = 500$, and vary the dimension $p \in \{50, 100, 150, 200, 250\}$. 
\end{enumerate}

\begin{figure*}
\centering
\begin{subfigure}[b]{0.45\columnwidth}
\includegraphics[width=1.0\columnwidth]{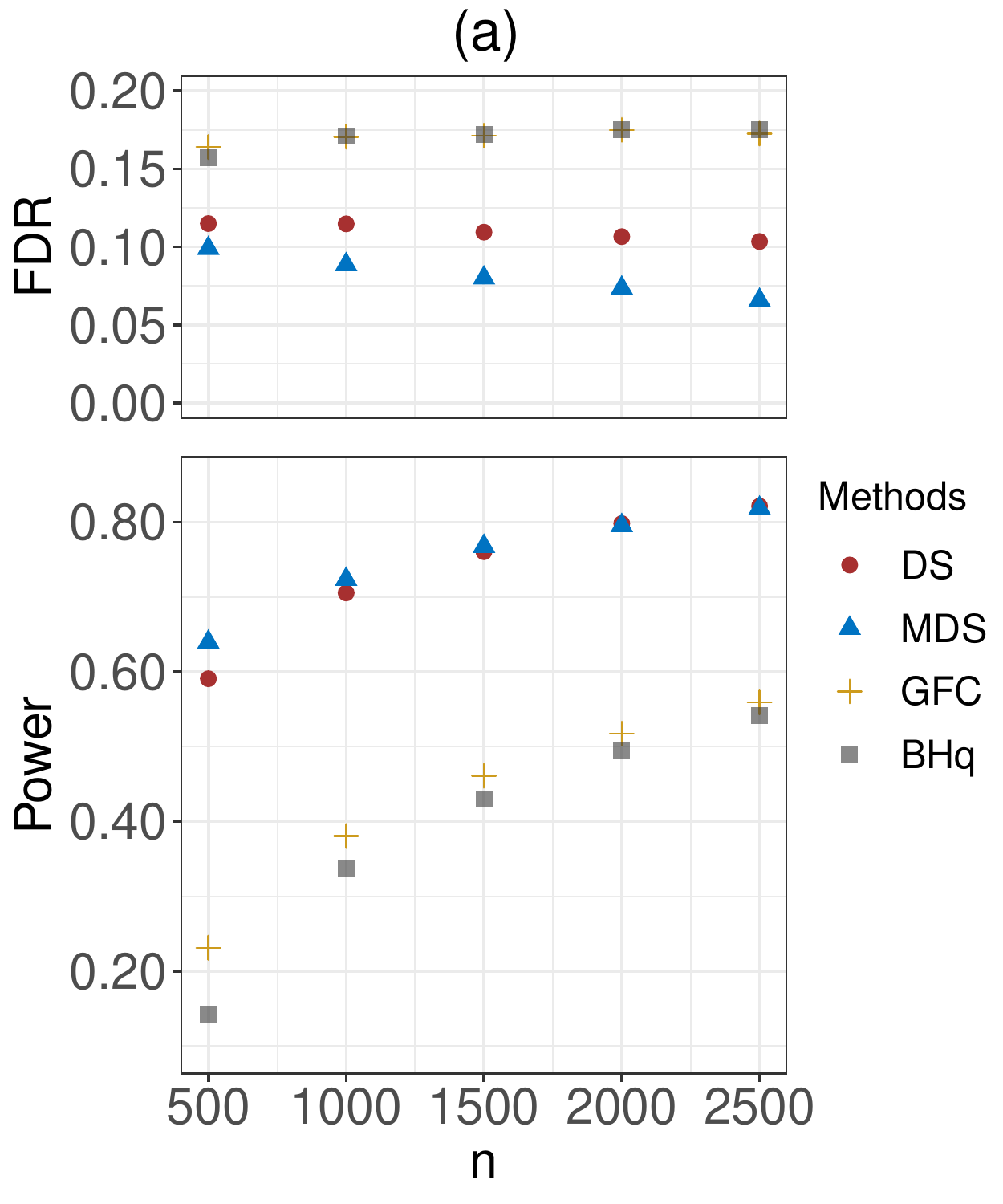}
\end{subfigure}%
\begin{subfigure}[b]{0.45\columnwidth}
\includegraphics[width=1.0\columnwidth]{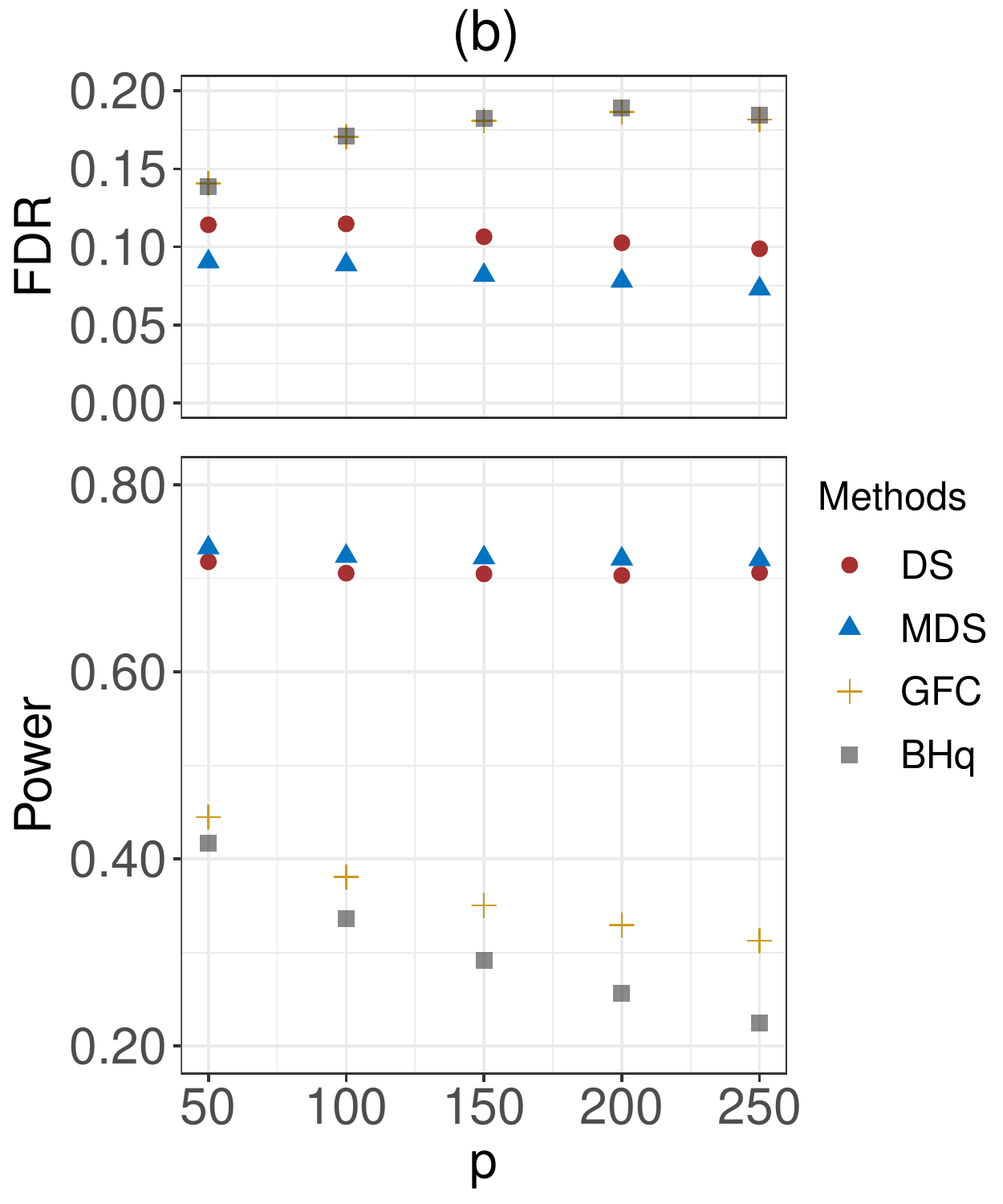}
\end{subfigure}\\
\begin{subfigure}[b]{0.45\columnwidth}
\centering
\includegraphics[width=1.0\columnwidth]{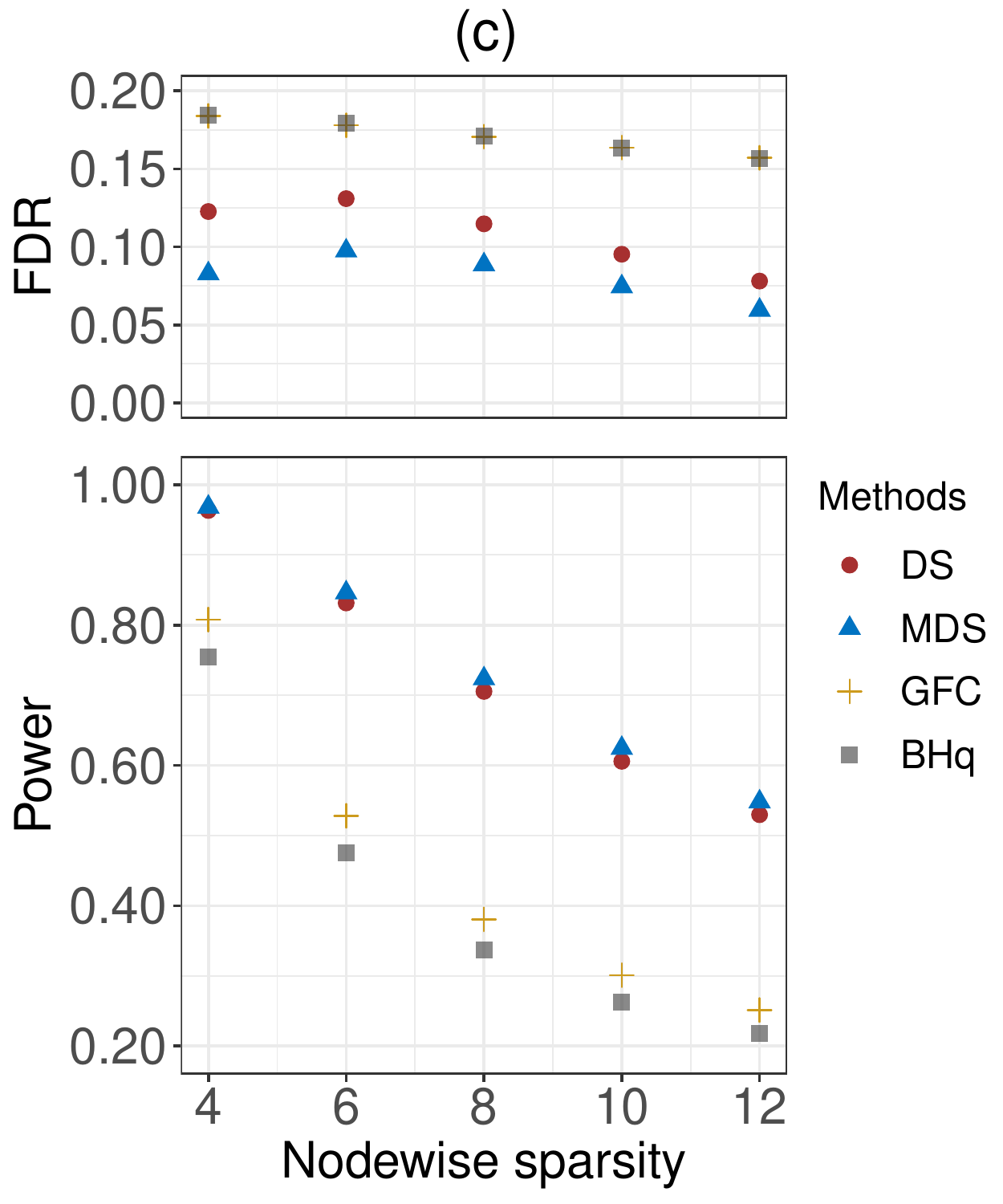}
\end{subfigure}%
\begin{subfigure}[b]{0.45\columnwidth}
\centering
\includegraphics[width=1.0\columnwidth]{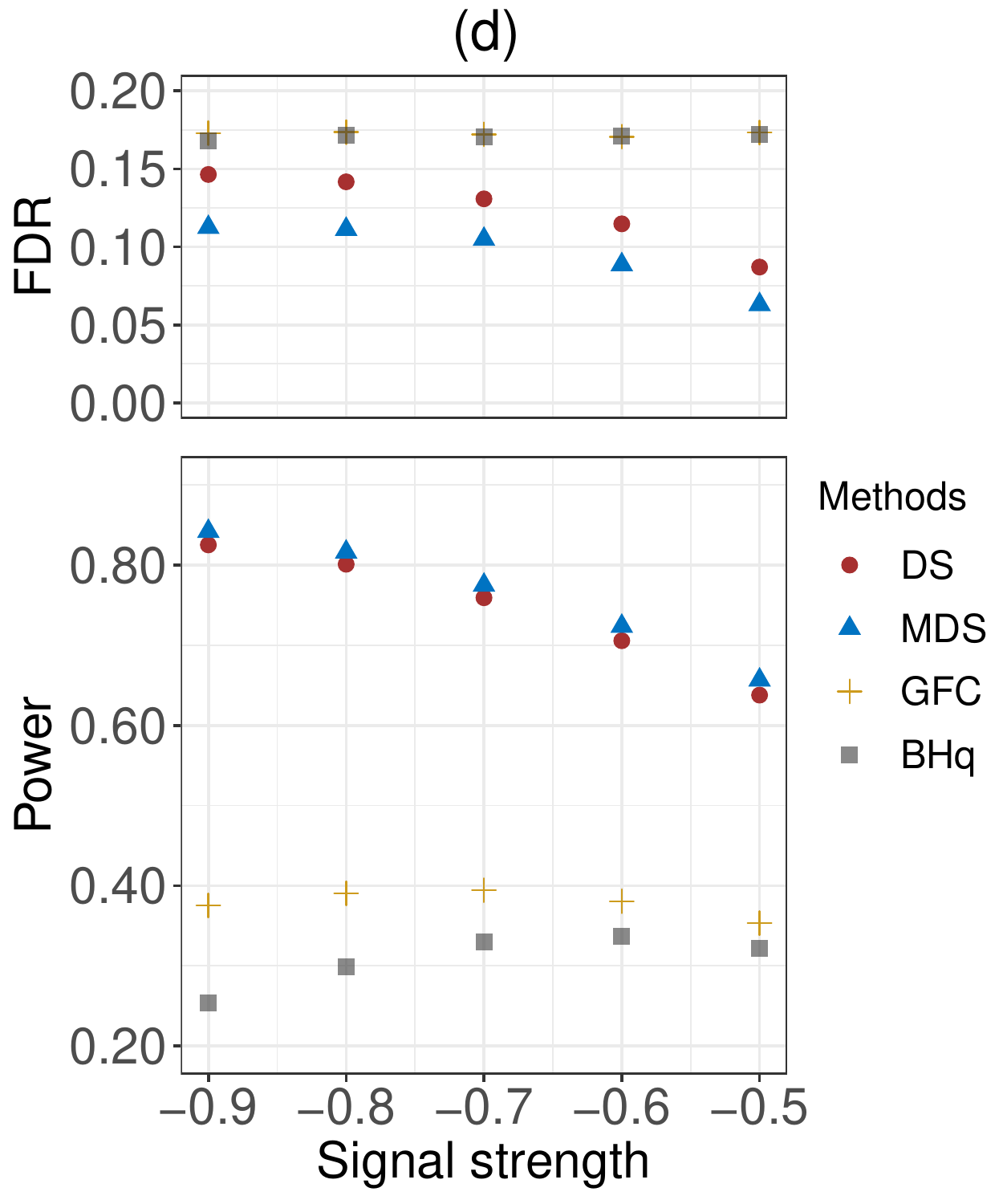}
\end{subfigure}\\
\caption{Empirical FDRs and powers for the banded graphs. The designated FDR control level is $q = 0.2$. Each dot in the figure represents the average from 50 independent runs.}
\label{fig:banded-graph}
\end{figure*}

\begin{figure*}
\begin{center}
\includegraphics[width=0.45\columnwidth]{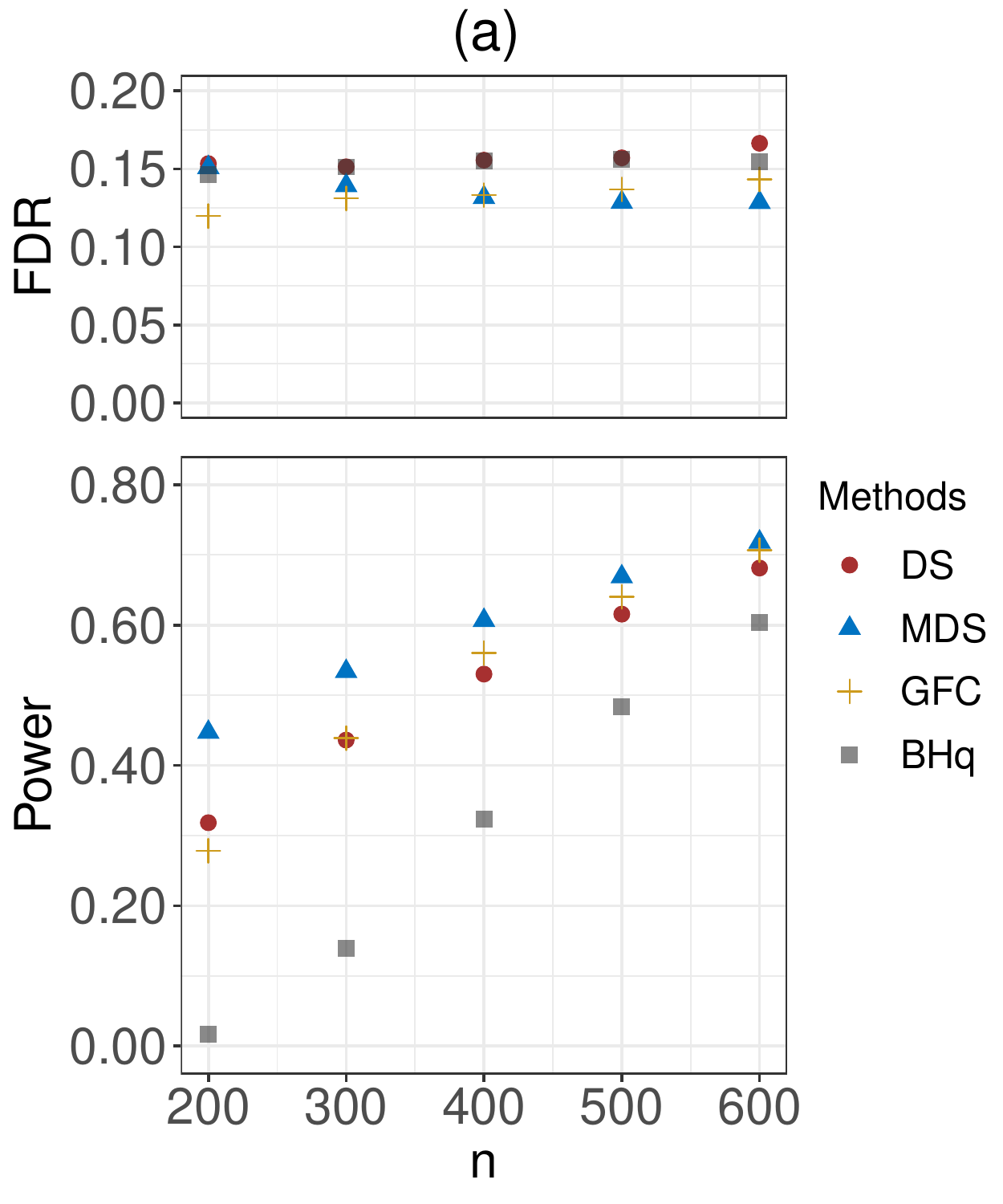}
\includegraphics[width=0.45\columnwidth]{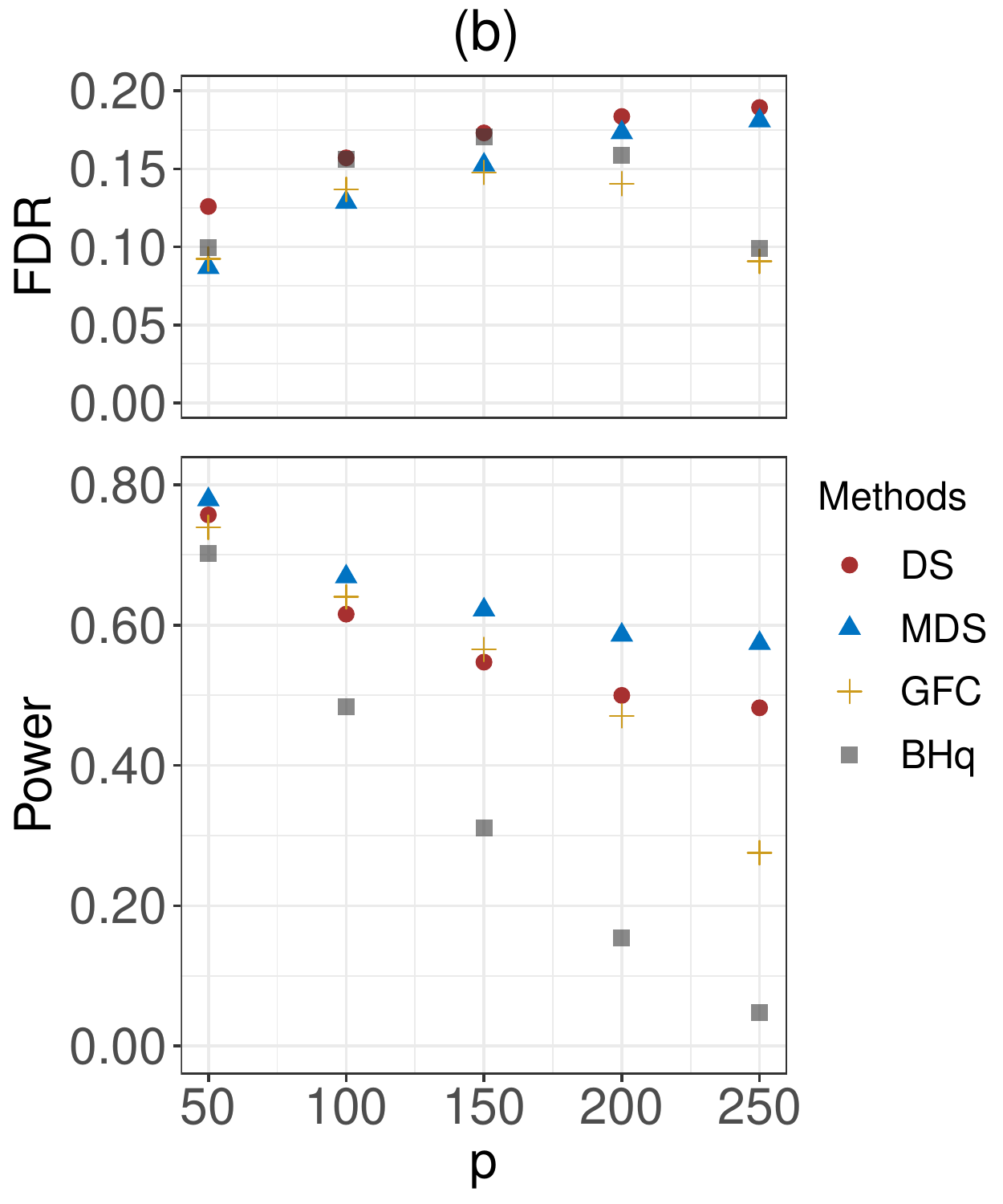}
\end{center}
\caption{
Empirical FDRs and powers for the blockwise diagonal graphs. The designated FDR control level is $q = 0.2$. Each dot in the figure represents the average from 50 independent runs.}
\label{fig:blockwise-graph}
\end{figure*}

The results for the banded graphs and the blockwise diagonal graphs are summarized in Figures \ref{fig:banded-graph} and \ref{fig:blockwise-graph}, respectively. 
We see that all the methods achieve FDR control at the designated level across different scenarios. For the banded graphs, DS and MDS are the two leading methods with significantly higher powers and also lower FDRs compared to the other two competing methods.
GFC and BHq perform similarly, of which GFC has a slightly higher power when $p$ is large or the signal strength is strong. 
In panel (d) of Figure \ref{fig:banded-graph}, the power of BHq exhibits an opposite trend compared to the other methods. One possible reason is that the pairwise correlation decreases when we increase $a$ from -0.9 to -0.5. Thus, the power of BHq increases as the p-values become less correlated.
For the blockwise diagonal graphs, MDS performs the best across all scenarios, enjoying a higher power and also a lower FDR compared to DS.
GFC performs similarly as DS in most scenarios, except for the case when $p$ is large, in which the power of GFC drops significantly.

\subsection{Real data application: HIV drug resistance}
\label{sec:HIV}
We apply DS and MDS to detect mutations in the Human Immunodeficiency Virus Type 1 (HIV-1) that are associated with drug resistance. The data set, which has also been analyzed in \citet{rhee2006genotypic}, \citet{barber2015controlling}, and \citet{lu2018deeppink}, contains resistance measurements of seven drugs for protease inhibitors (PIs), six drugs for nucleoside reverse-transcriptase inhibitors (NRTIs), and three drugs for nonnucleoside reverse transcriptase inhibitors (NNRTIs). We focus on the first two classes of inhibitors, PI and NRTI. 

\begin{figure*}
\begin{center}
\includegraphics[width=0.27\columnwidth]{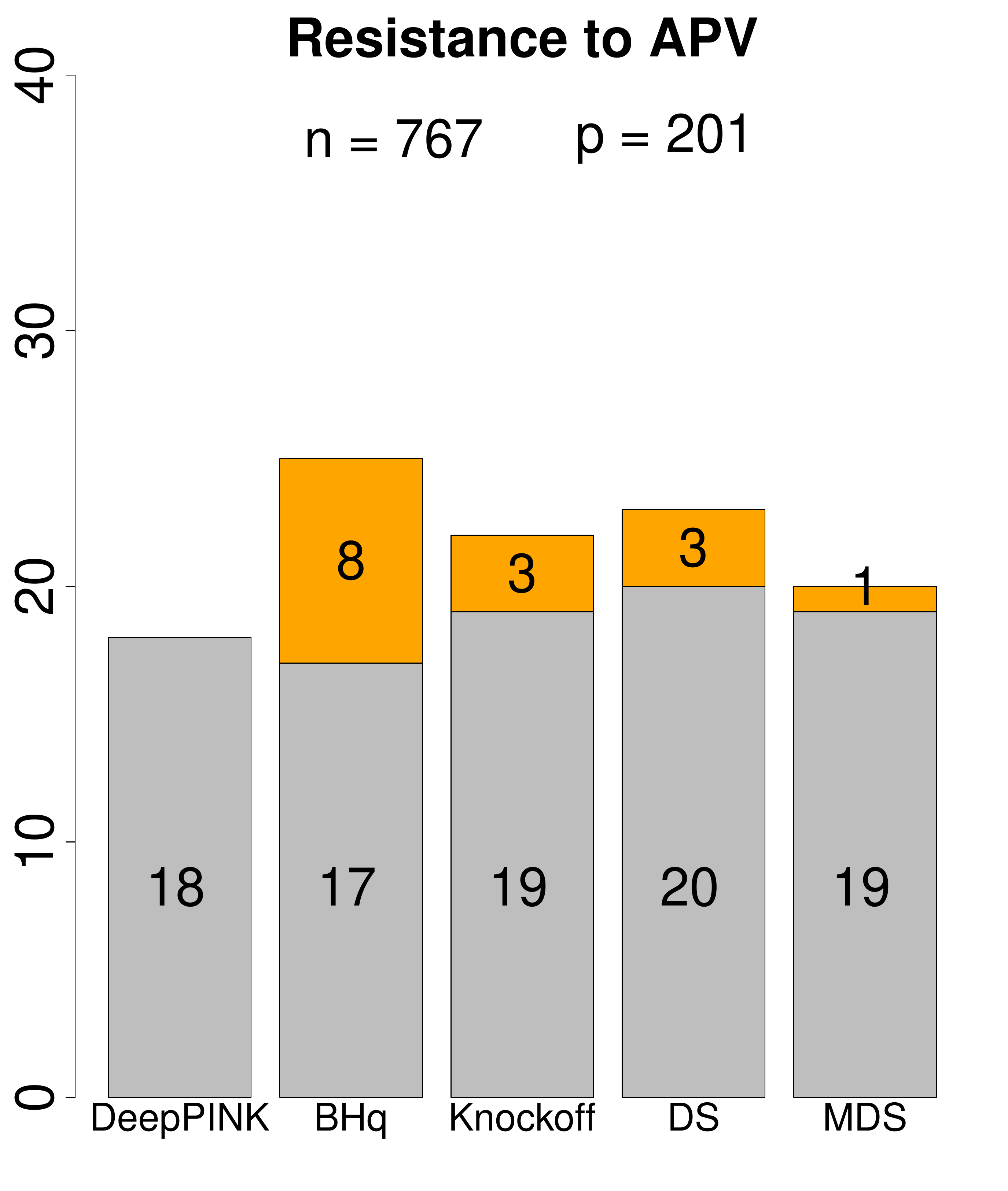}
\includegraphics[width=0.27\columnwidth]{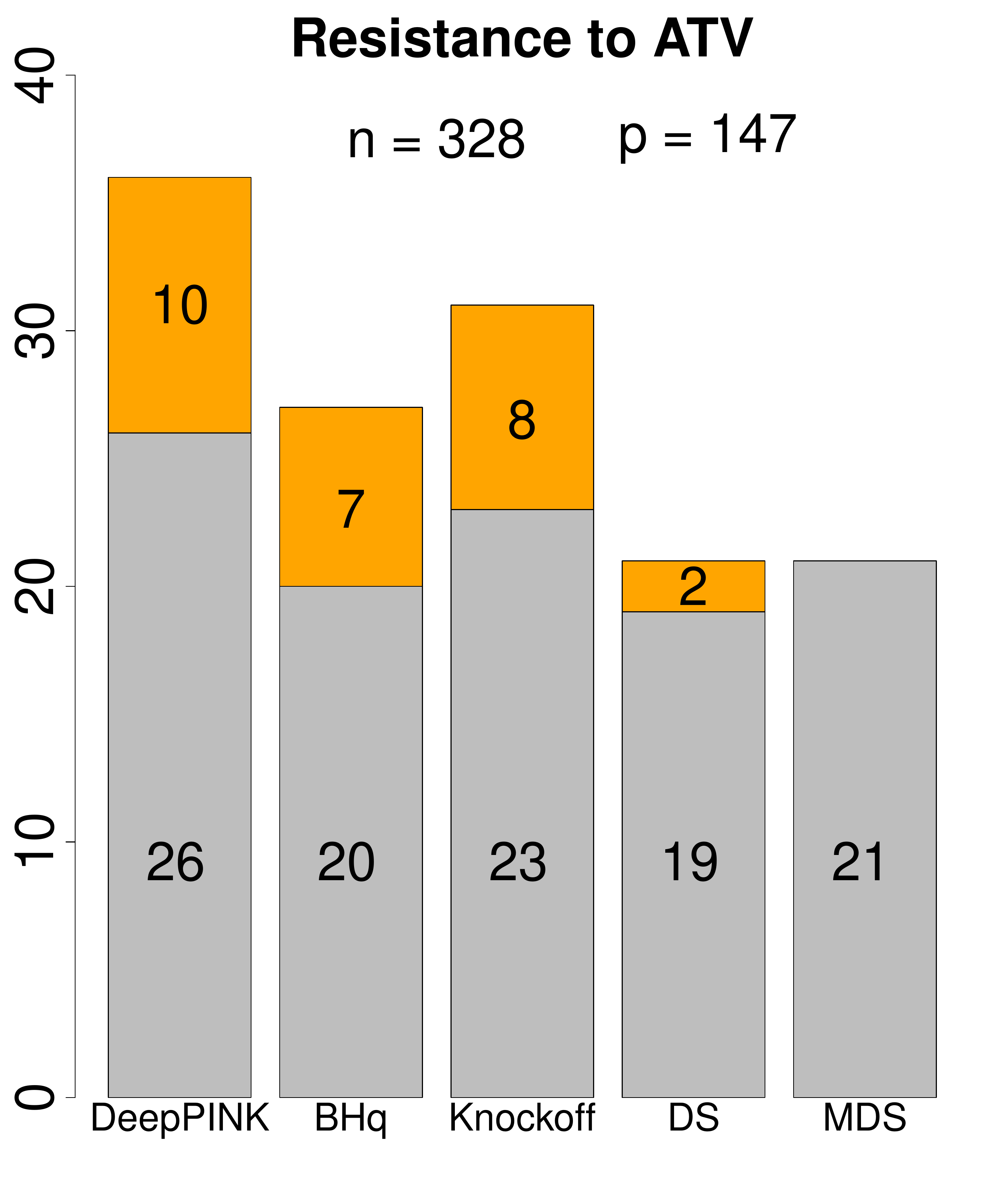}
\includegraphics[width=0.27\columnwidth]{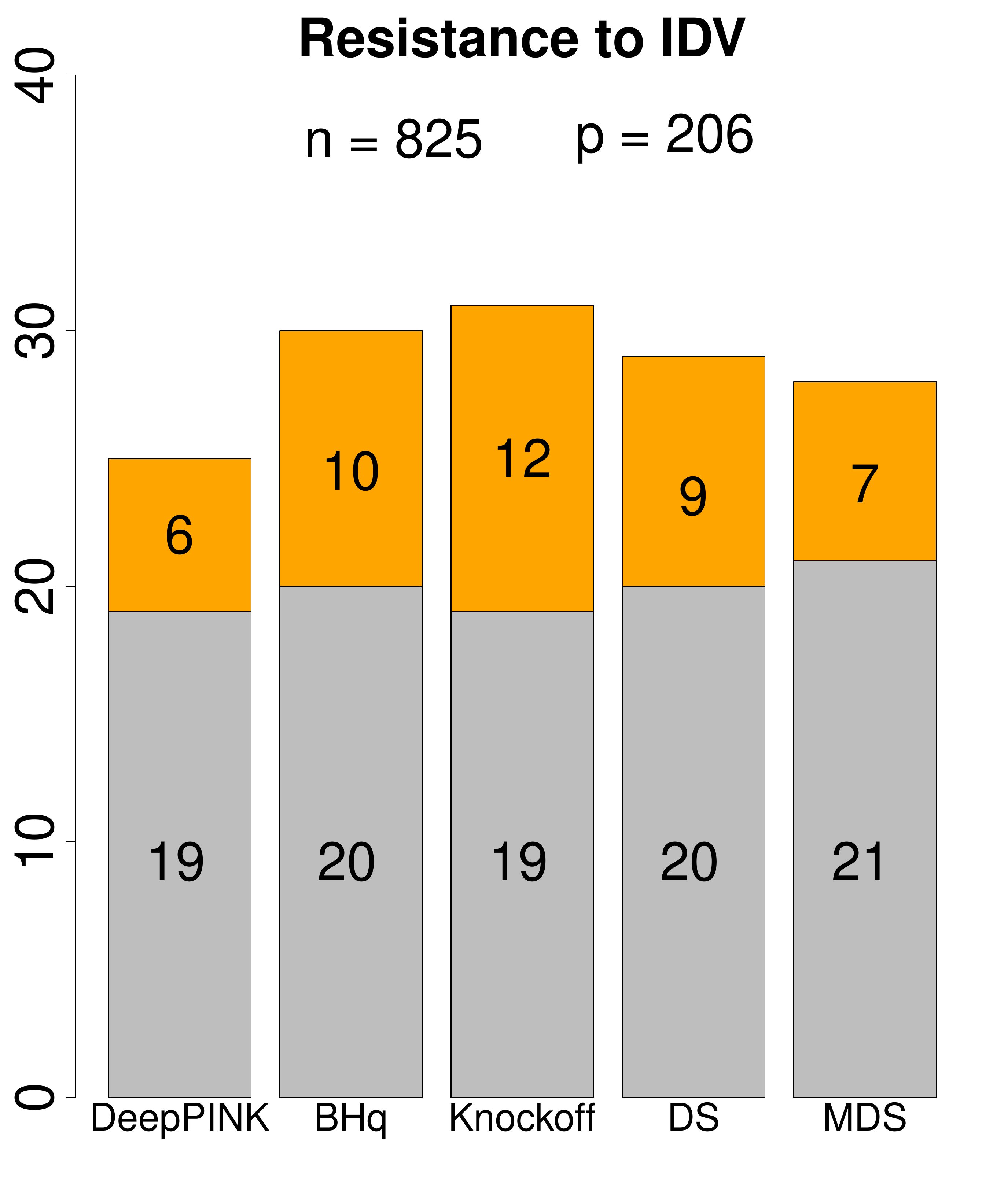}\\
\includegraphics[width=0.27\columnwidth]{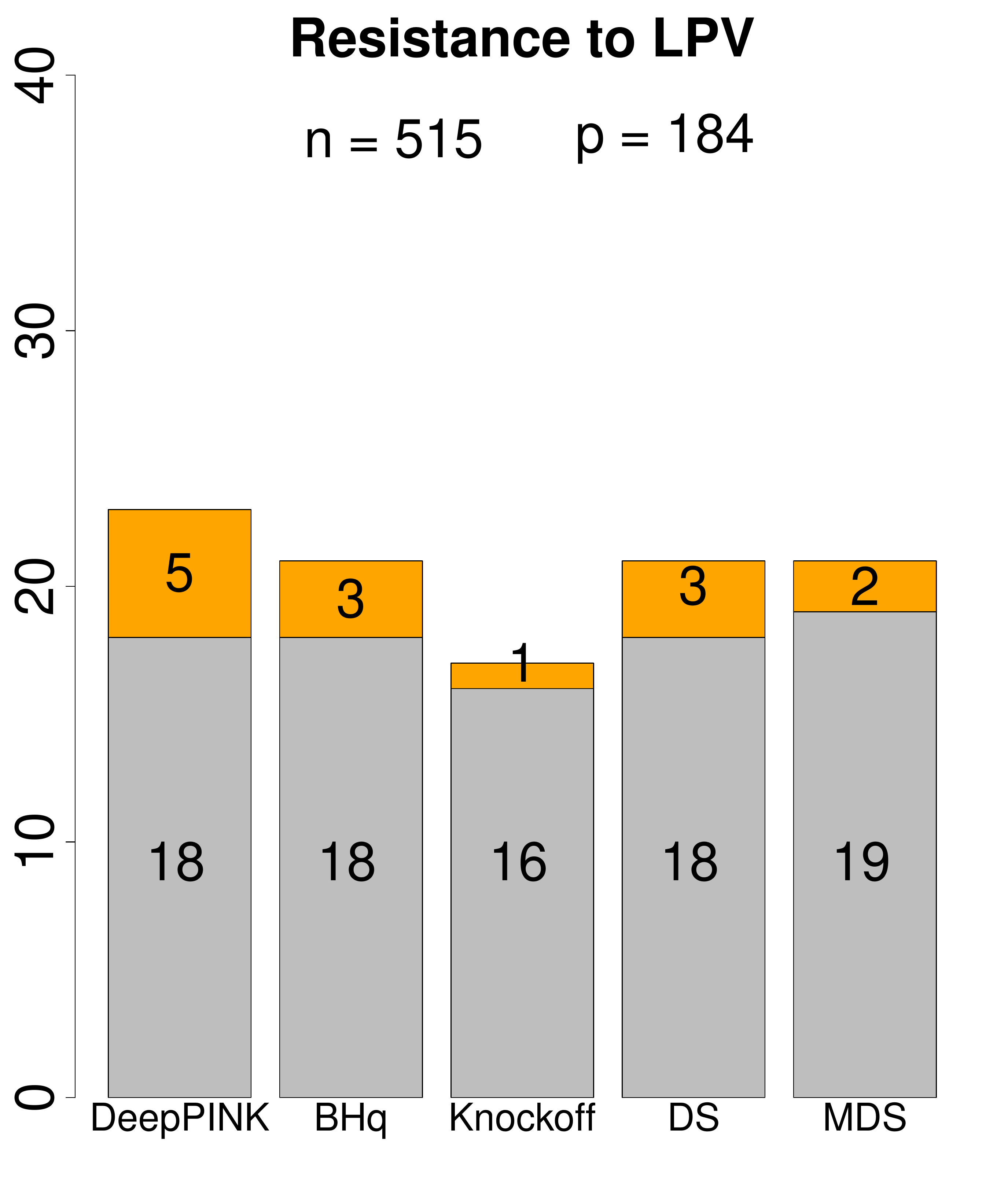}
\includegraphics[width=0.27\columnwidth]{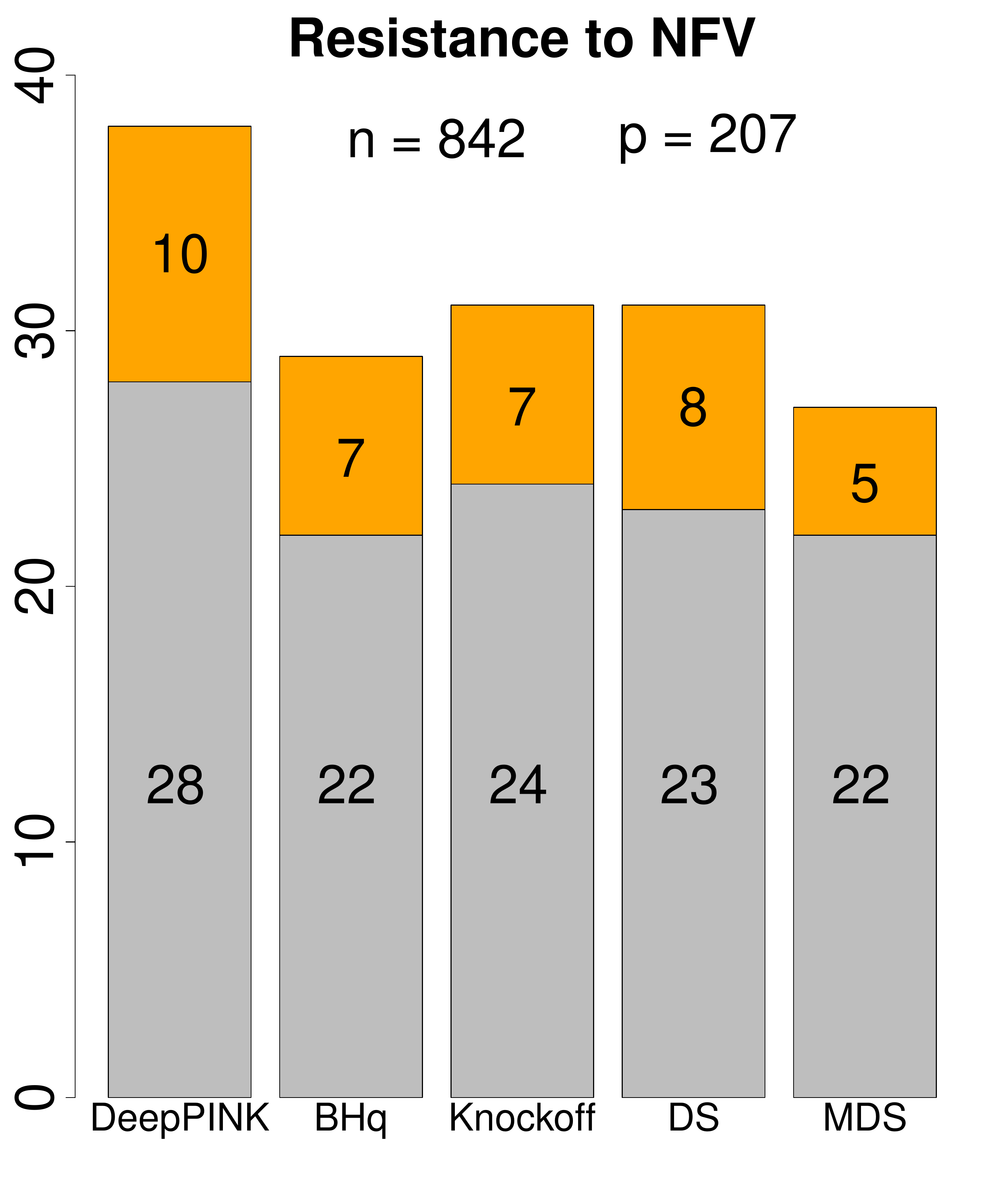}
\includegraphics[width=0.27\columnwidth]{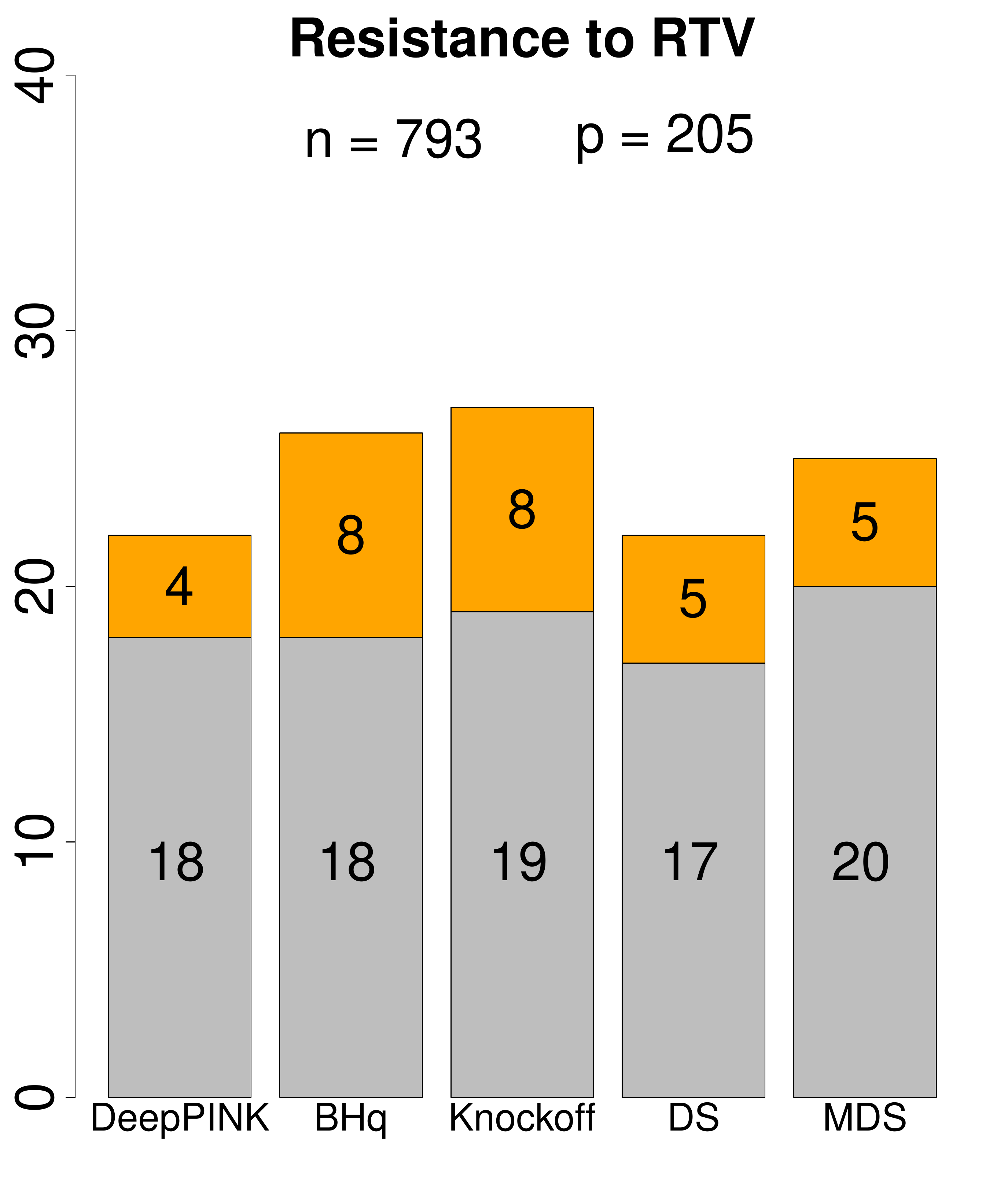}\\
\end{center}
\hspace{1.4cm}
\includegraphics[width=0.27\columnwidth]{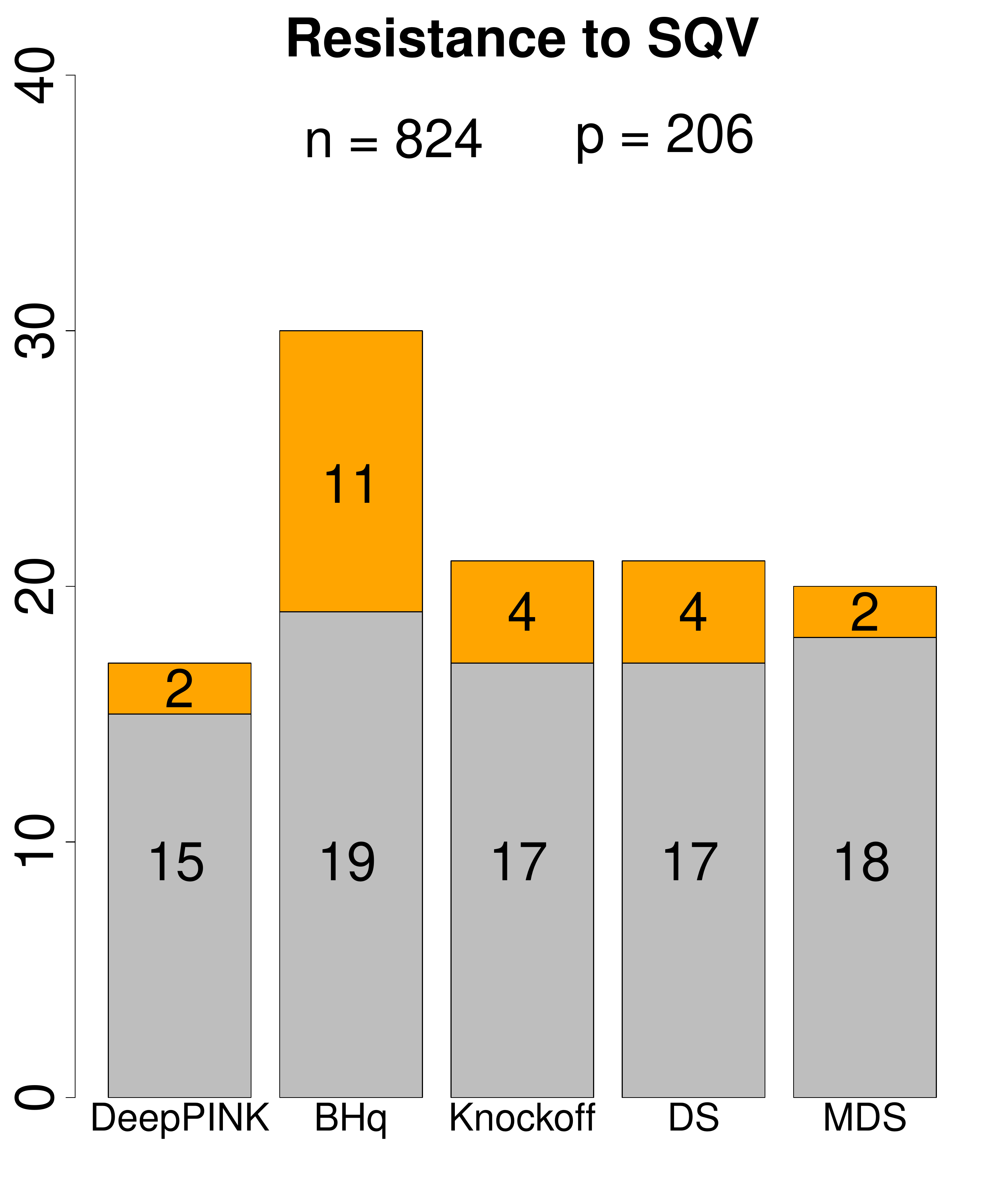}
\caption{Numbers of the discovered mutations for the seven PI drugs. The grey  and orange bars represent the numbers of true and false positives, respectively. The designated FDR control level is $q = 0.2$.}
\label{fig:HIV-A}
\end{figure*}

\begin{figure*}
\begin{center}
\includegraphics[width=0.27\columnwidth]{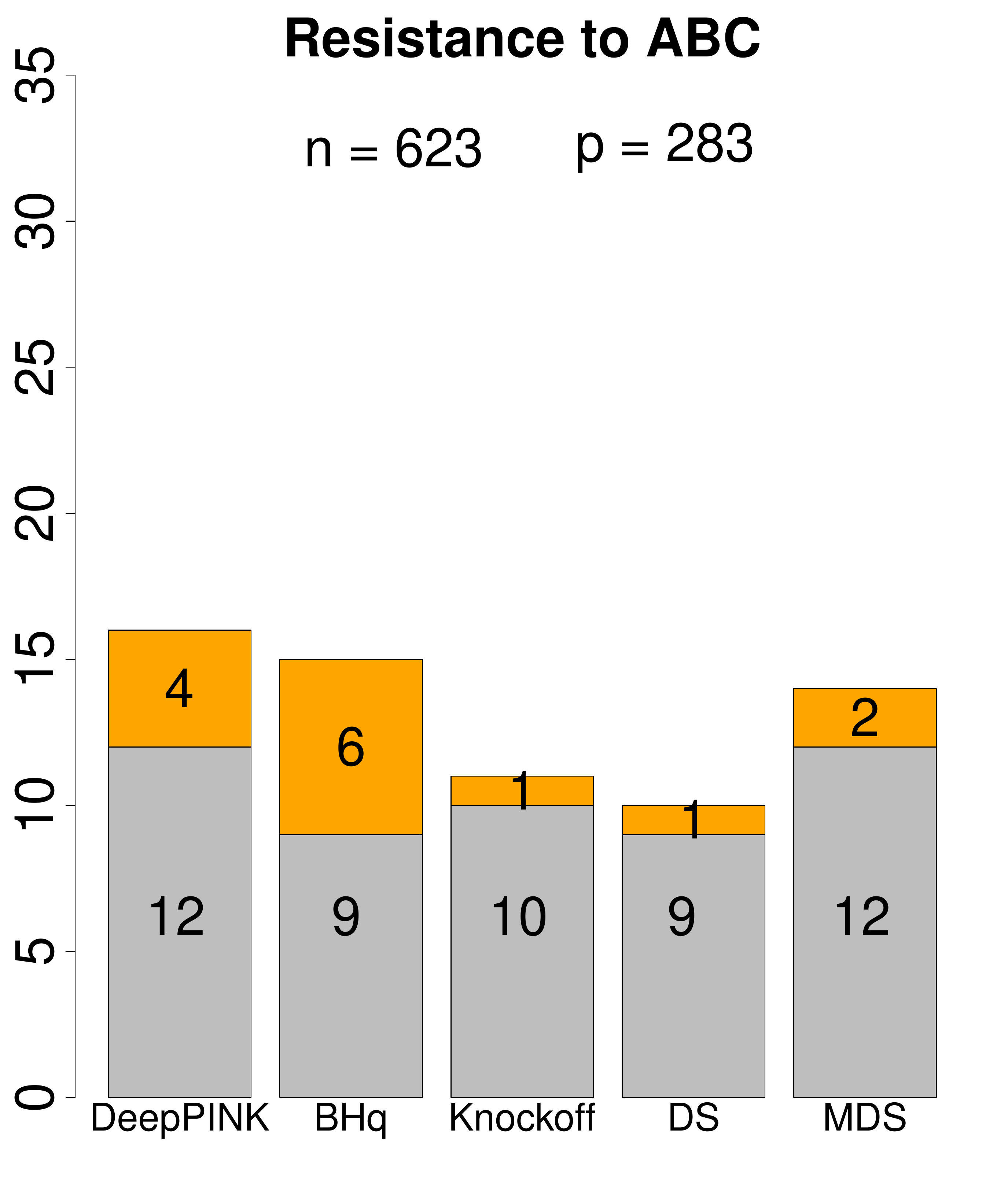}
\includegraphics[width=0.27\columnwidth]{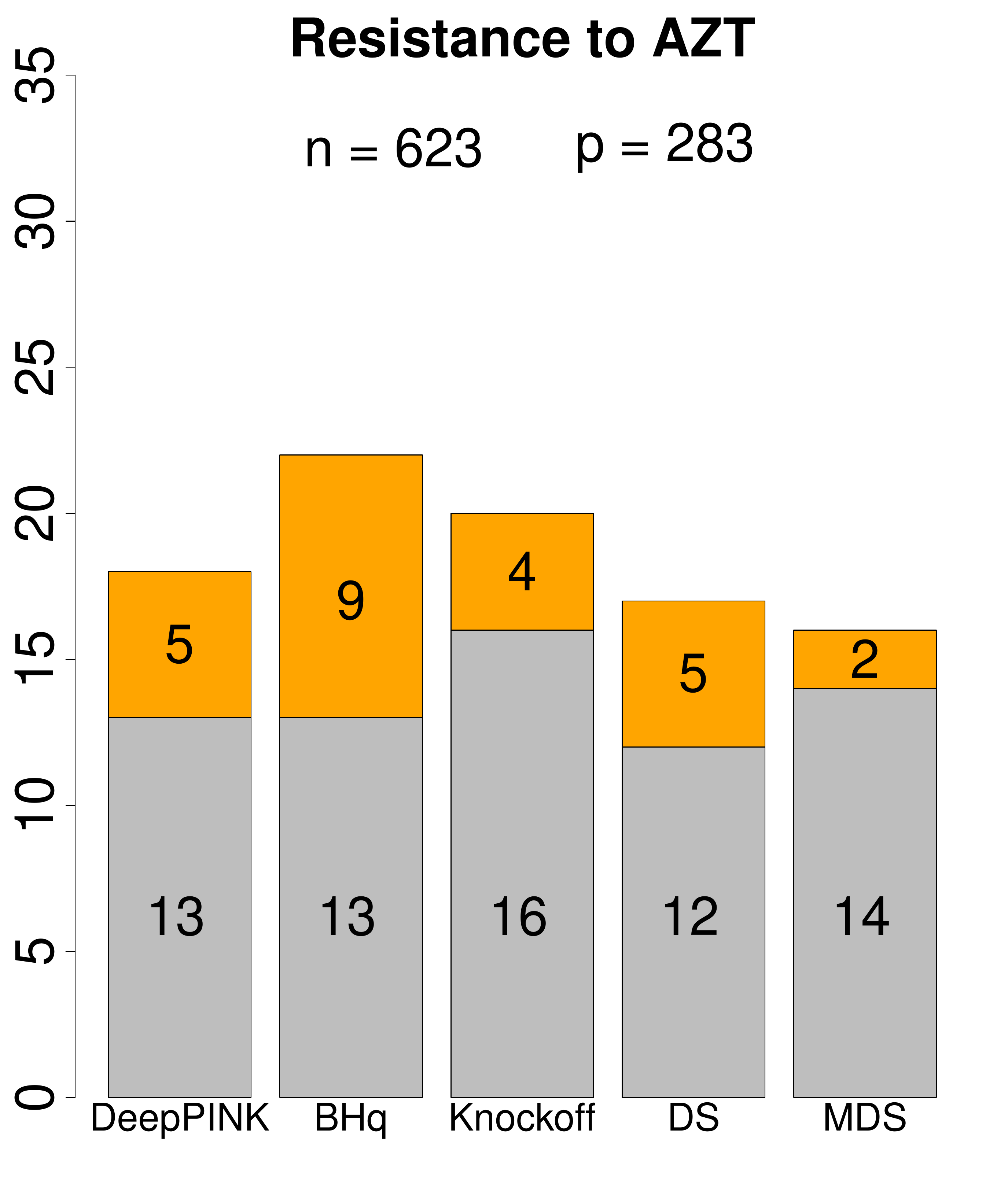}
\includegraphics[width=0.27\columnwidth]{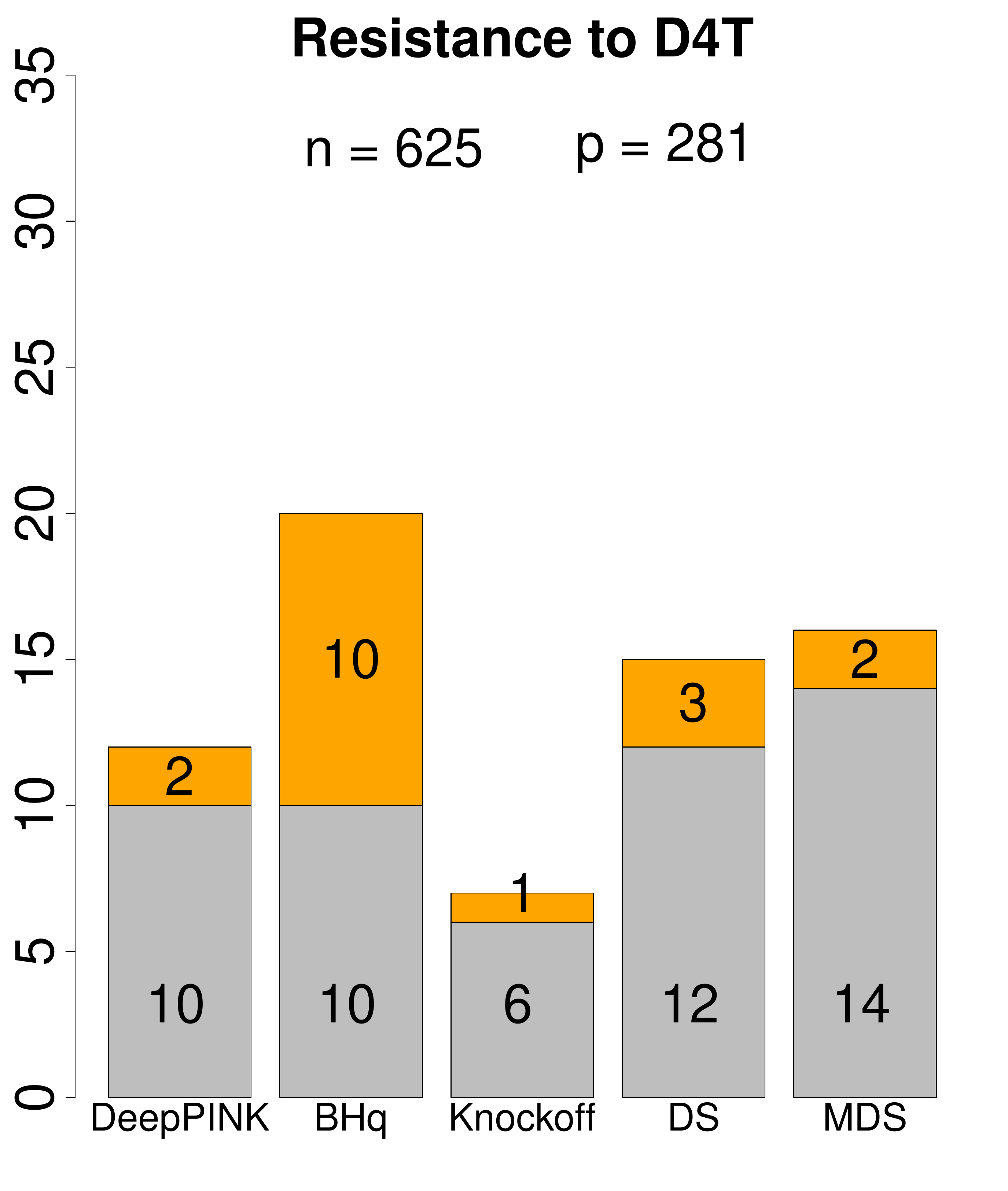}\\
\includegraphics[width=0.27\columnwidth]{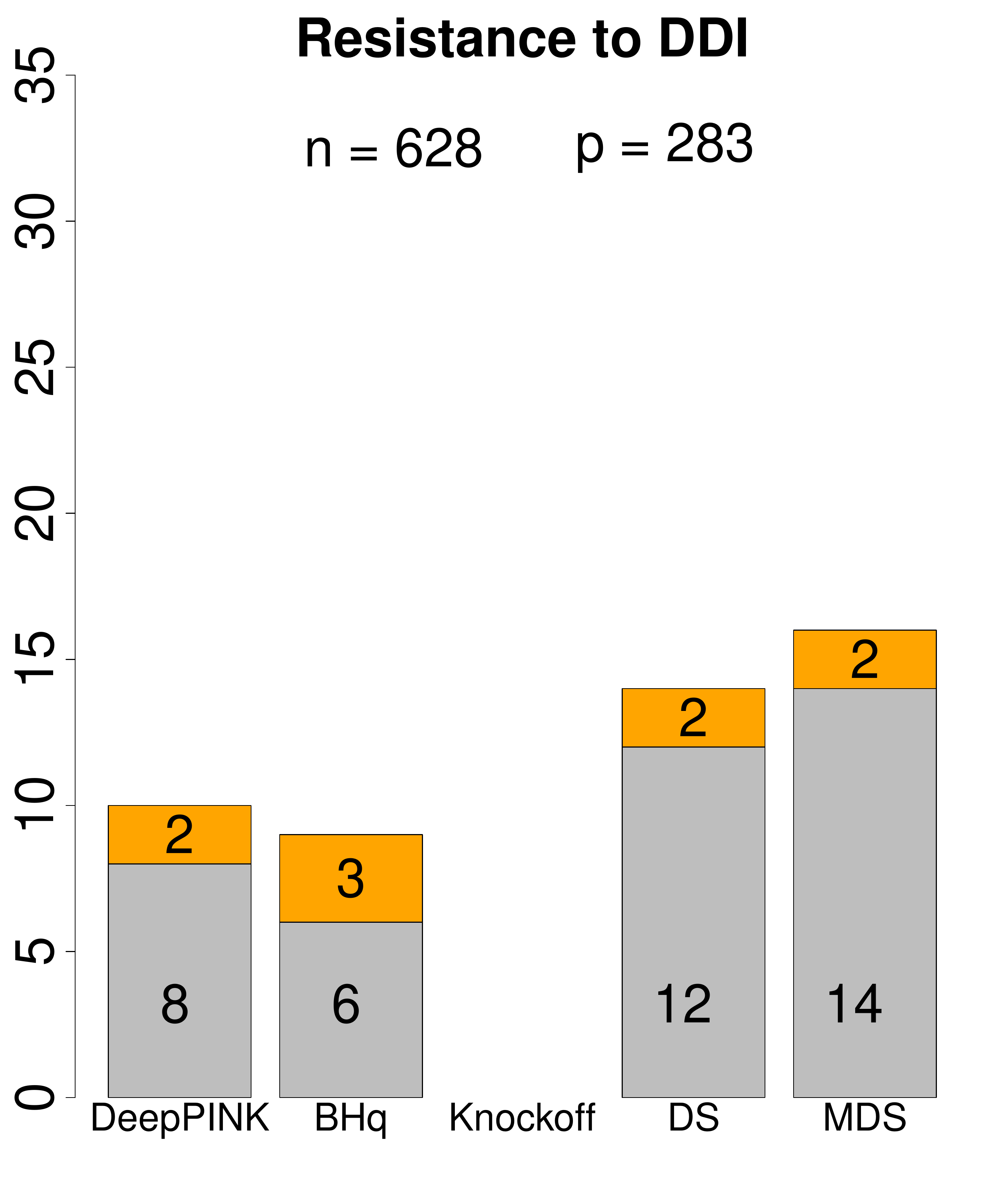}
\includegraphics[width=0.27\columnwidth]{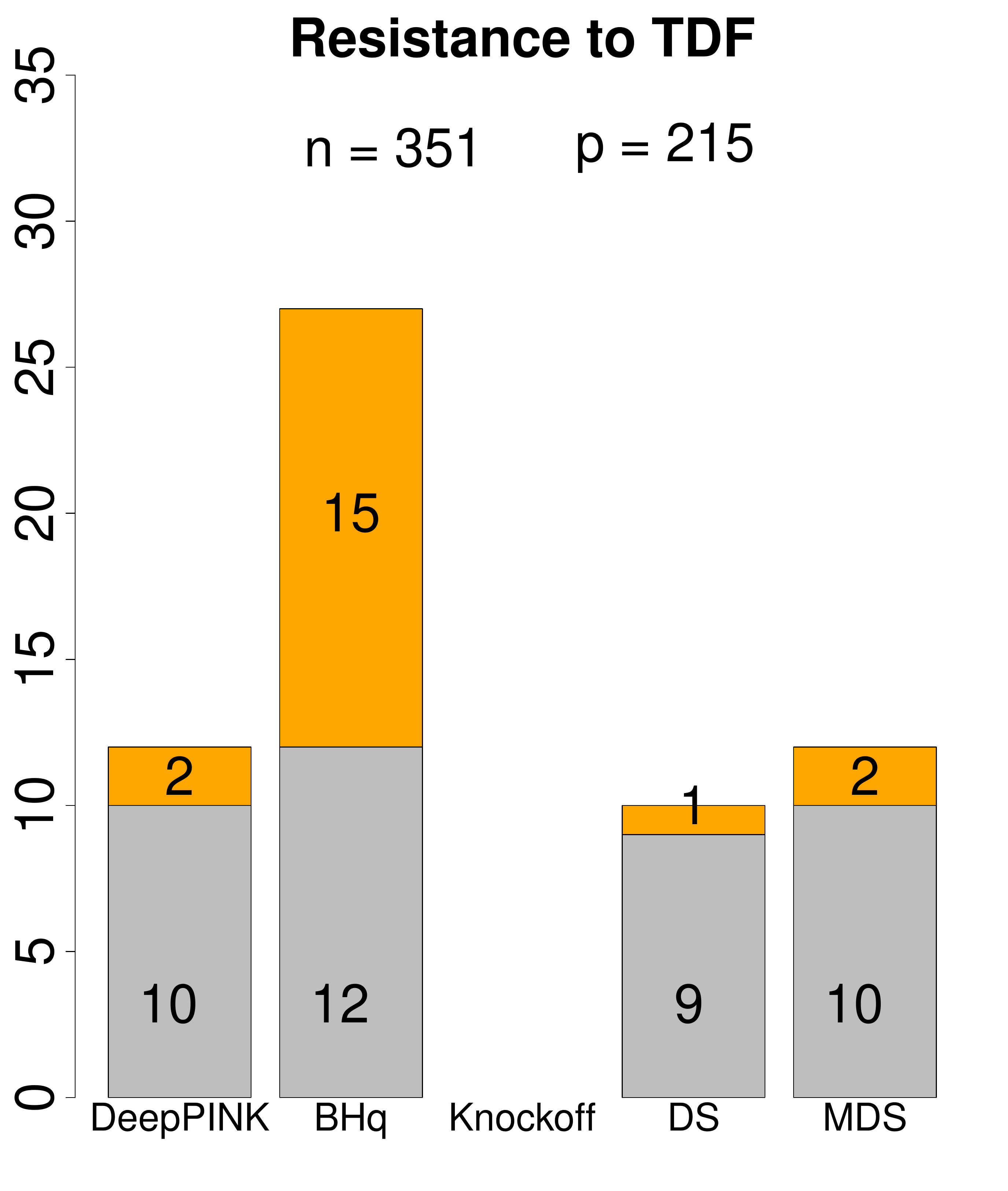}
\includegraphics[width=0.27\columnwidth]{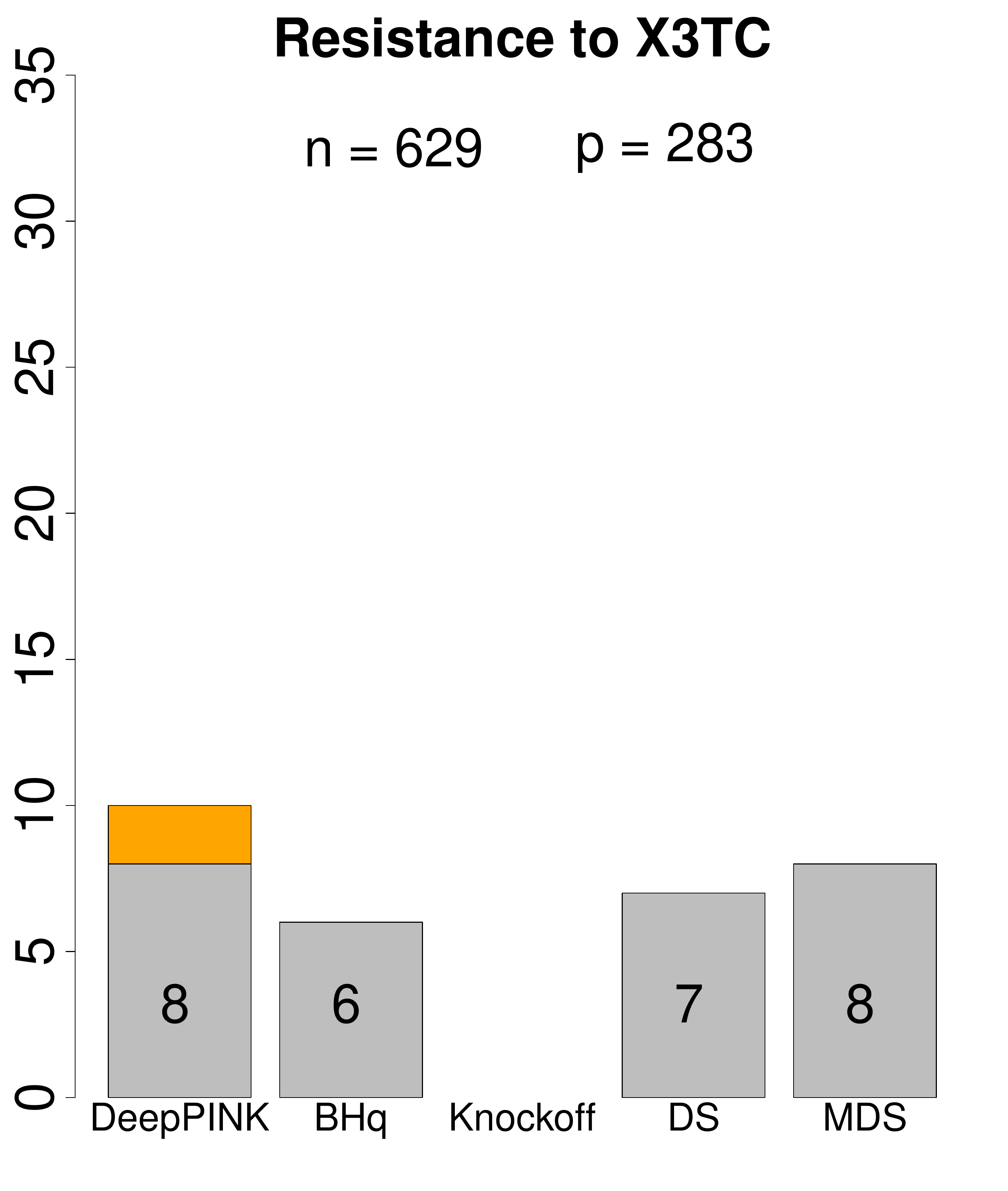}
\end{center}
\caption{Numbers of the discovered mutations for the six NRTI drugs. The grey and orange bars represent the numbers of true and false positives, respectively. The designated FDR control level is $q = 0.2$.}
\label{fig:HIV-B}
\end{figure*}

The response vector $\bm{y}$ calibrates the log-fold-increase of the lab-tested drug resistance. The design matrix $\bm{X}$ is binary, in which the $j$-th column indicates the presence or absence of the $j$-th mutation. The task is to select relevant mutations for each inhibitor against different drugs.
The data is preprocessed as follows. First, we remove the patients with missing drug resistance information. Second, we exclude those mutations that appear fewer than three times across all patients. The sample size $n$ and the number of mutations $p$ vary from drug to drug, but are all in hundreds with $n/p$ ranging from 1.5 to 4 (see Figures \ref{fig:HIV-A} and \ref{fig:HIV-B}). We assume a linear model between the response and features with no interactions.

Five methods are compared, including DeepPINK with the model-X knockoffs \citep{lu2018deeppink}, the fixed-design knockoff filter \citep{barber2015controlling}, BHq, DS, and MDS. For DeepPINK, the knockoff filter, and BHq, we report the selection results obtained in \citet{lu2018deeppink}. The designated FDR control level is $q = 0.2$ throughout.
As in \citet{barber2015controlling}, we treat the existing treatment-selected mutation (TSM) panels \citep{rhee2005hiv} as the ground truth. 

For PI, the number of discovered mutations for each drug, including the number of true and false positives, are summarized in Figure \ref{fig:HIV-A}. We see that MDS performs the best for three out of seven PI drugs, including ATV, LPV and SQV. For drugs APV, IDV, and RTV, MDS is comparable to DeepPINK, and both perform better than the knockoff filter and BHq. For drug NFV, MDS and the knockoff filter are the two leading methods. 
Figure \ref{fig:HIV-B} shows the corresponding results for the NRTI drugs. Among the six NRTI drugs, MDS performs the best in four, including ABC, D4T, DDI, and X3TC. For drug AZT, MDS and the knockoff filter perform the best. For drug TDF, MDS is comparable to DeepPINK, and both are much better than BHq and the knockoff filter. In particular, we see that  the knockoff filter has no power and does not select any mutations for  drugs DDI, TDF, and X3TC. 

\section{Concluding Remarks}
\label{sec:conclusion}
We have described a general data-splitting framework to control the FDR for feature selection in high-dimensional regressions. We theoretically prove and empirically demonstrate that the proposed approaches (DS and MDS) allow us to asymptotically control the FDR in canonical statistical models including linear and Gaussian graphical models. MDS is shown to be a particularly attractive strategy as it  helps stabilize the selection result and improves the power. Both DS and MDS  require little knowledge on the joint distribution of features in consideration, and are conceptually simple and easy to implement based upon existing softwares for high-dimensional regression methods. 

We conclude by pointing out several directions for future work. First, for linear models, an interesting extension of the Lasso + OLS procedure is to consider features with a group structure. A natural strategy is to substitute Lasso with group Lasso. However, unlike Lasso, group Lasso can potentially select more than $n$ features ($n$ is the sample size), thus the companion OLS step, which guarantees the symmetric assumption, may not be easily applied. Second, we would like to investigate the applicability and the theoretical properties of DS and MDS  for dealing with  neural networks and other nonlinear models, in order to handle more complex data such as images and natural languages. Third, it is of interest to investigate the multiple testing problem in sparse high-dimensional covariance matrix estimation, where some thresholding estimators are typically employed. Our proposed framework is applicable as long as the estimator of any zero covariance entry is symmetric about 0. Last but not the least, extensions of  the data-splitting  framework to handle data containing dependent observations or having hierarchical structures are of immediate interest.

\section*{Acknowledge}
We thank Lucas Janson, Wenshuo Wang, and Dongming Huang for many helpful discussions and constructive suggestions. We also thank the three anonymous referees for their detailed critiques, especially the helpful comments on implementing the model-X knockoff filter.
This research is supported in part by the National Science Foundation grants   DMS-1903139 and DMS-2015411.

\bibliographystyle{chicago}
\bibliography{FDR.bib}

\begin{thebibliography}{}

\bibitem[\protect\citeauthoryear{Azriel and Schwartzman}{Azriel and
  Schwartzman}{2015}]{azriel2015empirical}
Azriel, D. and A.~Schwartzman (2015).
\newblock The empirical distribution of a large number of correlated {N}ormal
  variables.
\newblock {\em Journal of the American Statistical Association\/}~{\em
  110\/}(511), 1217--1228.

\bibitem[\protect\citeauthoryear{Barber and Cand{\`e}s}{Barber and
  Cand{\`e}s}{2015}]{barber2015controlling}
Barber, R.~F. and E.~J. Cand{\`e}s (2015).
\newblock Controlling the false discovery rate via knockoffs.
\newblock {\em The Annals of Statistics\/}~{\em 43\/}(5), 2055--2085.

\bibitem[\protect\citeauthoryear{Barber and Cand{\`e}s}{Barber and
  Cand{\`e}s}{2019}]{barber2019knockoff}
Barber, R.~F. and E.~J. Cand{\`e}s (2019).
\newblock A knockoff filter for high-dimensional selective inference.
\newblock {\em The Annals of Statistics\/}~{\em 47\/}(5), 2504--2537.

\bibitem[\protect\citeauthoryear{Barber, Cand{\`e}s, and Samworth}{Barber
  et~al.}{2020}]{barber2020}
Barber, R.~F., E.~J. Cand{\`e}s, and R.~J. Samworth (2020).
\newblock Robust inference with knockoffs.
\newblock {\em The Annals of Statistics\/}~{\em 48\/}(3), 1409--1431.

\bibitem[\protect\citeauthoryear{Bates, Cand{\'e}s, Janson, and Wang}{Bates
  et~al.}{2020}]{2020wenshuo}
Bates, S., E.~J. Cand{\'e}s, L.~Janson, and W.~Wang (2020).
\newblock Metropolized knockoff sampling.
\newblock {\em Journal of the American Statistical Association\/}.

\bibitem[\protect\citeauthoryear{Benjamini and Gavrilov}{Benjamini and
  Gavrilov}{2009}]{2009Benjaminiforward}
Benjamini, Y. and Y.~Gavrilov (2009).
\newblock A simple forward selection procedure based on false discovery rate
  control.
\newblock {\em The Annals of Applied Statistics\/}~{\em 3\/}(1), 179--198.

\bibitem[\protect\citeauthoryear{Benjamini and Hochberg}{Benjamini and
  Hochberg}{1995}]{benjamini1995controlling}
Benjamini, Y. and Y.~Hochberg (1995).
\newblock Controlling the false discovery rate: a practical and powerful
  approach to multiple testing.
\newblock {\em Journal of the Royal Statistical Society: Series B (Statistical
  Methodology)\/}~{\em 57\/}(1), 289--300.

\bibitem[\protect\citeauthoryear{Benjamini and Yekutieli}{Benjamini and
  Yekutieli}{2001}]{benjamini2001control}
Benjamini, Y. and D.~Yekutieli (2001).
\newblock The control of the false discovery rate in multiple testing under
  dependency.
\newblock {\em The Annals of Statistics\/}~{\em 29\/}(4), 1165--1188.

\bibitem[\protect\citeauthoryear{Bickel, Ritov, and Tsybakov}{Bickel
  et~al.}{2009}]{bickel2009simultaneous}
Bickel, P.~J., Y.~Ritov, and A.~B. Tsybakov (2009).
\newblock Simultaneous analysis of {L}asso and {D}antzig selector.
\newblock {\em The Annals of Statistics\/}~{\em 37\/}(4), 1705--1732.

\bibitem[\protect\citeauthoryear{Bogdan, Berg, Sabatti, Su, and
  Cand{\`e}s}{Bogdan et~al.}{2015}]{2015Bogdan}
Bogdan, M., E.~Berg, C.~Sabatti, W.~Su, and E.~J. Cand{\`e}s (2015).
\newblock Slope -- adaptive variable selection via convex optimization.
\newblock {\em The Annals of Applied Statistics\/}~{\em 9\/}(3), 1103--1150.

\bibitem[\protect\citeauthoryear{Cand{\`e}s, Fan, Janson, and Lv}{Cand{\`e}s
  et~al.}{2018}]{candes2018panning}
Cand{\`e}s, E.~J., Y.~Fan, L.~Janson, and J.~Lv (2018).
\newblock Panning for gold: ‘model-{X}’ knockoffs for high dimensional
  controlled variable selection.
\newblock {\em Journal of the Royal Statistical Society: Series B (Statistical
  Methodology)\/}~{\em 80\/}(3), 551--577.

\bibitem[\protect\citeauthoryear{Clarke and Hall}{Clarke and
  Hall}{2009}]{clarke2009robustness}
Clarke, S. and P.~Hall (2009).
\newblock Robustness of multiple testing procedures against dependence.
\newblock {\em The Annals of Statistics\/}~{\em 37\/}(1), 332--358.

\bibitem[\protect\citeauthoryear{Cox}{Cox}{1975}]{cox1975note}
Cox, D.~R. (1975).
\newblock A note on data-splitting for the evaluation of significance levels.
\newblock {\em Biometrika\/}~{\em 62\/}(2), 441--444.

\bibitem[\protect\citeauthoryear{Dai, Lin, Xing, and Liu}{Dai
  et~al.}{2020}]{dai2020scale}
Dai, C., B.~Lin, X.~Xing, and J.~S. Liu (2020).
\newblock A scale-free approach for false discovery rate control in generalized
  linear models.
\newblock {\em arXiv preprint: 2007.01237\/}.

\bibitem[\protect\citeauthoryear{Dezeure, B{\"u}hlmann, Meier, and
  Meinshausen}{Dezeure et~al.}{2015}]{dezeure2015high}
Dezeure, R., P.~B{\"u}hlmann, L.~Meier, and N.~Meinshausen (2015).
\newblock High-dimensional inference: confidence intervals, p-values and
  {R}-software hdi.
\newblock {\em Statistical Science\/}, 533--558.

\bibitem[\protect\citeauthoryear{Efron}{Efron}{2005}]{2005localfdr}
Efron, B. (2005).
\newblock Local false discovery rates.
\newblock {\em Technical report\/}.

\bibitem[\protect\citeauthoryear{Efron, Tibshirani, Storey, and Tusher}{Efron
  et~al.}{2001}]{2001localfdr}
Efron, B., R.~Tibshirani, J.~D. Storey, and V.~Tusher (2001).
\newblock Empirical {B}ayes analysis of a microarray experiment.
\newblock {\em Journal of the American Statistical Association\/}~{\em
  96\/}(456), 1151--1160.

\bibitem[\protect\citeauthoryear{Efroymson}{Efroymson}{1960}]{efroymson1960multiple}
Efroymson, M. (1960).
\newblock Multiple regression analysis.
\newblock {\em Mathematical Methods for Digital Computers\/}, 191--203.

\bibitem[\protect\citeauthoryear{Fan, Demirkaya, Li, and Lv}{Fan
  et~al.}{2020}]{rank2020fan}
Fan, Y., E.~Demirkaya, G.~Li, and J.~Lv (2020).
\newblock Rank: large-scale inference with graphical nonlinear knockoffs.
\newblock {\em Journal of the American Statistical Association\/}~{\em
  115\/}(529), 362--379.

\bibitem[\protect\citeauthoryear{Hoffman, Papas, Trotter, and Archer}{Hoffman
  et~al.}{2020}]{hoffman2020single}
Hoffman, J.~A., B.~N. Papas, K.~W. Trotter, and T.~K. Archer (2020).
\newblock Single-cell {RNA} sequencing reveals a heterogeneous response to
  glucocorticoids in breast cancer cells.
\newblock {\em Communications Biology\/}~{\em 3\/}(1), 1--11.

\bibitem[\protect\citeauthoryear{Huang and Janson}{Huang and
  Janson}{2020}]{huang2020relaxing}
Huang, D. and L.~Janson (2020).
\newblock Relaxing the assumptions of knockoffs by conditioning.
\newblock {\em The Annals of Statistics\/}~{\em 48\/}(5), 3021--3042.

\bibitem[\protect\citeauthoryear{Ignatiadis, Klaus, Zaugg, and
  Huber}{Ignatiadis et~al.}{2016}]{ignatiadis2016data}
Ignatiadis, N., B.~Klaus, J.~B. Zaugg, and W.~Huber (2016).
\newblock Data-driven hypothesis weighting increases detection power in
  genome-scale multiple testing.
\newblock {\em Nature Methods\/}~{\em 13\/}(7), 577.

\bibitem[\protect\citeauthoryear{Javanmard and Javadi}{Javanmard and
  Javadi}{2019}]{javanmard2019false}
Javanmard, A. and H.~Javadi (2019).
\newblock False discovery rate control via debiased {L}asso.
\newblock {\em Electronic Journal of Statistics\/}~{\em 13\/}(1), 1212--1253.

\bibitem[\protect\citeauthoryear{Javanmard and Montanari}{Javanmard and
  Montanari}{2014}]{javanmard2014confidence}
Javanmard, A. and A.~Montanari (2014).
\newblock Confidence intervals and hypothesis testing for high-dimensional
  regression.
\newblock {\em The Journal of Machine Learning Research\/}~{\em 15\/}(1),
  2869--2909.

\bibitem[\protect\citeauthoryear{Jordon, Yoon, and Schaar}{Jordon
  et~al.}{2019}]{Jordon2019KnockoffGANGK}
Jordon, J., J.~Yoon, and M.~V.~D. Schaar (2019).
\newblock Knockoff{GAN}: generating knockoffs for feature selection using
  generative adversarial networks.
\newblock {\em The International Conference on Learning Representations\/}.

\bibitem[\protect\citeauthoryear{Katsevich and Sabatti}{Katsevich and
  Sabatti}{2019}]{katsevich2019multilayer}
Katsevich, E. and C.~Sabatti (2019).
\newblock Multilayer knockoff filter: controlled variable selection at multiple
  resolutions.
\newblock {\em The Annals of Applied Statistics\/}~{\em 13\/}(1), 1--33.

\bibitem[\protect\citeauthoryear{Ke, Liu, and Ma}{Ke
  et~al.}{2020}]{ke2020power}
Ke, Z.~T., J.~S. Liu, and Y.~Ma (2020).
\newblock {Power of FDR control methods: the impact of ranking algorithm,
  tampered design, and symmetric statistic}.
\newblock {\em arXiv preprint: 2010.08132\/}.

\bibitem[\protect\citeauthoryear{Kim}{Kim}{2015}]{kim2015ppcor}
Kim, S. (2015).
\newblock ppcor: an {R} package for a fast calculation to semi-partial
  correlation coefficients.
\newblock {\em Communications for Statistical Applications and Methods\/}~{\em
  22\/}(6), 665.

\bibitem[\protect\citeauthoryear{Kotz, Balakrishnan, and Johnson}{Kotz
  et~al.}{2000}]{kotz2000bivariate}
Kotz, S., N.~Balakrishnan, and N.~L. Johnson (2000).
\newblock Bivariate and trivariate {N}ormal distributions.
\newblock {\em Continuous {M}ultivariate {D}istributions\/}~{\em 1}, 251--348.

\bibitem[\protect\citeauthoryear{Lauritzen}{Lauritzen}{1996}]{lauritzengraphical}
Lauritzen, S.~L. (1996).
\newblock Graphical models.

\bibitem[\protect\citeauthoryear{Lee, Sun, Sun, and Taylor}{Lee
  et~al.}{2016}]{lee2016exact}
Lee, J.~D., D.~L. Sun, Y.~Sun, and J.~E. Taylor (2016).
\newblock Exact post-selection inference, with application to the {L}asso.
\newblock {\em The Annals of Statistics\/}~{\em 44\/}(3), 907--927.

\bibitem[\protect\citeauthoryear{Li and Maathuis}{Li and
  Maathuis}{2019}]{li2019nodewise}
Li, J. and M.~H. Maathuis (2019).
\newblock Nodewise knockoffs: false discovery rate control for {G}aussian
  graphical models.
\newblock {\em arXiv preprint: 1908.11611\/}.

\bibitem[\protect\citeauthoryear{Liu and Rigollet}{Liu and
  Rigollet}{2019}]{liu2019power}
Liu, J. and P.~Rigollet (2019).
\newblock Power analysis of knockoff filters for correlated designs.
\newblock {\em Advances in Neural Information Processing Systems 32\/},
  15446--15455.

\bibitem[\protect\citeauthoryear{Liu}{Liu}{2013}]{liu2013gaussian}
Liu, W. (2013).
\newblock Gaussian graphical model estimation with false discovery rate
  control.
\newblock {\em The Annals of Statistics\/}~{\em 41\/}(6), 2948--2978.

\bibitem[\protect\citeauthoryear{Lu, Fan, Lv, and Noble}{Lu
  et~al.}{2018}]{lu2018deeppink}
Lu, Y., Y.~Fan, J.~Lv, and W.~S. Noble (2018).
\newblock Deep{PINK}: reproducible feature selection in deep neural networks.
\newblock {\em Advances in Neural Information Processing Systems\/},
  8676--8686.

\bibitem[\protect\citeauthoryear{Ma, Cai, and Li}{Ma
  et~al.}{2020}]{ma2020global}
Ma, R., T.~T. Cai, and H.~Li (2020).
\newblock Global and simultaneous hypothesis testing for high-dimensional
  {L}ogistic regression models.
\newblock {\em Journal of the American Statistical Association\/}, 1--15.

\bibitem[\protect\citeauthoryear{Meinshausen and B{\"u}hlmann}{Meinshausen and
  B{\"u}hlmann}{2006}]{meinshausen2006high}
Meinshausen, N. and P.~B{\"u}hlmann (2006).
\newblock High-dimensional graphs and variable selection with the {L}asso.
\newblock {\em The Annals of Statistics\/}~{\em 34\/}(3), 1436--1462.

\bibitem[\protect\citeauthoryear{Meinshausen and B{\"u}hlmann}{Meinshausen and
  B{\"u}hlmann}{2010}]{meinshausen2010stability}
Meinshausen, N. and P.~B{\"u}hlmann (2010).
\newblock Stability selection.
\newblock {\em Journal of the Royal Statistical Society: Series B (Statistical
  Methodology)\/}~{\em 72\/}(4), 417--473.

\bibitem[\protect\citeauthoryear{Meinshausen, Meier, and
  B{\"u}hlmann}{Meinshausen et~al.}{2009}]{meinshausen2009p}
Meinshausen, N., L.~Meier, and P.~B{\"u}hlmann (2009).
\newblock {p}-values for high-dimensional regression.
\newblock {\em Journal of the American Statistical Association\/}~{\em
  104\/}(488), 1671--1681.

\bibitem[\protect\citeauthoryear{Meinshausen and Yu}{Meinshausen and
  Yu}{2009}]{meinshausen2009}
Meinshausen, N. and B.~Yu (2009).
\newblock Lasso-type recovery of sparse representations for high-dimensional
  data.
\newblock {\em The Annals of Statistics\/}~{\em 37\/}(1), 246--270.

\bibitem[\protect\citeauthoryear{Moran}{Moran}{1973}]{moran1973dividing}
Moran, P. A.~P. (1973).
\newblock Dividing a sample into two parts a statistical dilemma.
\newblock {\em Sankhy{\=a}: The Indian Journal of Statistics, Series A\/},
  329--333.

\bibitem[\protect\citeauthoryear{O'Hara, Sillanp{\"a}{\"a}, et~al.}{O'Hara
  et~al.}{2009}]{o2009review}
O'Hara, R.~B., M.~J. Sillanp{\"a}{\"a}, et~al. (2009).
\newblock A review of {B}ayesian variable selection methods: what, how and
  which.
\newblock {\em Bayesian {A}nalysis\/}~{\em 4\/}(1), 85--117.

\bibitem[\protect\citeauthoryear{Raskutti, Wainwright, and Yu}{Raskutti
  et~al.}{2010}]{raskutti2010restricted}
Raskutti, G., M.~J. Wainwright, and B.~Yu (2010).
\newblock Restricted eigenvalue properties for correlated {G}aussian designs.
\newblock {\em Journal of Machine Learning Research\/}~{\em 11\/}(8),
  2241--2259.

\bibitem[\protect\citeauthoryear{Rhee, Fessel, Zolopa, Hurley, Liu, Taylor,
  Nguyen, Slome, Klein, and Horberg}{Rhee et~al.}{2005}]{rhee2005hiv}
Rhee, S.~Y., W.~J. Fessel, A.~R. Zolopa, L.~Hurley, T.~Liu, J.~Taylor, D.~P.
  Nguyen, S.~Slome, D.~Klein, and M.~Horberg (2005).
\newblock {HIV}-1 protease and reverse-transcriptase mutations: correlations
  with antiretroviral therapy in subtype {B} isolates and implications for
  drug-resistance surveillance.
\newblock {\em The Journal of Infectious Diseases\/}~{\em 192\/}(3), 456--465.

\bibitem[\protect\citeauthoryear{Rhee, Taylor, Wadhera, Ben-Hur, Brutlag, and
  Shafer}{Rhee et~al.}{2006}]{rhee2006genotypic}
Rhee, S.~Y., J.~Taylor, G.~Wadhera, A.~Ben-Hur, D.~L. Brutlag, and R.~W. Shafer
  (2006).
\newblock Genotypic predictors of human immunodeficiency virus type 1 drug
  resistance.
\newblock {\em Proceedings of the National Academy of Sciences\/}~{\em
  103\/}(46), 17355--17360.

\bibitem[\protect\citeauthoryear{Romano and DiCiccio}{Romano and
  DiCiccio}{2019}]{romano2019multiple}
Romano, J.~P. and C.~DiCiccio (2019).
\newblock Multiple data splitting for testing.

\bibitem[\protect\citeauthoryear{Romano, Sesia, and Cand{\`e}s}{Romano
  et~al.}{2019}]{Romano_2019}
Romano, Y., M.~Sesia, and E.~J. Cand{\`e}s (2019).
\newblock Deep knockoffs.
\newblock {\em Journal of the American Statistical Association\/}.

\bibitem[\protect\citeauthoryear{Rubin, Dudoit, and der Laan}{Rubin
  et~al.}{2006}]{rubin2006method}
Rubin, D., S.~Dudoit, and M.~V. der Laan (2006).
\newblock A method to increase the power of multiple testing procedures through
  sample splitting.
\newblock {\em Statistical Applications in Genetics and Molecular
  Biology\/}~{\em 5\/}(1).

\bibitem[\protect\citeauthoryear{{Rudelson} and {Zhou}}{{Rudelson} and
  {Zhou}}{2013}]{6471235}
{Rudelson}, M. and S.~{Zhou} (2013).
\newblock Reconstruction from anisotropic random measurements.
\newblock {\em IEEE Transactions on Information Theory\/}~{\em 59\/}(6),
  3434--3447.

\bibitem[\protect\citeauthoryear{Sarkar}{Sarkar}{2002}]{2002Sanat}
Sarkar, S.~K. (2002).
\newblock Some results on false discovery rate in stepwise multiple testing
  procedures.
\newblock {\em The Annals of Statistics\/}~{\em 30\/}(1), 239--257.

\bibitem[\protect\citeauthoryear{Sesia, Katsevich, Bates, Cand{\`e}s, and
  Sabatti}{Sesia et~al.}{2020}]{2020Sesiamulti}
Sesia, M., E.~Katsevich, S.~Bates, E.~J. Cand{\`e}s, and C.~Sabatti (2020).
\newblock Multi-resolution localization of causal variants across the genome.
\newblock {\em Nature Communications\/}~{\em 11\/}(1).

\bibitem[\protect\citeauthoryear{Sesia, Sabatti, and Cand{\`e}s}{Sesia
  et~al.}{2018}]{2018genehunting}
Sesia, M., C.~Sabatti, and E.~J. Cand{\`e}s (2018).
\newblock {Gene hunting with hidden Markov model knockoffs}.
\newblock {\em Biometrika\/}~{\em 106\/}(1), 1--18.

\bibitem[\protect\citeauthoryear{Stone}{Stone}{1974}]{stone1974cross}
Stone, M. (1974).
\newblock Cross-validatory choice and assessment of statistical predictions.
\newblock {\em Journal of the Royal Statistical Society: Series B (Statistical
  Methodology)\/}~{\em 36\/}(2), 111--133.

\bibitem[\protect\citeauthoryear{Storey}{Storey}{2003}]{2003Storeyqvalue}
Storey, J.~D. (2003).
\newblock The positive false discovery rate: a {B}ayesian interpretation and
  the q-value.
\newblock {\em The Annals of Statistics\/}~{\em 31\/}(6), 2013--2035.

\bibitem[\protect\citeauthoryear{Storey, Taylor, and Siegmund}{Storey
  et~al.}{2004}]{storey2004strong}
Storey, J.~D., J.~E. Taylor, and D.~Siegmund (2004).
\newblock Strong control, conservative point estimation and simultaneous
  conservative consistency of false discovery rates: a unified approach.
\newblock {\em Journal of the Royal Statistical Society: Series B (Statistical
  Methodology)\/}~{\em 66\/}(1), 187--205.

\bibitem[\protect\citeauthoryear{Tibshirani}{Tibshirani}{1996}]{tibshirani1996regression}
Tibshirani, R. (1996).
\newblock Regression shrinkage and selection via the {L}asso.
\newblock {\em Journal of the Royal Statistical Society: Series B (Statistical
  Methodology)\/}~{\em 58\/}(1), 267--288.

\bibitem[\protect\citeauthoryear{Tibshirani, Taylor, Lockhart, and
  Tibshirani}{Tibshirani et~al.}{2016}]{tibshirani2016exact}
Tibshirani, R.~J., J.~Taylor, R.~Lockhart, and R.~Tibshirani (2016).
\newblock Exact post-selection inference for sequential regression procedures.
\newblock {\em Journal of the American Statistical Association\/}~{\em
  111\/}(514), 600--620.

\bibitem[\protect\citeauthoryear{Van~de Geer and B{\"u}hlmann}{Van~de Geer and
  B{\"u}hlmann}{2009}]{vandegeer2009}
Van~de Geer, S.~A. and P.~B{\"u}hlmann (2009).
\newblock On the conditions used to prove oracle results for the lasso.
\newblock {\em Electron. J. Statist.\/}~{\em 3}, 1360--1392.

\bibitem[\protect\citeauthoryear{{Van de Geer}, B{\"u}hlmann, Ritov, and
  Dezeure}{{Van de Geer} et~al.}{2014}]{van2014asymptotically}
{Van de Geer}, S.~A., P.~B{\"u}hlmann, Y.~Ritov, and R.~Dezeure (2014).
\newblock On asymptotically optimal confidence regions and tests for
  high-dimensional models.
\newblock {\em The Annals of Statistics\/}~{\em 42\/}(3), 1166--1202.

\bibitem[\protect\citeauthoryear{van~de Wiel, Berkhof, and van
  Wieringen}{van~de Wiel et~al.}{2009}]{van2009testing}
van~de Wiel, M.~A., J.~Berkhof, and W.~N. van Wieringen (2009).
\newblock Testing the prediction error difference between two predictors.
\newblock {\em Biostatistics\/}~{\em 10\/}(3), 550--560.

\bibitem[\protect\citeauthoryear{Wainwright}{Wainwright}{2019}]{wainwright2019high}
Wainwright, M.~J. (2019).
\newblock {\em High-dimensional statistics: a non-asymptotic viewpoint},
  Volume~48.
\newblock Cambridge University Press.

\bibitem[\protect\citeauthoryear{Wang and Janson}{Wang and
  Janson}{2020}]{WW-LJ:2020}
Wang, W. and L.~Janson (2020).
\newblock A power analysis of the conditional randomization test and knockoffs.
\newblock {\em arXiv preprint: 2010.02304\/}.

\bibitem[\protect\citeauthoryear{Wasserman and Roeder}{Wasserman and
  Roeder}{2009}]{wasserman2009high}
Wasserman, L. and K.~Roeder (2009).
\newblock High dimensional variable selection.
\newblock {\em The Annals of Statistics\/}~{\em 37\/}(5A), 2178.

\bibitem[\protect\citeauthoryear{Weinstein, Barber, and Cand{\`e}s}{Weinstein
  et~al.}{2017}]{weinstein2017power}
Weinstein, A., R.~F. Barber, and E.~J. Cand{\`e}s (2017).
\newblock A power and prediction analysis for knockoffs with {L}asso
  statistics.

\bibitem[\protect\citeauthoryear{Weinstein, Su, Bogdan, Barber, and
  Cand{\`e}s}{Weinstein et~al.}{2020}]{weinstein2020power}
Weinstein, A., W.~J. Su, M.~Bogdan, R.~F. Barber, and E.~J. Cand{\`e}s (2020).
\newblock A power analysis for knockoffs with the {L}asso
  coefficient-difference statistic.

\bibitem[\protect\citeauthoryear{Wu}{Wu}{2008}]{2008WeibiaoWu}
Wu, W.~B. (2008).
\newblock On false discovery control under dependence.
\newblock {\em The Annals of Statistics\/}~{\em 36}, 364–--380.

\bibitem[\protect\citeauthoryear{Xing, Zhao, and Liu}{Xing
  et~al.}{2019}]{xin2019GM}
Xing, X., Z.~Zhao, and J.~S. Liu (2019).
\newblock Controlling false discovery rate using {G}aussian mirrors.
\newblock {\em arXiv preprint: 1911.09761\/}.

\bibitem[\protect\citeauthoryear{Zhang and Zhang}{Zhang and
  Zhang}{2014}]{zhang2014confidence}
Zhang, C.~H. and S.~S. Zhang (2014).
\newblock Confidence intervals for low dimensional parameters in high
  dimensional linear models.
\newblock {\em Journal of the Royal Statistical Society: Series B (Statistical
  Methodology)\/}~{\em 76\/}(1), 217--242.

\bibitem[\protect\citeauthoryear{Zhang, Ren, and Chen}{Zhang
  et~al.}{2018}]{zhang2018silggm}
Zhang, R., Z.~Ren, and W.~Chen (2018).
\newblock {SILGGM}: an extensive {R} package for efficient statistical
  inference in large-scale gene networks.
\newblock {\em PLoS Computational Biology\/}~{\em 14\/}(8).

\end{thebibliography}

\newpage
\section{Supplementary Materials}

\subsection{Proofs}
Throughout, we consider the general form of the mirror statistic defined in \eqref{eq:mirror-statistic},
in which function $f(u,v)$ is non-negative, symmetric about $u$ and $v$, and monotonically increasing in both $u$ and $v$.
For any $t > 0$ and $u\geq 0$, let 
\begin{equation}\nonumber
\mathcal{I}_t(u) = \inf\{v\geq 0: f(u, v) > t\}
\end{equation}
with the convention $\inf\varnothing = +\infty$. The CDF and pdf of the standard Normal distribution are denoted as $\Phi$ and $\phi$, respectively.

\subsubsection{Proof of Proposition \ref{prop:optimality-mirror-statistics}}
Let $Z_1, Z_2$ follow $N(0,1)$, $Z_3, Z_4$ follow $N(\omega, 1)$, all of which are independent. 
Without loss of generality, we assume that the designated FDR control level $q\in(0,1)$ satisfies $rq/(1 - q) < 1$, otherwise selecting all features would maximize the power and also achieve asymptotic FDR control.
Let $f^\text{opt}(u, v)$ be the optimal choice, and let $\widehat{S}^{\text{opt}}$ be the optimal selection result that achieves asymptotic FDR control. By the law of large numbers, we have
\begin{equation}
\label{eq:optimality-LLN}
\lim_{p\to\infty}\frac{\#\{j: j \in S_0, j \in \widehat{S}^{\text{opt}}\}}{\# \{j: j \in \widehat{S}^{\text{opt}}\}} = \frac{\mathbbm{P}(j \in \widehat{S}^{\text{opt}}\mid  j\in S_0)}{\mathbbm{P}(j \in \widehat{S}^{\text{opt}} \mid  j\in S_0) + r\mathbbm{P}(j \in \widehat{S}^{\text{opt}} \mid j\in S_1)} \leq q,
\end{equation}
in which the numerator is the type-I error. More precisely,
\begin{equation}\nonumber
\begin{aligned}
& \mathbbm{P}(j \in \widehat{S}^{\text{opt}}\mid  j\in S_0) =  \mathbbm{P}(\text{sign}(Z_1Z_2)f^{\text{opt}}(|Z_1|, |Z_2|) > t^{\text{opt}}),\\
& \mathbbm{P}(j \in \widehat{S}^{\text{opt}}\mid  j\in S_1) =  \mathbbm{P}(\text{sign}(Z_3Z_4)f^{\text{opt}}(|Z_3|, |Z_4|) > t^{\text{opt}}).
\end{aligned}
\end{equation}
$t^{\text{opt}} > 0$ is the cutoff that maximizes the power $\mathbbm{P}(j \in \widehat{S}^{\text{opt}}\mid  j\in S_1)$, under the constraint that Equation \eqref{eq:optimality-LLN} holds.

We now consider testing whether $X_j$ is a null feature, with the significance level $\alpha$ specified as:
\begin{equation}\nonumber
\alpha = \frac{rq}{1 - q}\mathbbm{P}(j \in \widehat{S}^{\text{opt}} \mid j\in S_1) < 1.
\end{equation}
We have two observations $\widehat{\beta}_j^{(1)}$ and $\widehat{\beta}_j^{(2)}$, which independently follow $N(0,1)$ or $N(\omega, 1)$ if $X_j$ is a null feature or a relevant feature, respectively. By Equation \eqref{eq:optimality-LLN}, the test which rejects the null hypothesis (i.e., $j\in \widehat{S}^{\text{opt}} $) if
\begin{equation}\nonumber
\text{sign}(\widehat{\beta}_j^{(1)}\widehat{\beta}_j^{(2)})f^{\text{opt}}(|\widehat{\beta}_j^{(1)}|, |\widehat{\beta}_j^{(2)}|) > t^{\text{opt}}
\end{equation}
achieves the significance level $\alpha$. 

In the following, $f$ refers to $f(u, v) = u + v$. We consider the following rejection rule,
\begin{equation}\nonumber
\text{sign}(\widehat{\beta}_j^{(1)}\widehat{\beta}_j^{(2)})f(|\widehat{\beta}_j^{(1)}|, |\widehat{\beta}_j^{(2)}|) > t^{\text{lik}},
\end{equation}
in which $t^{\text{lik}} > 0$ satisfies
\begin{equation}
\label{eq:lik-cutoff}
    \mathbbm{P}(f(|Z_1|, |Z_2|) > t^{\text{lik}} \mid \text{sign}(Z_1) = \text{sign}(Z_2)) = 2\alpha.
\end{equation}
Let $\widehat{S}^{\text{lik}}$ be the corresponding selection set. We first show that this rejection rule controls the type-I error below $\alpha$. Indeed, 
\begin{equation}
\label{eq:optimality-error-bound}
\begin{aligned}
\mathbbm{P}(j \in \widehat{S}^{\text{lik}} \mid j\in S_0) = \frac{1}{2} \mathbbm{P}(f(|\widehat{\beta}_j^{(1)}|, |\widehat{\beta}_j^{(2)}|) > t^{\text{lik}} \mid j\in S_0,\ \text{sign}(\widehat{\beta}_j^{(1)}) = \text{sign}(\widehat{\beta}_j^{(2)})) = \alpha.
\end{aligned}
\end{equation}
In terms of power, we have 
\begin{equation}
\label{eq:optimality-power-bound}
\begin{aligned}
\mathbbm{P}(j \in \widehat{S}^{\text{lik}}\mid  j\in S_1) &  =  p_w\mathbbm{P}(f(|\widehat{\beta}_j^{(1)}|, |\widehat{\beta}_j^{(2)}|) > t^{\text{lik}}\mid  j\in S_1,\ \text{sign}(\widehat{\beta}_j^{(1)}) = \text{sign}(\widehat{\beta}_j^{(2)}))\\
& \geq p_w\mathbbm{P}(f^{\text{opt}}(|\widehat{\beta}_j^{(1)}|, |\widehat{\beta}_j^{(2)}|)>t^{\text{opt}} \mid  j\in S_1,\ \text{sign}(\widehat{\beta}_j^{(1)}) = \text{sign}(\widehat{\beta}_j^{(2)})) \\
& = \mathbbm{P}(j \in \widehat{S}^{\text{opt}}\mid  j\in S_1),
\end{aligned}
\end{equation}
in which $p_w = \mathbbm{P}(\text{sign}(\widehat{\beta}_j^{(1)}) = \text{sign}(\widehat{\beta}_j^{(2)}) \mid j\in S_1)$. The inequality in the second line is a direct consequence of the Neymann-Pearson lemma. 

To see this, suppose we only observe $|\widehat{\beta}_j^{(1)}|, |\widehat{\beta}_j^{(2)}|$ and also know that  $\text{sign}(\widehat{\beta}_j^{(1)}) = \text{sign}(\widehat{\beta}_j^{(2)})$. Then the rejection rule $f(|\widehat{\beta}_j^{(1)}|, |\widehat{\beta}_j^{(2)}|) > t^{\text{lik}}$ controls the type-I error below $2\alpha$ by the definition of $t^{\text{lik}}$ in \eqref{eq:lik-cutoff}. Further, the  rejection rule $f(|\widehat{\beta}_j^{(1)}|, |\widehat{\beta}_j^{(2)}|) > t^{\text{opt}}$ also controls the type-I error below $2\alpha$ by Equation \eqref{eq:optimality-LLN}.
The likelihood ratio (LR) is given by
\begin{equation}\nonumber
    \text{LR} = \frac{\phi_1(|\widehat{\beta}_j^{(1)}|, |\widehat{\beta}_j^{(2)}|)}{\phi_0(|\widehat{\beta}_j^{(1)}|, |\widehat{\beta}_j^{(2)}|)} \propto \cosh(w(\widehat{\beta}_j^{(1)}+\widehat{\beta}_j^{(2)})) ,
\end{equation}
where $\phi_1$ and $\phi_0$ are the joint densities of $(Z_3, Z_4)$ and $(Z_1, Z_2)$, respectively, conditioning on $\text{sign}(Z_1) = \text{sign}(Z_2)$ and $\text{sign}(Z_3) = \text{sign}(Z_4)$. Note that this is a monotone function of $|\widehat{\beta}_j^{(1)}+\widehat{\beta}_j^{(2)}|$, which equals to $|\widehat{\beta}_j^{(1)}|+|\widehat{\beta}_j^{(2)}|$ under the condition $\text{sign}(\widehat{\beta}_j^{(1)}) = \text{sign}(\widehat{\beta}_j^{(2)})$.

Combining Equations \eqref{eq:optimality-error-bound} and \eqref{eq:optimality-power-bound}, it follows that the selection set $\widehat{S}^{\text{lik}}$ achieves asymptotic FDR control since
\begin{equation}\nonumber
\begin{aligned}
\lim_{p\to\infty}\frac{\#\{j: j \in S_0, j \in \widehat{S}^{\text{lik}}\}}{\# \{j: j \in \widehat{S}^{\text{lik}}\}} & = \frac{\mathbbm{P}(j \in \widehat{S}^{\text{lik}}\mid  j\in S_0)}{\mathbbm{P}(j \in \widehat{S}^{\text{lik}} \mid  j\in S_0) + r\mathbbm{P}(j \in \widehat{S}^{\text{lik}} \mid j\in S_1)} \leq q.
\end{aligned}
\end{equation}
As $f^{\text{opt}}$ is optimal, by Equation \eqref{eq:optimality-power-bound}, $f^{\text{lik}}$ is also optimal. This concludes the proof of Proposition \ref{prop:optimality-mirror-statistics}.

\subsubsection{Proof of Proposition \ref{prop:FDR}}
\label{subsub:proof-proposition-FDR}
For the ease of presentation, we introduce the following notations. For $t \in \mathbbm{R}$, denote
\begin{equation}
\label{eq:fp-tp-proportion}
\begin{aligned}
&\widehat{G}^0_{p}(t) = \frac{1}{p_0}\sum_{j \in S_0}\mathbbm{1}(M_j > t),\ \ \
G^0_{p}(t) = \frac{1}{p_0}\sum_{j \in S_0}\mathbbm{P}(M_j > t),\\
&\widehat{G}^1_{p}(t) = \frac{1}{p_1}\sum_{j \in S_1}\mathbbm{1}(M_j > t),\hspace{0.37cm} 
\widehat{V}^0_{p}(t) = \frac{1}{p_0}\sum_{j \in S_0}\mathbbm{1}(M_j < - t). 
\end{aligned}
\end{equation}
Let $r_{p} = p_1/p_0$. In addition, denote
\begin{equation}\nonumber
\text{FDP}_p(t)  = \frac{\widehat{G}_{p}^0(t)}{\widehat{G}_{p}^0(t) + r_{p}\widehat{G}_{p}^1(t)},\ \ \ 
\text{FDP}^\dagger_p(t)  = \frac{\widehat{V}_{p}^0(t)}{\widehat{G}_{p}^0(t) + r_{p}\widehat{G}_{p}^1(t)},\ \ \ 
\overline{\text{FDP}}_p(t)  = \frac{G_{p}^0(t)}{G_{p}^0(t) + r_{p}\widehat{G}_{p}^1(t)}.
\end{equation}

\begin{lemma}
\label{lemma:FDR}
Under Assumption \ref{assump:weak-dependency}, if $p_0\to\infty$ as $p\to\infty$, we have in probability,
\begin{equation}\nonumber
\sup_{t\in\mathbbm{R}} \left|\widehat{G}^0_{p}(t) - G^0_{p}(t)\right| \longrightarrow 0,\ \ \ \ \
\sup_{t\in\mathbbm{R}} \left|\widehat{V}^0_{p}(t) - G^0_{p}(t)\right| \longrightarrow 0.
\end{equation}
\end{lemma}
\textit{Proof of Lemma \ref{lemma:FDR}}. For any $\epsilon \in (0, 1)$, denote $-\infty = \alpha^{p}_0 < \alpha^{p}_1 < \cdots < \alpha^{p}_{N_\epsilon} = \infty$ with $N_\epsilon = \lceil2/\epsilon\rceil$, such that $G^0_{p}(\alpha^{p}_{k - 1}) - G^0_{p}(\alpha^{p}_k) \leq \epsilon /2$ for $k=1,\ldots,N_\epsilon$. By Assumption \ref{assump:weak-dependency}, such a sequence $\{\alpha_k^p\}$ exists since $G_p^0(t)$ is a continuous function for $t\in \mathbbm{R}$.
We have
\begin{equation}
\label{eq:lemma-2-1}
\begin{aligned}
\mathbbm{P}\left(\sup_{t\in\mathbbm{R}}\widehat{G}^0_{p}(t) - G^0_{p}(t) > \epsilon\right) & \leq \mathbbm{P}\left(\bigcup_{k = 1}^{N_\epsilon}\sup_{t \in\left[\alpha^{p}_{k - 1}, \alpha^{p}_k\right)}\widehat{G}^0_{p}(t) - G^0_{p}(t) > \epsilon\right)\\
& \leq \sum_{k = 1}^{N_{\epsilon}}\mathbbm{P}\left(\sup_{t \in\left[\alpha^{p}_{k - 1}, \alpha^{p}_k\right)}\widehat{G}^0_{p}(t) - G^0_{p}(t) > \epsilon\right).
\end{aligned}
\end{equation}
We note that both $\widehat{G}^0_{p}(t)$ and $G^0_{p}(t)$ are monotonically decreasing. Therefore, $\forall \ k \in \{1,\ldots,N_\epsilon\}$, we have
\begin{equation}\nonumber
\sup_{t \in\left[\alpha^{p}_{k - 1}, \alpha^{p}_k\right)}\widehat{G}^0_{p}(t) - G^0_{p}(t) \leq \widehat{G}^0_{p}(\alpha^{p}_{k - 1}) - G^0_{p}(\alpha^{p}_{k}) \leq \widehat{G}^0_{p}(\alpha^{p}_{k - 1}) - G^0_{p}(\alpha^{p}_{k - 1}) + \epsilon/2.
\end{equation}
By Equation \eqref{eq:lemma-2-1}, Assumption \ref{assump:weak-dependency}, and the Chebyshev's inequality, it follows that
\begin{equation}\nonumber
\mathbbm{P}\left(\sup_{t\in\mathbbm{R}}\widehat{G}^0_{p}(t) - G^0_{p}(t) > \epsilon\right) \leq \sum_{k = 1}^{N_{\epsilon}}\mathbbm{P}\left(\widehat{G}^0_{p}(\alpha^{p}_{k - 1}) - G^0_{p}(\alpha^{p}_{k - 1}) > \frac{\epsilon}{2}\right)\leq \frac{4cN_\epsilon}{p_0^{2-\alpha}\epsilon^2} \to 0,\ \ \textnormal{as}\ \ p\to\infty.
\end{equation}
Similarly, we can show that
\begin{equation}\nonumber
\mathbbm{P}\left(\inf_{t\in\mathbbm{R}}\widehat{G}^0_{p}(t) - G^0_{p}(t) < -\epsilon\right) \leq \sum_{k = 1}^{N_{\epsilon}}\mathbbm{P}\left(\widehat{G}^0_{p}(\alpha^{p}_k) - G^0_{p}(\alpha^{p}_k) < -\frac{\epsilon}{2}\right)\leq \frac{4cN_\epsilon}{p_0^{2-\alpha}\epsilon^2} \to 0,\ \ \ \ \textnormal{as}\ \ p\to\infty.
\end{equation}
This concludes the proof of the first claim in Lemma \ref{lemma:FDR}. The second claim follows similarly using the symmetric property of the mirror statistics $M_j$'s for $j \in S_0$.

\vspace{0.5cm}
\noindent\textit{Proof of Proposition \ref{prop:FDR}}. We first show that for any $\epsilon \in (0, q)$, we have
\begin{equation}\nonumber
\mathbbm{P}(\tau_q\leq t_{q-\epsilon}) \geq 1 - \epsilon,
\end{equation}
in which $t_{q-\epsilon} > 0$ satisfying $\mathbbm{P}(\text{FDP}(t_{q-\epsilon}) \leq q - \epsilon) \to 1$. 
Since the variances of the mirror statistics are upper bounded and also bounded away from 0, by Lemma \ref{lemma:FDR}, we have 
\begin{equation}\nonumber
\sup_{0 < t\leq c}|\text{FDP}^\dagger_p(t)-\text{FDP}_p(t)| \overset{p}{\to} 0
\end{equation}
for any constant $c > 0$.
By the definition of $\tau_q$, i.e., $\tau_q = \inf\{t > 0: \text{FDP}^\dagger_p(t)\leq q\}$, we have
\begin{equation}\nonumber
\begin{aligned}
\mathbbm{P}(\tau_q\leq t_{q-\epsilon}) & \geq \mathbbm{P}(\text{FDP}^\dagger_p(t_{q-\epsilon}) \leq q)\\
& \geq \mathbbm{P}(|\text{FDP}^\dagger_p(t_{q-\epsilon}) -\text{FDP}_p(t_{q-\epsilon})| \leq \epsilon,\ \text{FDP}(t_{q-\epsilon}) \leq q - \epsilon)\\
& \geq 1 - \epsilon
\end{aligned}
\end{equation}
for $p$ large enough. 
Conditioning on the event $\tau_q\leq t_{q-\epsilon}$, we have
\begin{equation}\nonumber
\begin{aligned}
\limsup_{p\to\infty}
\mathbbm{E}\left[\text{FDP}_p\left(\tau_q\right)\right] 
& \leq \limsup_{p\to\infty}
\mathbbm{E}\left[\text{FDP}_p\left(\tau_q\right)\mid \tau_q\leq t_{q-\epsilon}\right] \mathbbm{P}(\tau_q\leq t_{q-\epsilon}) + \epsilon\\
& \leq \limsup_{p\to\infty} \mathbbm{E}\Big[\big|\text{FDP}_p\left(\tau_q\right) - \overline{\text{FDP}}_p\left(\tau_q\right)\big|\ \big\vert\ \tau_q\leq t_{q-\epsilon}\Big]\mathbbm{P}(\tau_q\leq t_{q-\epsilon}) \\
& \hspace{0.05cm} + \limsup_{p\to\infty} \mathbbm{E}\left[\big|\text{FDP}^\dagger_p\left(\tau_q\right) - \overline{\text{FDP}}_p\left(\tau_q\right)\big|\ \big\vert\ \tau_q\leq t_{q-\epsilon}\right]\mathbbm{P}(\tau_q\leq t_{q-\epsilon})\\
& \hspace{0.05cm} + \limsup_{p\to\infty} \mathbbm{E}\left[\text{FDP}^\dagger_p\left(\tau_q\right) \ \big\vert\ \tau_q\leq t_{q-\epsilon}\right]\mathbbm{P}(\tau_q\leq t_{q-\epsilon}) + \epsilon\\
& \leq \limsup_{p\to\infty} \mathbbm{E}\Big[\sup_{0 < t \leq t_{q-\epsilon}}\left|\text{FDP}_p(t) - \overline{\text{FDP}}_p(t)\right|\Big]\\
& \hspace{0.05cm}  + \limsup_{p\to\infty} \mathbbm{E}\Big[\sup_{0<t\leq t_{q-\epsilon}}\left|\text{FDP}^\dagger_p(t) - \overline{\text{FDP}}_p(t)\right|\Big] \\
& \hspace{0.05cm} + \limsup_{p\to\infty} \mathbbm{E}\left[\text{FDP}^\dagger_p\left(\tau_q\right)\right] + \epsilon.
\end{aligned}
\end{equation}
The first two terms are 0 based on Lemma \ref{lemma:FDR} and the dominated convergence theorem. For the third term, we have
$\text{FDP}^\dagger_p\left(\tau_q\right)\leq q$ 
by the definition of $\tau_q$. This concludes the proof of Proposition \ref{prop:FDR}.

\subsubsection{Proof of Proposition \ref{prop:MDS-FDR}}
We first establish a probabilistic upper bound and a lower bound for the data-dependent cutoff $\ell$ used in MDS. We start with the upper bound.

\begin{lemma}
\label{lemma:ell-upper-bound}
Under  the assumptions in Proposition~\ref{prop:MDS-FDR}, in both the sparse and the non-sparse regimes, as $n,p\to\infty$, we have 
\begin{equation}\nonumber
\mathbbm{P}(\ell \leq p - cp_1) \to 1
\end{equation}
for some constant $c > 0$, in which $\ell$ is defined in Algorithm \ref{alg:multiple-splits}.
\end{lemma}

\noindent
\textit{Proof of Lemma \ref{lemma:ell-upper-bound}.}
$\forall\epsilon>0$, we first show that 
\begin{equation}
\label{eq:inclusion-rate-null-uppr-bound}
\mathbbm{P}\big(\sum_{j\in S_0} I_j\leq q+\epsilon\big) \to 1
\end{equation}
as $n,p\to\infty$. 
Let $B$ (as a function of $n$) be the total number of different sample splits. For any sample split $b\in\{1,\ldots, B\}$, let 
\begin{equation}\nonumber
\text{FDP}_b = \frac{\sum_{j\in S_0}\mathbbm{1}(j\in\widehat{S}_b)}{|\widehat{S}_b|\vee 1},
\end{equation}
in which $\widehat{S}_b$ is the index set of the selected features by DS.
Consider the proportion of sample splits with FDP larger than $q + \epsilon/2$, i.e.,
\begin{equation}\nonumber
U = \frac{1}{B}\sum_{b = 1}^B\mathbbm{1}\left(\text{FDP}_b > q + \epsilon/2\right).
\end{equation}
Since DS achieves an asymptotic FDP control, we have 
\begin{equation}\nonumber
\mathbbm{E}[U] = \mathbbm{P}\left(\text{FDP}_b > q + \epsilon/2\right) \to 0
\end{equation}
as $n,p\to\infty$. Therefore, by the Markov's inequality, we have
\begin{equation}\nonumber
\mathbbm{P}\left(U \leq  \epsilon/2\right) \to 1.
\end{equation}
We note that the event $U \leq \epsilon/2$ implies the event $\sum_{j\in S_0} I_j\leq q+\epsilon$. To see this,
\begin{equation}\nonumber
\begin{aligned}
\sum_{j\in S_0}I_j & = \frac{1}{B}\sum_{b = 1}^B \text{FDP}_b  \leq \frac{1}{B}\sum_{ b: \text{FDP}_b\leq q + \epsilon/2}\text{FDP}_b + \frac{1}{B}\sum_{b: \text{FDP}_b > q + \epsilon/2}1\leq q + \epsilon.
\end{aligned}
\end{equation}
Thus, the claim in \eqref{eq:inclusion-rate-null-uppr-bound} holds.

We then establish a probabilistic upper bound and a lower bound for the sum of the inclusion rates over the selected relevant features by MDS, denoted as $\Delta = \sum_{j\in S_1} I_j\mathbbm{1}(I_j > I_{(\ell)})$.
For the lower bound, by the definition of $\ell$ in Algorithm \ref{alg:multiple-splits}, we have
\begin{equation}\nonumber
\sum_{k > \ell}^{p} I_{(k)} \geq 1-q. 
\end{equation}
Combining it with \eqref{eq:inclusion-rate-null-uppr-bound}, we have
\begin{equation}
\label{eq:sum-inclusion-lower-bound}
\mathbbm{P}\Big(\Delta \geq 1-2q-\epsilon\Big) \to 1.
\end{equation}

The upper bound relies on the assumption that the power of DS is lower bounded by some constant $\kappa > 0$ with probability approaching 1. Let 
$$\mathcal{B}_1 = \left\{b\in\{1,\ldots,B\}:\ \sum_{j\in S_1}\mathbbm{1}(j\in\widehat{S}_b) > \kappa p_1\ \ \ \textnormal{and}\ \ \ \textnormal{FDP}_b\leq q+\epsilon \right\}$$
and $\mathcal{B}_2 = \{1,\ldots, B\}\setminus \mathcal{B}_1$. Then we have 
\begin{equation}
\label{eq:B1}
|\mathcal{B}_2|/B = o_p(1)\ \ \ \text{and}\ \ \ |\widehat{S}_b|\leq \frac{p_1}{1-q-\epsilon}\ \ \text{for}\ \ b\in \mathcal{B}_1.
\end{equation}
With respect to the sets $\mathcal{B}_1$ and $\mathcal{B}_2$, we can decompose $\Delta = \Delta_1 + \Delta_2$ and $\sum_{j\in S_1} I_j = \Omega_1 + \Omega_2$, in which
\begin{equation}\nonumber
\begin{aligned}
\Delta_1 & = \frac{1}{B}\sum_{b\in \mathcal{B}_1}\sum_{j\in S_1, I_j > I_{(\ell)}} \mathbbm{1}(j\in\widehat{S}_b)/|\widehat{S}_b|,\ \ \ \Delta_2 = \frac{1}{B}\sum_{b\in \mathcal{B}_2}\sum_{j\in S_1, I_j > I_{(\ell)}} \mathbbm{1}(j\in\widehat{S}_b)/|\widehat{S}_b|,\\
\Omega_1 & = \frac{1}{B}\sum_{b\in \mathcal{B}_1}\sum_{j\in S_1} \mathbbm{1}(j\in\widehat{S}_b)/|\widehat{S}_b|,\ \ \ \ \ \ \ \ \ \ \ \ \Omega_2 = \frac{1}{B}\sum_{b\in \mathcal{B}_2}\sum_{j\in S_1} \mathbbm{1}(j\in\widehat{S}_b)/|\widehat{S}_b|.
\end{aligned}
\end{equation}
Then we have $\Delta_1 \leq \Omega_1$, $\Delta_2 \leq \Omega_2 \leq |\mathcal{B}_2|/B$, and
\begin{equation}\nonumber
\Delta_1 \leq \frac{1}{B}\sum_{b\in \mathcal{B}_1} \frac{p - \ell}{|\widehat{S}_b|},\ \ \ \ \ \Omega_1 \geq \frac{1}{B}\sum_{b\in \mathcal{B}_1} \frac{\kappa p_1}{|\widehat{S}_b|}.
\end{equation}
It follows that
\begin{equation}
\label{eq:sum-inclusion-upper-bound1}
\begin{aligned}
\Delta  \leq \frac{\Delta}{\sum_{j\in S_1} I_j} & = \frac{\Delta_1 + \Delta_2}{\Omega_1 + \Omega_2}  \leq \frac{\Delta_1 + \Delta_2}{\Omega_1 + \Delta_2} \leq \frac{\Delta_1 + |\mathcal{B}_2|/B}{\Omega_1 + |\mathcal{B}_2|/B}\\
& \leq  \frac{\sum_{b\in \mathcal{B}_1} (p-\ell)/|\widehat{S}_b|+|\mathcal{B}_2|}{\sum_{b\in \mathcal{B}_1}\kappa p_1/|\widehat{S}_b|+|\mathcal{B}_2|}   \leq \frac{p-\ell}{\kappa p_1}+ \frac{|\mathcal{B}_2|}{\kappa(1-q-\epsilon)|\mathcal{B}_1|}.
\end{aligned}
\end{equation}
The first inequality follows from the fact that $\sum_{j\in S_1}I_j\leq \sum_{j=1}^p I_j = 1$. 
Combining Equations \eqref{eq:sum-inclusion-lower-bound}, \eqref{eq:B1} and \eqref{eq:sum-inclusion-upper-bound1}, we conclude the proof of Lemma \ref{lemma:ell-upper-bound}.\\

The following lemma establishes a lower bound for the cutoff $\ell$.

\begin{lemma}
\label{lemma:MDS-FDR}
$\forall \epsilon>0$, under  the assumptions in Proposition~\ref{prop:MDS-FDR}, for the cutoff $\ell$ defined in Algorithm \ref{alg:multiple-splits}, we have
\begin{enumerate}
\item $\mathbbm{P}\big(\ell  \geq (1 - \epsilon)p_0\big)\to 1$ in the non-sparse regime;
\item $\mathbbm{P}\big(\ell \geq p_0 - 1\big)\to 1$ in the sparse regime.
\end{enumerate}
\end{lemma}

\noindent
\textit{Proof of Lemma \ref{lemma:MDS-FDR}.}
We first prove the claim in the non-sparse regime. Suppose $\mathbbm{P}(\ell < p_0) \nrightarrow 0$.
It is sufficient to show that $\forall \epsilon > 0$, as $n,p\to\infty$,
\begin{equation}\nonumber
\mathbbm{P}\big(\ell  \geq (1 - \epsilon)p_0 \mid \ell < p_0\big)\to 1.
\end{equation}
Choose $\epsilon^\prime > 0$ such that $\epsilon\geq \epsilon^\prime/(q + \epsilon^\prime) + 1/p_0$ when $p_0$ is large enough. By \eqref{eq:inclusion-rate-null-uppr-bound}, we have
\begin{equation}\nonumber
\mathbbm{P}\big(\sum_{j\in S_0} I_j\leq q+\epsilon^\prime \mid \ell < p_0 \big) \to 1.
\end{equation}
We note that conditioning on $\ell < p_0$, the high probability event $\sum_{j\in S_0} I_j\leq q+\epsilon^\prime$ implies the event $\ell \geq (1 - \epsilon)p_0$. To see this, we require the following simple observation,
\begin{equation}
\label{eq:lemma-MDS-FDR-obs}
\frac{q}{\ell + 1} < \frac{\sum_{j  \leq \ell + 1}I_{(j)}}{\ell + 1} \leq \frac{\sum_{j\in S_0}I_j}{p_0} \leq \frac{q + \epsilon^\prime}{p_0}.
\end{equation}
The first inequality follows from the definition of $\ell$, i.e., $\ell$ is the largest index such that $\sum_{j\leq \ell}I_{(j)} \leq q$. The second inequality follows from the fact that when $\ell < p_0$, the mean of the smallest $\ell + 1$ inclusion rates should be no larger than the mean of the $p_0$ inclusion rates associated with the null features. Equation \eqref{eq:lemma-MDS-FDR-obs} immediately implies that
\begin{equation}\nonumber
\ell > \frac{qp_0}{q+\epsilon^\prime} - 1 \geq (1 - \epsilon)p_0,
\end{equation}
and thus the claim in Lemma \ref{lemma:MDS-FDR} for the non-sparse regime holds. For the sparse regime, we asymptotically control the FDP of DS at some level $q'<q$. 
Similar as \eqref{eq:inclusion-rate-null-uppr-bound}, we have
\begin{equation}\nonumber
\mathbbm{P}\big(\sum_{j\in S_0} I_j\leq q \big) \to 1
\end{equation}
as $n,p\to\infty$. 
Conditioning on $\ell < p_0$, by the similar argument as in Equation \eqref{eq:lemma-MDS-FDR-obs}, we can show that the high probability event $\sum_{j\in S_0} I_j\leq q$ implies $\ell \geq p_0 - 1$. This completes the proof of Lemma \ref{lemma:MDS-FDR}.

\vspace{0.5cm}
\noindent\textit{Proof of Proposition \ref{prop:MDS-FDR}}. For MDS, let $k$ be the number of false positives. 
\begin{enumerate}[(a)]
\item The non-sparse regime.
For any pair $(i, j)$ with $i\in S_1$ and $j\in S_0$, we refer to it as a ``falsely ranked pair" if 
$I_i < I_j$.
Let $N$ be the total number of falsely ranked pairs. 
Given that the mirror statistics are consistent at ranking features, we have
\begin{equation}\nonumber
\frac{1}{p_0p_1}\mathbbm{E}\left[N\right] = \frac{1}{p_0p_1}\sum_{i\in S_1, j\in S_0}\mathbbm{P}(I_i < I_j) \leq \sup_{i\in S_1, j\in S_0}\mathbbm{P}(I_i < I_j) \to 0,
\end{equation}
which further implies $N = o_p(p_0p_1)$ by the Markov's inequality.
There are $[p_1-(p-\ell-k)]_+$\footnote{$[x]_+ = \max(x, 0)$.} relevant features not selected by MDS. By Lemma \ref{lemma:MDS-FDR}, with probability approaching 1, we have
\begin{equation}\nonumber
N \geq k[p_1-(p-\ell-k)]_+ \geq k[k-\epsilon p_0]_+.
\end{equation}
Thus, $k[k-\epsilon p_0]_+ = o_p(p_0p_1)$, which further implies $k = o_p(p_0)$. Together with Lemma \ref{lemma:ell-upper-bound}, we complete the proof for the non-sparse regime.

\item The sparse regime. 
For $i\in S_1$, we refer to it as a ``falsely ranked signal" if $I_i < \max_{j\in S_0}I_j$.
Let $N$ be the total number of falsely ranked signals. 
Given that the mirror statistics are strongly consistent at ranking features, we have
\begin{equation}\nonumber
\frac{1}{p_1}\mathbbm{E}\left[N\right] = \frac{1}{p_1}\sum_{i\in S_1}\mathbbm{P}(I_i < \max_{j\in S_0}I_j) \leq \sup_{i\in S_1}\mathbbm{P}(I_i < \max_{j\in S_0} I_j) \to 0,
\end{equation}
which further implies $N = o_p(p_1)$ by the Markov's inequality.
By Lemma \ref{lemma:MDS-FDR}, with probability approaching 1, we have
\begin{equation}\nonumber
N \geq [p_1-(p-\ell-k)]_+ \geq [k-1]_+.
\end{equation}
Thus, $[k-1]_+ = o_p(p_1)$, which further implies $k = o_p(p_1)$. Together with Lemma \ref{lemma:ell-upper-bound}, we complete the proof for the sparse regime.
\end{enumerate}

\subsubsection{Proof of Proposition \ref{prop:rank-DS-MDS}}
Let $Z_j = \bar{\bm{X}}^{(1)}_j - \bar{\bm{X}}^{(2)}_j$ for $j\in\{1,\ldots, p\}$. Since $Z_j \perp \bar{\bm{X}}_j$, given $\bar{\bm{X}}_j$, the variance of $M_j$ is in the order of $1/n$. This implies the first claim in Proposition \ref{prop:rank-DS-MDS}.

We proceed to prove the second claim. 
We have
\begin{equation}\nonumber
\begin{aligned}
I_i-I_j = \frac{1}{B}\sum_{b = 1}^B \frac{\mathbbm{1}(i\in \widehat S_b)-\mathbbm{1}(j\in \widehat S_b)}{|\widehat S_b|} = \frac{1}{B}\sum_{b\in \mathcal{B}^{+}}\frac{1}{|\widehat S_b|}-\frac{1}{B}\sum_{b\in  \mathcal{B}^{-}}\frac{1}{|\widehat S_b|},
\end{aligned}
\end{equation}
in which $B$ (as a function of $n$) denotes the total number of different sample splits, $b$ indexes one specific sample split, $\widehat{S}_b$ denotes the set of rejected hypotheses,
$\mathcal{B}^+ = \{b: i\in \widehat S_b,\ j\notin\widehat S_b\}$, and $ \mathcal{B}^- = \{b: j\in \widehat S_b,\ i\notin\widehat S_b\}$.

To characterize the event $\{i\in \widehat S,\ j\notin\widehat S\}$, we first define $\textnormal{Pivot}_1$ as follows: 
\begin{equation}\nonumber
\textnormal{Pivot}_1 = \min _{k\neq i,j}\bigg\{|M_k|: M_k < 0,\ \frac{ \# \{\ell: \ell\neq i,j, M_\ell < M_k\}}{2 +\# \{\ell: \ell\neq i,j, M_\ell > |M_k|\}} \leq q\bigg\}.
\end{equation}
We see that (1) $\{i,j\in\widehat S\}$ if $M_i > \textnormal{Pivot}_1$ and $M_j > \textnormal{Pivot}_1$; (2) $\{i,j\notin\widehat S\}$ if $M_i \leq \textnormal{Pivot}_1$ and $M_j \leq \textnormal{Pivot}_1$. Therefore, the event $\{i\in \widehat S,\ j\notin\widehat S\}$ implies $M_i > \textnormal{Pivot}_1$ and $M_j \leq \textnormal{Pivot}_1$. However, the reverse may not be true, and requires more delicate analysis.
We further define $\textnormal{Pivot}_2$ as follows:
\begin{equation}\nonumber
\textnormal{Pivot}_2 = \min _{k\neq i,j}\bigg\{|M_k|: M_k< 0,\ \frac{ \# \{\ell: \ell\neq i,j, M_\ell < M_k\}}{ 1 + \# \{\ell: \ell\neq i,j, M_\ell > |M_k|\}} \leq q\bigg\}.
\end{equation}
We see that if $M_j \in[-\textnormal{Pivot}_2, \textnormal{Pivot}_1]$, $M_i> \textnormal{Pivot}_2$ implies the event $\{i\in \widehat S,\ j\notin\widehat S\}$. Otherwise, 
suppose there are $m$ mirror statistics smaller than $-\textnormal{Pivot}_2$, and we sort them as 
$$\textnormal{Pivot}_2<|M_{(1)}|<|M_{(2)}|<\ldots<|M_{(m)}|.$$
Let $M_{(0)} = -\textnormal{Pivot}_2$ and $M_{(m+1)} = -\infty$. 
For $h\in\{0,\ldots,m\}$, if $M_j \in [M_{(h+1)}, M_{(h)})$, $M_i> \textnormal{Pivot}_{h + 3}$ implies the event $\{i\in \widehat S,\ j\notin\widehat S\}$, in which
\begin{equation}\nonumber
    \textnormal{Pivot}_{h+3} = \min _{k\neq i,j}\left\{|M_k|: M_k<0,\ \frac{ \# \{\ell: \ell\neq i,j, M_\ell < M_k\}+\mathbbm{1}(M_k\geq M_{(h)})}{ 1+\# \{\ell: \ell\neq i,j, M_\ell > |M_k|\}} \leq q \right\}.
\end{equation}
Note that all the pivotal quantities only depend on $\{M_k, k\neq i, j\}$. Similarly, the event $\{i\notin \widehat S,\ j\in\widehat S\}$ is also characterized by these pivotal quantities.

We can thus define new partitions of $\mathcal{B}^+$/$\mathcal{B}^-$ using these pivotal quantities.
Denote $M_{(-1)} = \text{Pivot}_1$. For $h \in \{0,\ldots,m+1\}$, let 
$$\mathcal{B}^+_h = \{b:M_j \in [M_{(h)}, M_{(h-1)}), M_i>\textnormal{Pivot}_{h+2}\}.$$
Similarly, we can define $\mathcal{B}^-_h$ by simply exchanging the indexes $i, j$. Then we have $ \mathcal{B}^+ = \bigcup_{h=0}^{m +1} \mathcal{B}^+_h$ and $ \mathcal{B}^- = \bigcup_{h=0}^{m +1} \mathcal{B}^-_h$.
In addition, $|\widehat S_b|$ remains a constant for $b\in \mathcal{B}^+_h\bigcup \mathcal{B}^-_h$. Therefore, it is sufficient for us to show that for any $h \in \{0,\ldots,m+1\}$, with probability approaching 1, $|\mathcal{B}^+_h| - |\mathcal{B}^-_h| \geq 0$.

We have
\begin{equation}\nonumber
\begin{aligned}
     \frac{1}{B}|\mathcal{B}^+_h|-\frac{1}{B}|\mathcal{B}^-_h| &= \mathbbm{P}(2|\bar{\bm{X}}_j|-M_{(h-1)}<|Z_j|<2|\bar{\bm{X}}_j|-M_{(h)})\mathbbm{P}(|Z_i|<2|\bar{\bm{X}}_i|-\textnormal{Pivot}_{h+2})\\
& \hspace{0.08cm} - \mathbbm{P}(2|\bar{\bm{X}}_i|-M_{(h-1)}<|Z_i|<2|\bar{\bm{X}}_i|-M_{(h)})\mathbbm{P}(|Z_j|<2|\bar{\bm{X}}_j|-\textnormal{Pivot}_{h+2})+o_p(1).
\end{aligned}
\end{equation}
Since $|\bar{\bm{X}}_i|-|\bar{\bm{X}}_j| = O(1/\sqrt{n})$ and the variances of $Z_i$ and $Z_j$ are in the order of $1/n$, we have
\begin{equation}
\label{eq:inclusion-inequality-one}
    \begin{aligned}
    \mathbbm{P}(|Z_i|<2|\bar{\bm{X}}_i|-\textnormal{Pivot}_{h+2}) & = \mathbbm{P}(|Z_i|<2|\bar{\bm{X}}_j|-\textnormal{Pivot}_{h+2}+2|\bar{\bm{X}}_i|-2|\bar{\bm{X}}_j|)\\
    & = \mathbbm{P}(\sqrt{n}|Z_i|<\sqrt{n}(2|\bar{\bm{X}}_j|-\textnormal{Pivot}_{h+2})+2\sqrt{n}(|\bar{\bm{X}}_i|-|\bar{\bm{X}}_j|))\\
    & >\mathbbm{P}(|Z_j|<2|\bar{\bm{X}}_j|-\textnormal{Pivot}_{h+2})+O(1),
    \end{aligned}
\end{equation}
and
\begin{equation}\nonumber
    \begin{aligned}
    &\mathbbm{P}(2|\bar{\bm{X}}_j|- M_{(h-1)}<|Z_j|<2|\bar{\bm{X}}_j|-M_{(h)}) > \mathbbm{P}(2|\bar{\bm{X}}_i|-M_{(h-1)}<|Z_i|<2|\bar{\bm{X}}_i|-M_{(h)}).
    \end{aligned}
\end{equation}
The above inequality follows from the following simple fact: suppose $Z$ follows the standard Normal distribution, then $\mathbbm{P}(a<|Z|<b)> \mathbbm{P}(a+c<|Z|<b+c)$ for any constants $a,\ b,\ c > 0$. This concludes the proof of Proposition \ref{prop:rank-DS-MDS}.
We remark that if $|\bar{\bm{X}}_i|-|\bar{\bm{X}}_j| = o(1/\sqrt{n})$, the $O(1)$ term in Equation \eqref{eq:inclusion-inequality-one} would be replaced by an $o(1)$ term. Since the approximation error is also in the order of $o_p(1)$, the rankings of $I_i$ and $I_j$ are not necessarily aligned with the rankings by the p-values.

\subsubsection{Proof of Proposition \ref{prop:linear-FDR}}
We first prove the claims for DS. 
By Assumption \ref{assump:linear}, as $n,p\to\infty$, the event $E_1 = \{S_1 \subseteq \widehat{S}^{(1)}\}$ holds with probability approaching 1.
In the following, we implicitly condition on the desired $(\bm{X}^{(1)}, \bm{y}^{(1)})$ such that $E_1$ holds.
Let $\widehat{S}_0 = \widehat{S}^{(1)} \bigcap S_0$, $\hat{p}_0 = |\widehat{S}_0|$ and $\hat p = |\widehat{S}^{(1)}|$. 
Since $p_1 \to \infty$, by the sure screening property, $\hat{p}\to\infty$. Without loss of generality, we assume $\hat{p}_0\to\infty$, otherwise the FDR control problem becomes trivial.
Define $R$ and its normalized version $R^0$ as
\begin{equation}\nonumber
R = \left(\frac{1}{n/2}\ \bm{X}_{\widehat{S}^{(1)}}^{(2)}{}^\intercal 
\bm{X}_{\widehat{S}^{(1)}}^{(2)}\right)^{-1}, \ \ \ R^0_{ij} = \frac{R_{ij}}{\sqrt{R_{ii}R_{jj}}},
\end{equation}
in which $R^0$ characterizes the correlation structure of the OLS regression coefficients $\widehat{\bm{\beta}}^{(2)}$.
Let $||R^0_{\widehat S_0}||_1 = \sum_{i, j\in \widehat{S}_0}|R^0_{ij}|$ and $||R^0_{\widehat S_0}||_2 = (\sum_{i, j\in \widehat{S}_0}|R^0_{ij}|^2)^{1/2}$.
We have the following probabilistic bound.

\begin{lemma}
\label{lemma:linear-covariance-bound}
Under Assumption \ref{assump:linear}, we have 
$||R^0_{\widehat S_0}||_1 = O_p(\hat p_0^{3/2})$.
\end{lemma}
\textit{Proof of Lemma \ref{lemma:linear-covariance-bound}.}  
We first show that $|\widehat S^{(1)}| = o_p(n)$.
We define the $m$-sparse minimal eigenvalue $\phi_{\min}(m)$ and the $m$-sparse maximal eigenvalue $\phi_{\max}(m)$ of the covariance matrix $\Sigma$ as follows:
$$\phi_{\min}(m) = \min_{\bm\beta:||\bm\beta||_{0}\leq m}\frac{\bm\beta^\intercal \Sigma\bm\beta}{\bm\beta^\intercal\bm\beta}\ \  \textnormal{and}\ \  \phi_{\max}(m) = \max_{\bm\beta:||\bm\beta||_{0}\leq m}\frac{\bm\beta^\intercal \Sigma\bm\beta}{\bm\beta^\intercal\bm\beta}.$$ 
For the sample covariance matrix $\widehat{\Sigma}$, we denote them as $\hat{\phi}_{\min}(m)$ and $\hat{\phi}_{\max}(m)$. By Corollary 3.3 in \citet{6471235}, we have
\begin{equation}
\label{eq:spectral-difference1}
\hat{\phi}_{\min}(p_1\log n) \geq \frac{1}{2} \phi_{\min}(p_1\log n) \geq  \frac{1}{2}\lambda_{\min}(\Sigma) \geq \frac{1}{2c}
\end{equation}
with probability approaching $1$.
Following the argument in Section 2.2 of \citet{meinshausen2009}, we can show that $\hat\phi_{\max}(p_1+\min\{n,p\})$ is upper bounded with probability approaching $1$.
Thus, the required conditions in Corollary 1 in \citet{meinshausen2009} hold with probability approaching 1, which further implies that Lasso selects at most $O(p_1)$ features (see their discussions in Section 2.4). Since $p_1 = o(n)$, we have $|\widehat S^{(1)}| = o_p(n)$. Consequently, 
\begin{equation}\nonumber
\begin{aligned}
&\lambda_{\min}(R) \geq 1/\lambda_{\max }(\Sigma_{\widehat S^{(1)}})-o_p(1) \geq 1/c - o_p(1),\\
&\lambda_{\max}(R)\leq 1/\lambda_{\min}(\Sigma_{\widehat S^{(1)}})+o_p(1)\leq c + o_p(1).
\end{aligned}
\end{equation}
It follows that
\begin{equation}
\begin{aligned}\nonumber
    ||R_{\widehat S_0}^0||_1 & \leq 1/\lambda_{\min}(R_{\widehat S_0})||R_{\widehat S_0}||_1\leq \hat p_0/\lambda_{\min}(R_{\widehat S_0}) ||R_{\widehat S_0}||_2\\
    & \leq \hat p^{3/2}_0\lambda_{\max}(R_{\widehat S_0})/\lambda_{\min}(R_{\widehat S_0}) = O_p(\hat p_0^{3/2}).
\end{aligned}
\end{equation}
The first inequality follows from the fact that for any positive definite matrix $A\in \mathbbm{R}^{m\times m}$, $\lambda_{\min}(A)\leq A_{ii}\leq \lambda_{\max}(A)$ for $i\in\{1\ldots,m\}$. The second inequality follows from the Cauchy-Schwartz inequality. The third inequality is based on the following fact:
\begin{equation}\nonumber
   \sum_{i,j}A_{ij}^2 = \text{tr}(A^\intercal A) = \sum_{i = 1}^m\lambda_{i}^2(A).
\end{equation}
The proof is thus completed.
\vspace{0.5cm}

By Lemma \ref{lemma:linear-covariance-bound}, we can further condition on the desired $\bm{X}^{(2)}$ such that the event $E_2 = \{||R^0_{\widehat S_0}||_1\leq c_1 \hat{p}_0^{3/2}\}$ holds for some constant $c_1 > 0$. Define $\widehat{G}^0_{p}(t),\ G^0_{p}(t),\ \widehat{V}^0_{p}(t)$ in analogy to \eqref{eq:fp-tp-proportion}, by replacing $p_0$, $p$, $S_0$ with $\hat p_0$, $\hat p$, $\widehat S_0$, respectively. Similar to Lemma \ref{lemma:FDR}, we have the following Lemma \ref{lemma:linear-fixed-design}. 

\begin{lemma}
\label{lemma:linear-fixed-design}
For any $t \in \mathbbm{R}$, under Assumption \ref{assump:linear}, we have in probability
\begin{equation}\nonumber
\left|\widehat{G}^0_{ p}(t) - G^0_{ p}(t)\right| \longrightarrow 0,\ \ \ \left|\widehat{V}^0_{ p}(t) - G^0_{ p}(t)\right| \longrightarrow 0.
\end{equation}
\end{lemma}

\noindent\textit{Proof of Lemma \ref{lemma:linear-fixed-design}.} We prove the first claim by conditioning on the desired $(\bm{X}^{(1)}, \bm{y}^{(1)})$ and $\bm{X}^{(2)}$ such that the high probability event $E_1\bigcap E_2$ holds. The second claim follows similarly. 
We have the following decomposition,
\begin{equation}\nonumber
\textnormal{Var}\big(\widehat{G}^0_{p}(t)\big)  = \frac{1}{\hat p_0^2}\sum_{j\in \widehat S_0}\text{Var}(\mathbbm{1}(M_j > t)) + \frac{1}{\hat p_0^2}\sum_{i\neq j\in \hat S_0}\text {Cov}(\mathbbm{1}(M_i > t), \mathbbm{1}(M_j > t)).
\end{equation}

The first term is bounded by $1/\hat p_0$. We proceed to bound the covariance in the second term for each pair $(i,j)$. Without loss of generality, we assume $\widehat{\beta}_i^{(1)} > 0$ and $\widehat{\beta}_j^{(1)} > 0$. 
Note that $(\widehat{\beta}_i^{(2)}, \widehat{\beta}_j^{(2)})$ follows a bivariate Normal distribution with correlation $R^0_{ij}$.
Using  Mehler's identity \citep{kotz2000bivariate}, i.e., for any $t_1, t_2 \in\mathbbm{R}$,
\begin{equation}\nonumber
\Phi_r(t_1, t_2) = \Phi(t_1)\Phi(t_2) + \sum_{n = 1}^\infty \frac{r^n}{n!}\phi^{(n - 1)}(t_1)\phi^{(n - 1)}(t_2),
\end{equation}
together with Lemma 1 in \citet{azriel2015empirical}, i.e.,
\begin{equation}\nonumber
\sum_{n = 1}^\infty \frac{\left[\sup_{t\in\mathbbm{R}}\phi^{(n - 1)}(t)\right]^2}{n!} < \infty,
\end{equation}
we have
\begin{equation}\nonumber
\begin{aligned}
\mathbbm{P}(M_i > t, M_j > t) & = \mathbbm{P}\left(\widehat{\beta}_i^{(2)} > I_t\big(\widehat{\beta}_i^{(1)} \big), \widehat{\beta}_j^{(2)} > I_t\big(\widehat{\beta}_j^{(1)}\big)\right)\\
& \leq \mathbbm{P}\left(\widehat{\beta}_i^{(2)} > I_t\big(\widehat{\beta}_i^{(1)}\big)\right) \mathbbm{P}\left(\widehat{\beta}_j^{(2)} > I_t\big(\widehat{\beta}_j^{(1)}\big)\right) + O(|R^0_{ij}|).
\end{aligned}
\end{equation}
Conditioning on the event $E_2$, it follows that 
\begin{equation}\nonumber
\frac{1}{\hat p_0^2} \sum_{i\neq j\in S_0}\text{Cov}(\mathbbm{1}(M_i > t), \mathbbm{1}(M_j > t)) \leq \frac{c_2}{\hat p_0^2}||R^0_{\widehat S_0}||_1 \leq \frac{c_1c_2}{\sqrt{\hat{p}_0}}
\end{equation}
for some constant $c_2 >0$.
The proof of Lemma \ref{lemma:linear-fixed-design} concludes using the Markov's inequality.

\vspace{0.5cm}
Let $\hat{r}_p = p_1/\hat{p}_0$. In the following, we show that as $n,p\to\infty$, with probability approaching 1, $\hat{r}_p\widehat{G}_p^1(\tau_q)$ is bounded away from 0. 
First, $\hat{p}_0 = O_p(p_1)$ based on the arguments in the proof of Lemma \ref{lemma:linear-covariance-bound}, thus $\hat{r}_p$ is asymptotically bounded away from 0.
Second, following \citet{bickel2009simultaneous}, we have
\begin{equation}\nonumber
||\widehat{\bm\beta}^{(1)} - \bm{\beta}^\star||_\infty = O_p(\sqrt{p_1\log p/n}).
\end{equation}
For OLS, since the eigenvalues of $R$ are doubly bounded with high probability (see Lemma \ref{lemma:linear-covariance-bound}), we have 
\begin{equation}\nonumber
   ||\widehat{\bm\beta}^{(2)} - \bm{\beta}^\star||_\infty = O_p(\sqrt{\log \hat{p}/n})\leq O_p(\sqrt{\log n/n}). 
\end{equation}
Under the signal strength condition, as $n,p\to\infty$, with probability approaching 1, we have
\begin{equation}\nonumber
\min_{i\in S_1}|\widehat\beta^{(1)}_i| \geq \max_{j\in S_0}|\widehat \beta^{(1)}_j|\ \ \ \text{and}\ \ \ 
\min_{i\in S_1}|\widehat\beta^{(2)}_i| \geq \max_{j\in S_0}|\widehat \beta^{(2)}_j|,
\end{equation}
which further implies that
\begin{equation}
\label{eq:key-relation}
\min_{i\in S_1} |M_i|\geq \max_{j\in S_0} |M_j|.
\end{equation}  
Consequently, we have $\widehat{G}_p^1(\tau_q) \to 1$ with probability approaching 1. Thus, the power of DS asymptotically converges to 1. 
FDR control of DS then follows from the same arguments in the proof of Proposition \ref{prop:FDR}.

\vspace{0.5cm}
We proceed to prove the claims for MDS. 
By Equation \eqref{eq:key-relation}, as $n,p\to\infty$, with probability approaching 1, we have
\begin{equation}\nonumber
\min_{i\in S_1} I_i \geq \max_{j\in S_0} I_j.
\end{equation}
Therefore, the mirror statistics are strongly consistent at ranking features. FDR control of MDS then follows from Proposition \ref{prop:MDS-FDR}. 

The power of MDS is $(p - \ell)/p_1$, in which $\ell$ is defined in Algorithm \ref{alg:multiple-splits}, thus it is sufficient to establish a probabilistic upper bound for $\ell$.
By Lemma \ref{lemma:linear-fixed-design}, we have $|\textnormal{FDP}_p^{\dagger}(t)-\textnormal{FDP}_p(t)|\stackrel{p}{\to} 0$, $\forall t\in\mathbbm{R}$, thus $\textnormal{FDP}_p(\tau_q) = q + o_p(1)$. Using the same arguments in the proof of Lemma \ref{lemma:ell-upper-bound}, we have $\forall \delta > 0$, 
\begin{equation}\nonumber
\mathbbm{P}\big(\sum_{j\in S_0}I_j > q - \delta\big) \to 1.
\end{equation}
In the following, we implicitly condition on the following high probability events: 
\begin{center}
(1) $\sum_{j\in S_0}I_j > q - \delta$;\ \ \ (2) $\min_{i\in S_1} I_i \geq \max_{j\in S_0} I_j$; \ \ \ (3) DS achieves power 1. 
\end{center}
Since the event (2) implies that MDS has power 1 if $\ell \leq p_0$, throughout we assume $\ell > p_0$.
On the one hand, the events (1) and (3) imply that
\begin{equation}
\label{eq:MDS-power-ell-bound-1}
I_{(p_0 + 1)} = \cdots = I_{(p)} = \frac{1}{p_1}\sum_{i\in S_1}I_i \geq (1 - q + \delta)/{p_1}.
\end{equation}
On the other hand, by the definition of $\ell$, we have $\sum_{k = 1}^{\ell} I_{(k)}\leq q$, thus
\begin{equation}
\label{eq:MDS-power-ell-bound-2}
I_{(p_0+1)}+\ldots+I_{(\ell)}\leq \delta
\end{equation}
conditioning on the event (1).
Equations \eqref{eq:MDS-power-ell-bound-1} and \eqref{eq:MDS-power-ell-bound-2} together implies that $(\ell-p_0)(1-q+\delta)/p_1\leq \delta$. Consequently, as $n,p\to\infty$, we have
\begin{equation}\nonumber
\mathbbm{P}\bigg(\frac{p - \ell}{p_1} \geq  1 - \frac{\delta}{1 - q + \delta}\bigg) \to 1.
\end{equation}
The proof is thus completed.

\subsubsection{Proof of Proposition \ref{prop:gaussian-graphical-model}}
Throughout, $c$ refers to a general positive constant and may vary case by case. We first show that $\forall j$, 
the restricted eigenvalue condition (\citet{bickel2009simultaneous}) holds with probability  approaching 1 in the $j$-th nodewise regression.
For any $J_0\subseteq \{1,\ldots,p\}\backslash\{j\}$ with $|J_0| \leq s$ and any $\bm{v} \neq \bm{0}$ satisfying $||\bm{v}_{J_0^c}||_1 \leq ||\bm{v}_{J_0}||_1$, we have $||\bm{v}||_1 \leq 2||\bm{v}_{J_0}||_1 \leq 2\sqrt{s}||\bm{v}||_2$. By Theorem 1 in \citet{raskutti2010restricted} and the Cauchy interlacing theorem, with probability at least $1 - \exp(-cn)$, we have
\begin{equation}\nonumber
\begin{aligned}
\frac{||\bm{X}_{-j}^{(1)}\bm{v}||_2}{\sqrt{n/2}} & \geq \left(\frac{1}{4}\lambda_{\min}(\Sigma) - 18\max_{1\leq j\leq p}\sigma_{jj}\sqrt{\frac{s\log p}{n/2}}\right)||\bm{v}||_2.
\end{aligned}
\end{equation}
Under Assumption \ref{assump:graphical-model}, we have $\max_{j\in\{1,\ldots,p\}} \sigma_{jj}\leq \lambda_{\max}(\Sigma)\leq 1/c$, thus
\begin{equation}\nonumber
\max_{1\leq j\leq p}\sigma_{jj}\sqrt{\frac{s\log p}{n/2}} \to 0.
\end{equation}
Therefore, with high probability, the restricted eigenvalue condition is satisfied. It follows that the Lasso estimator $\bm{\widehat{\beta}}^j$ satisfies the following bound
\begin{equation}\nonumber
||\bm{\widehat{\beta}}^j-\bm{\beta}^j||_2\leq c\sqrt{\frac{s\log p}{n}}
\end{equation}
with probability at least $1-\exp(-cn)$.
By the union bound, we know that with probability approaching 1, both the restricted eigenvalue condition and the $\ell_2$-bound simultaneously hold in all $p$ nodewise regressions. 
Together with the signal strength condition in Assumption \ref{assump:graphical-model}, we prove that the sure screening property, thus the symmetric assumption, simultaneously holds in all $p$ nodewise regressions with probability approaching 1 as $n,p\to\infty$.

\subsubsection{Proof of Proposition \ref{prop:graphical-FDR}}
By Proposition \ref{prop:gaussian-graphical-model}, throughout we implicitly condition on the desired $\bm{X}^{(1)}$ such that the sure screening property simultaneously holds for all $p$ nodewise regressions. For $j\in\{1,\ldots,p\}$, let $\widehat{S}_j$ be the index set of the selected vertexes that are connected to vertex $X_j$. 
For the ease of presentation, we introduce the following notations. For $j\in\{1,\ldots,p\}$ and $t \in\mathbbm{R}$, denote $\widehat{ne}_j^c = \widehat{S}_j \bigcap ne_j^c$ and 
\begin{equation}\nonumber
\begin{aligned}
&\widehat{G}^0_{p, j}(t) = \frac{1}{|\widehat{ne}_j^c|}\sum_{i \in \widehat{ne}_j^c}\mathbbm{1}(M_{ji} > t),\ \ \
\widehat{V}^0_{p, j}(t) = \frac{1}{|\widehat{ne}_j^c|}\sum_{i \in \widehat{ne}_j^c}\mathbbm{1}(M_{ji} < - t),\\
&\widehat{G}^1_{p, j}(t) = \frac{1}{|ne_j|}\sum_{i \in ne_j}\mathbbm{1}(M_{ji} > t),\ \ \ G^0_{p, j}(t) = \frac{1}{|\widehat{ne}_j^c|}\sum_{i \in \widehat{ne}_j^c}\mathbbm{P}(M_{ji} > t). 
\end{aligned}
\end{equation}
Let $\hat\pi^0_{p, j} = |\widehat{ne}_j^c|/\sum_{j = 1}^p |\widehat{ne}_j^c|$, $\pi^1_{p, j} = |ne_j|/\sum_{j = 1}^p |ne_j|$, and $\hat{r}_{p,j} = \sum_{j = 1}^p|ne_j|/\sum_{j = 1}^p|\widehat{ne}^c_j|$. In addition, denote
\begin{equation}\nonumber
\begin{aligned}
\text{FDP}_p(t_1, \cdots, t_p) & = \frac{\sum_{j = 1}^p\hat\pi^0_{p, j}\widehat{G}_{p, j}^0(t_j)}{\sum_{j = 1}^p\hat\pi^0_{p, j}\widehat{G}_{p, j}^0(t_j) + \hat{r}_{p,j}\sum_{j = 1}^p\pi^1_{p, j}\widehat{G}_{p, j}^1(t_j)},\\
\text{FDP}^\dagger_p(t_1, \cdots, t_p) & = \frac{\sum_{j = 1}^p\hat\pi^0_{p, j}\widehat{V}_{p, j}^0(t_j)}{\sum_{j = 1}^p\hat\pi^0_{p, j}\widehat{G}_{p, j}^0(t_j) + \hat{r}_{p,j}\sum_{j = 1}^p\pi^1_{p, j}\widehat{G}_{p, j}^1(t_j)},\\
\overline{\text{FDP}}_p(t_1, \cdots, t_p) & = \frac{\sum_{j = 1}^p\hat\pi^0_{p, j}G_{p, j}^0(t_j)}{\sum_{j = 1}^p\hat\pi^0_{p, j}G_{p, j}^0(t_j) + \hat{r}_{p,j}\sum_{j = 1}^p\pi^1_{p, j}\widehat{G}_{p, j}^1(t_j)}.
\end{aligned}
\end{equation}

\begin{lemma}
\label{lemma:graphical-FDR}
Under Assumption \ref{assump:graphical-model}, as $n,p\to\infty$, we have in probability
\begin{equation}\nonumber
\begin{aligned}
& \sup_{t_1, \cdots, t_p} \left|\sum_{j = 1}^p \hat\pi^0_{p, j}\left(\widehat{G}^0_{p, j}(t_j) - G^0_{p, j}(t_j)\right)\right| \longrightarrow 0,\\
& \sup_{t_1, \cdots, t_p} \left|\sum_{j = 1}^p \hat\pi^0_{p, j}\left(\widehat{V}^0_{p, j}(t_j) - G^0_{p, j}(t_j)\right)\hspace{0.08cm}\right| \longrightarrow 0.
\end{aligned}
\end{equation}
\end{lemma}

\noindent\textit{Proof of Lemma \ref{lemma:graphical-FDR}}. The key step is to establish a concentration inequality for $\widehat G^0_{p,j}(t)$. 
Without loss of generality, we assume $\widehat\beta_{ji}^{(1)} > 0$ thus
\begin{equation}\nonumber
    \mathbbm{1}(M_{ji}>t) = \mathbbm{1}(\widehat\beta_{ji}^{(2)}>I_t(\widehat\beta_{ji}^{(1)})).
\end{equation}
Let $t_{ji} = I_t(\widehat\beta_{ji}^{(1)})$ for $i \in ne_j^c$. 
We use the following Lipschitz continuous function $\phi_{t, L}(x)$ to approximate the indicator function $\mathbbm{1}(\widehat\beta_{ji}^{(2)}>t)$, 
\begin{equation}\nonumber
    \phi_{t,L}(x) = \left\{
    \begin{array}{ll}
    0,     & x\leq t-1/L, \vspace{0.1cm} \\ \vspace{0.1cm}
    Lx-Lt+1,     & x\in(t-1/L, t),\\
    1, & x\geq t,
    \end{array}
    \right.
\end{equation}
in which $L$ will be specified later. 

We note that $\{\widehat\beta_{ji}^{(2)}, i\in \widehat{ne}_j^c\}$ jointly follow a centered multivariate Normal distribution. In addition, by repeating the arguments in the proof of Lemma \ref{lemma:linear-covariance-bound}, we can show that with high probability, the eigenvalues of the corresponding covariance matrix are doubly bounded, and Lasso selects at most $O_p(ne_j)$ edges in the $j$-th nodewise regression.\footnote{By repeating the arguments in the proof of Lemma 5 in \citet{meinshausen2009}, together with the union bound, we can show that with probability approaching 1, the eigenvalues of the $p$ covariance matrices of the OLS regression coefficients are simultaneously doubly bounded, and Lasso selects at most $O_p(\max_{j\in[p]}|ne_j|)$ edges across all $p$ nodewise regressions.}
Therefore, we can write
$$\frac{1}{|\widehat{ne}_j^c|}\sum_{i\in \widehat{ne}_j^c}\phi_{t_{ji}, L}(\widehat\beta_{ji}^{(2)}) = g(Z_1, \cdots, Z_{|\widehat{ne}_j^c|}),$$ 
in which $\bm{Z} = (Z_1, \cdots, Z_{|\widehat{ne}_j^c|})^\intercal$ follows the standard multivariate Normal distribution, and $g$ is a Lipschitz continuous funtion with a Lipschitz constant $c_1L/|\widehat{ne}_j^c|^{1/2}$ for some $c_1 > 0$.

Then we have $\forall\epsilon > 0$,
\begin{equation}\nonumber
\begin{aligned}
\mathbbm{P}\left(\widehat G^0_{p,j}(t)-G^0_{p,j}(t)>\epsilon\right) & \leq \mathbbm{P}\bigg(\frac{1}{|\widehat{ne}_j^c|}\sum_{i\in \widehat{ne}_j^c}\phi_{t_{ji}, L}(\widehat\beta_{ji}^{(2)})-G^0_{p,j}(t)>\epsilon\bigg)\\
& \leq \mathbbm{P}\bigg(\frac{1}{|\widehat{ne}_j^c|}\sum_{i\in \widehat{ne}_j^c}\phi_{t_{ji}, L}(\widehat\beta_{ji}^{(2)})-\frac{1}{|\widehat{ne}_j^c|}\sum_{i\in \widehat{ne}_j^c}\mathbbm{E}(\phi_{t_{ji}, L}(\widehat\beta_{ji}^{(2)}))>\epsilon-1/L\bigg)\\
& \leq 2\exp\bigg(-\frac{|\widehat{ne}_j^c|( \epsilon-1/L)^2}{2c^2_1L^2}\bigg).
\end{aligned}
\end{equation}
The first inequality follows from the fact that $\mathbbm{1}(x>t)\leq\phi_{t, L}(x)$ for any $t$ and $x$. The second inequality follows from the fact that $\phi_{t, L}(x)$ and $\mathbbm{1}(x>t)$ only differ on the interval $(t-1/L, t)$, thus 
$$|\mathbbm{E}[\phi_{t, L}(W)]-\mathbbm{P}(W>t)|<1/L,$$
in which $W$ follows the Normal distribution. The third inequality follows from the Gaussian concentration inequality (see Theorem 2.26 in \cite{wainwright2019high}).

It remains to choose a proper Lipschitz constant $L$. Without loss of generality, we assume $|\widehat{ne}_j^c| \geq c_2|ne_j|$ for $\forall j\in\{1,\ldots,p\}$ and some $c_2 > 0$.
Since  $\min|ne_j|/\log p \to \infty$, we can choose $L$ such that 
\begin{equation}\nonumber
-\frac{c_2\min|ne_j|( \epsilon-1/L)^2}{2c^2_1L^2} + \log 2N_\epsilon p < \log\epsilon,
\end{equation}
in which $N_\epsilon = \lceil2/\epsilon\rceil$.
Using the similar arguments in the proof of Lemma \ref{lemma:FDR}, by the union bound, we have
\begin{equation}\nonumber
\begin{aligned}
\mathbbm{P}\left(\sup_{t_1, \cdots, t_p} \left|\sum_{j = 1}^p \pi^0_{p, j}\left(\widehat{G}^0_{p, j}(t_j) - G^0_{p, j}(t_j)\right)\right| > \epsilon\right)  & \leq \mathbbm{P}\left(\bigcup_{j = 1}^p\sup_{t_j\in\mathbbm{R}}\left|\widehat{G}^0_{p, j}(t_j) - G^0_{p, j}(t_j)\right| > \epsilon\right)\\
& \leq 2N_\epsilon p\exp\left(-\frac{c_2\min|ne_j|( \epsilon-1/L)^2}{2c^2_1L^2}\right) < \epsilon.
\end{aligned}
\end{equation}
The first claim in Lemma \ref{lemma:graphical-FDR} thus holds, and the second claim follows similarly.

\vspace{0.5cm}
\noindent\textit{Proof of Proposition \ref{prop:graphical-FDR}}. We have the following decomposition.
\begin{equation}\nonumber
\begin{aligned}
\limsup_{p\to\infty}\text{FDR} & \leq \limsup_{p\to\infty} \mathbbm{E}\left[\text{FDP}_p\left(\tau^1_{q/2}, \cdots, \tau^p_{q/2}\right)\right] \\
& \leq \limsup_{p\to\infty} \mathbbm{E}\left|\text{FDP}_p\left(\tau^1_{q/2}, \cdots, \tau^p_{q/2}\right) - \overline{\text{FDP}}_p\left(\tau^1_{q/2}, \cdots, \tau^p_{q/2}\right)\right| \\
& \hspace{0.05cm} + \limsup_{p\to\infty} \mathbbm{E}\left|\text{FDP}^\dagger_p\left(\tau^1_{q/2}, \cdots, \tau^p_{q/2}\right) - \overline{\text{FDP}}_p\left(\tau^1_{q/2}, \cdots, \tau^p_{q/2}\right)\right| \\
& \hspace{0.05cm} + \limsup_{p\to\infty} \mathbbm{E}\left[\text{FDP}^\dagger_p\left(\tau^1_{q/2}, \cdots, \tau^p_{q/2}\right) \right]\\
& \leq \limsup_{p\to\infty} \mathbbm{E}\Big[\sup_{t_1, \cdots, t_p > 0}\left|\text{FDP}_p(t_1, \cdots, t_p) - \overline{\text{FDP}}_p(t_1, \cdots, t_p)\right|\Big] \\
& \hspace{0.05cm} + \limsup_{p\to\infty} \mathbbm{E}\Big[\sup_{t_1, \cdots, t_p > 0}\left|\text{FDP}^\dagger_p(t_1, \cdots, t_p) - \overline{\text{FDP}}_p(t_1, \cdots, t_p)\right|\Big] \\
& \hspace{0.05cm} + \limsup_{p\to\infty} \mathbbm{E}\left[\text{FDP}^\dagger_p\left(\tau^1_{q/2}, \cdots, \tau^p_{q/2}\right) \right].
\end{aligned}
\end{equation}
The first two terms are 0 based on Lemma \ref{lemma:graphical-FDR} and the dominated convergence theorem (similar as the proof of Proposition \ref{prop:linear-FDR}). For the last term, we have
\begin{equation}\nonumber
\limsup_{p\to\infty} \mathbbm{E}\left[\text{FDP}^\dagger_p\left(\tau^1_{q/2}, \cdots, \tau^p_{q/2}\right) \right] \leq \limsup_{p\to\infty}2\mathbbm{E}\left[\max_{1\leq p}\frac{\#\{i\in ne_j^c, M_{ji} < -\tau_{q/2}^j\}}{\#\{M_{ji} > \tau_{q/2}^j\}\vee 1}\right] \leq q
\end{equation}
following Equation \eqref{eq:graphical-intuition}. This establishes the FDR control property for DS. 
FDR control for MDS and the power guarantee for both DS and MDS follow similarly as the proof of  Proposition \ref{prop:linear-FDR}.

\subsection{The Normal means model}
\label{subsec:simulation-details}
\begin{figure}
\begin{center}
\includegraphics[width=0.45\columnwidth]{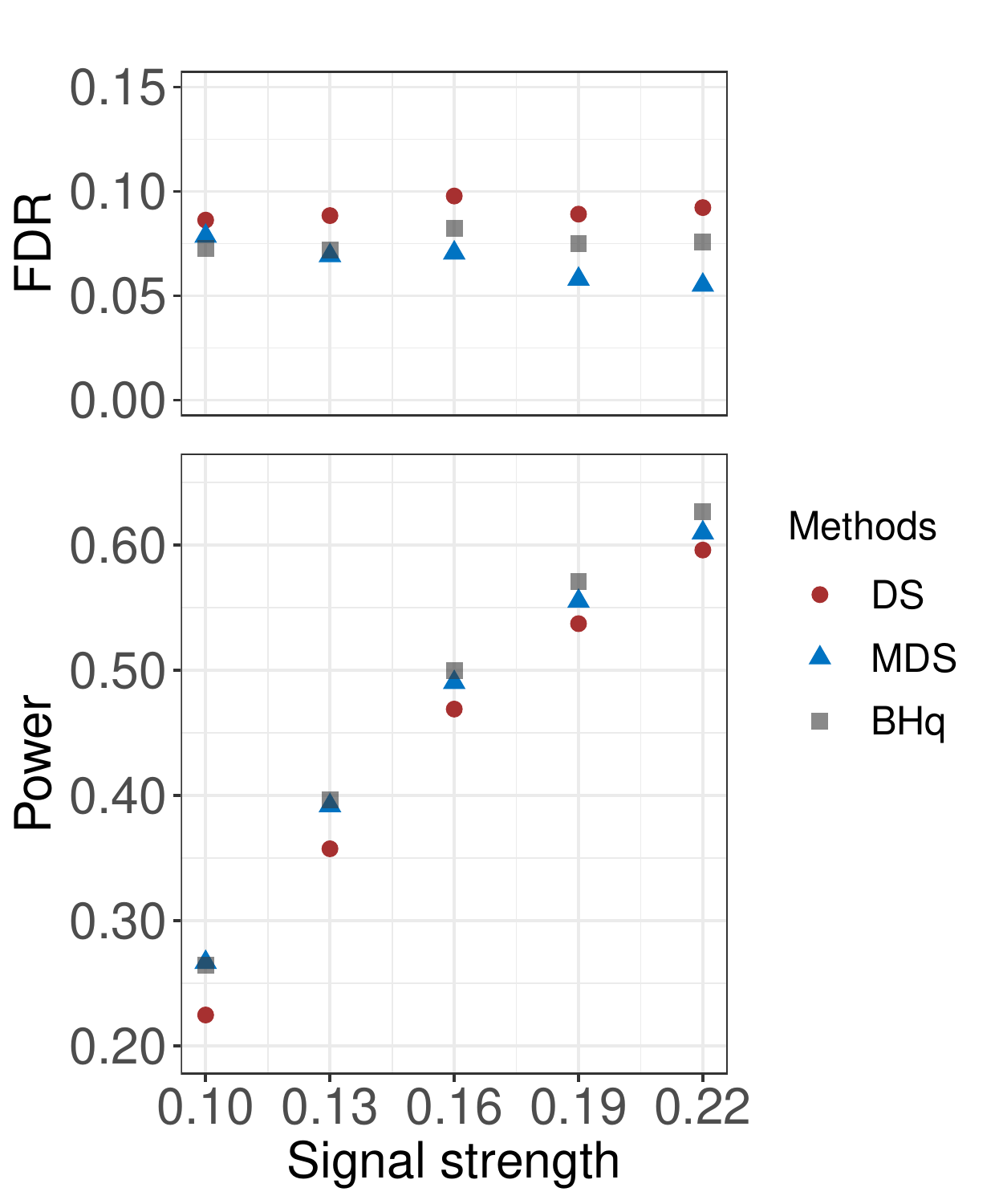}
\includegraphics[width=0.37\columnwidth]{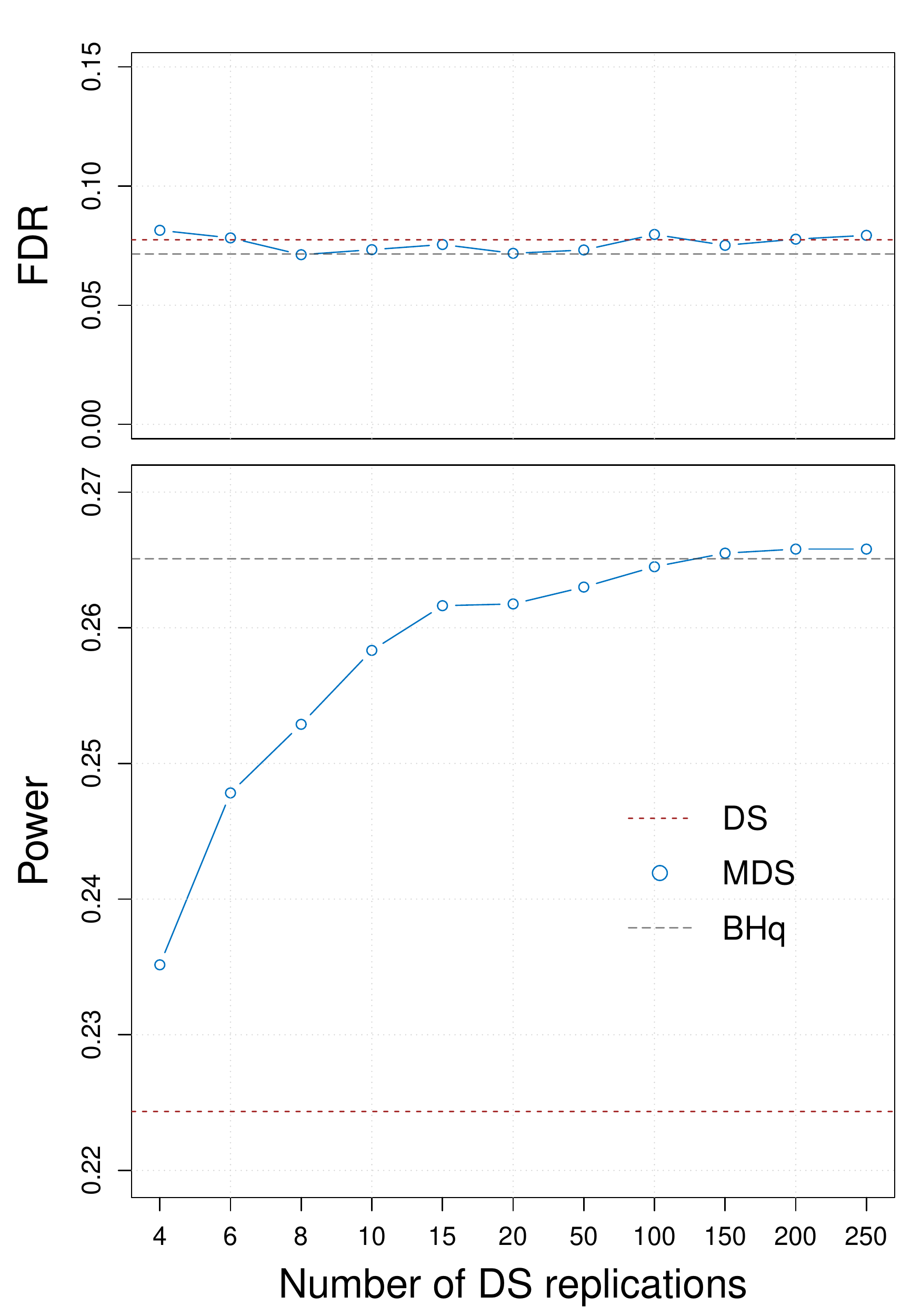}
\end{center}
\caption{Empirical FDRs and powers for the Normal means model. 
Throughout, we set $n = 500$, $p = 800$, and $p_1 = 160$.
For $j\in S_1$, the $\mu_j$'s are independent samples from $N(0, \delta^2)$.
In the left panels, we fix the number of DS replications in MDS at $m = 200$ and vary the signal strength $\delta$.
In the right panels, we fix the signal strength at $\delta = 0.08$ and vary the number of DS replications $m$.
The designated FDR control level is $q = 0.1$.
Each dot in the figure represents the average from 50 independent runs.}
\label{fig:normal-mean-power-splits}
\end{figure}

Figure \ref{fig:normal-mean-power-splits} (left panel) compares DS, MDS and BHq across various signal strengths. We see that DS controls the FDR slightly below the nominal level $q = 0.1$. For BHq, all the p-values are independent, and it controls the FDR at the theoretically predicted level $qp_1/p = 0.08$. MDS appears most conservative, but still enjoys a competitive power (comparable to BHq and higher than DS).
Figure \ref{fig:normal-mean-power-splits} (right panel) shows that the power of MDS becomes quite stable when the number of DS replications is larger than 100. 

\begin{figure}
\begin{center}
\includegraphics[width=0.45\columnwidth]{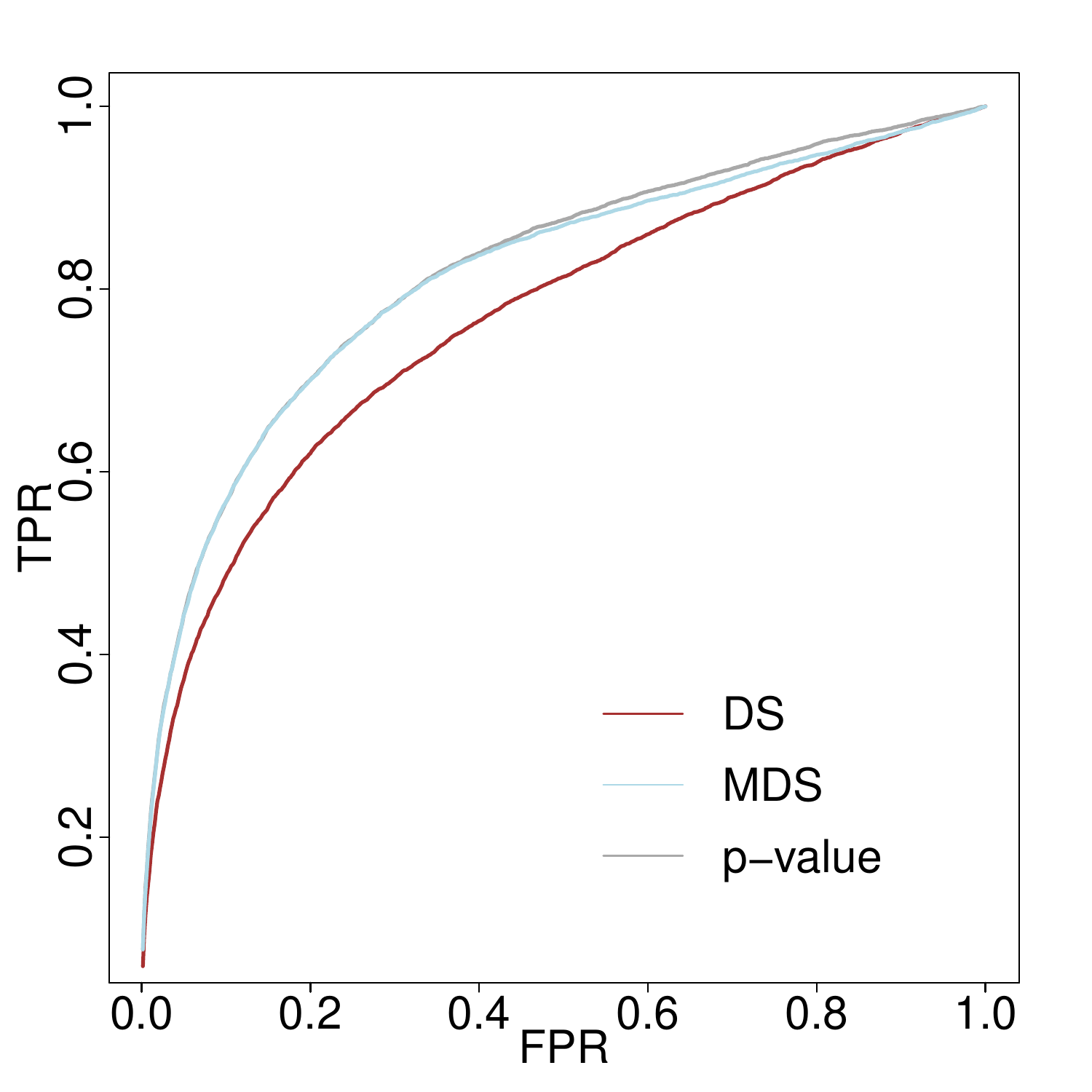}
\includegraphics[width=0.45\columnwidth]{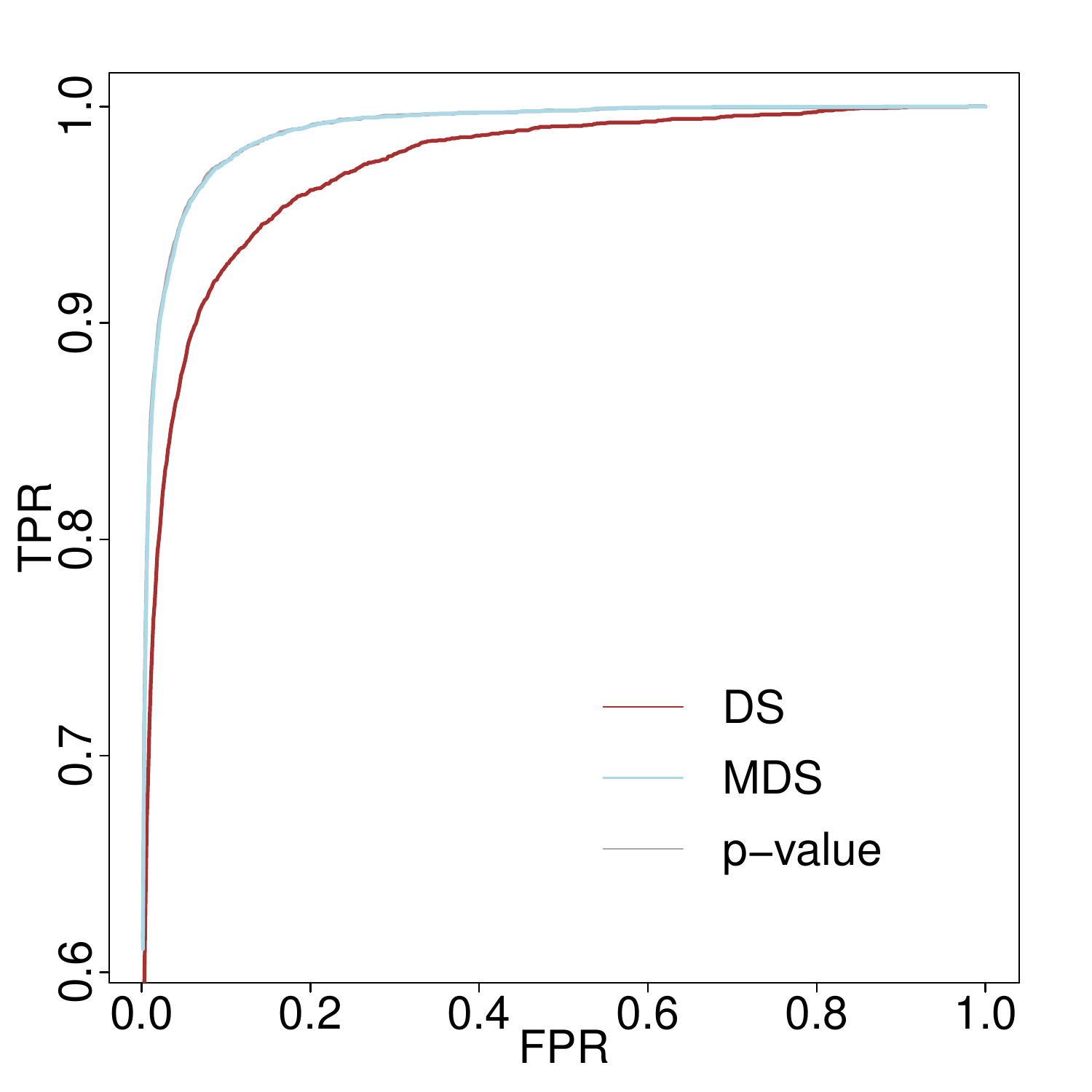}
\end{center}
\caption{ROC curves for the Normal means model. The $x$-axis and the $y$-axis denote the false positive rate (FPR) and the true positive rate (TPR), respectively. Throughout, we set $n = 500$, $p = 800$, and $p_1 = 160$. For $j\in S_1$, the $\mu_j$'s are set to be $\pm\delta$ with random signs. $\delta = 0.08$ and $0.16$ in the left and the right panel, respectively. We set the number of DS replications in MDS at $m = 1000$. The designated FDR control level is $q = 0.5$.
Each dot in the figure represents the average from 50 independent runs.}
\label{fig:normal-mean-ROC}
\end{figure}

Figure \ref{fig:normal-mean-ROC} plots the ROC curves of the rankings of features by the mirror statistics (DS), the inclusion rates (MDS) and the p-values, respectively. We set the FDR level to be slightly large, that is, at $q = 0.5$, so that a majority of features have nonzero inclusion rates. Compared to DS, MDS improves the rankings of features, and the ROC curve of MDS greatly overlaps with that of the p-values except at the right tail, which represents  features that have too low an inclusion rate for MDS to rank them properly. 

\subsection{Simulation details}
\label{subsec:simulation-details}
To complement Section \ref{subsec:sim-linear}, we first detail the blockwise diagonal Toeplitz covariance matrix, of which each block along the diagonal is set to be
\begin{equation}
\label{eq:blockwise-diagonal-formula}
\begin{bmatrix}
1 & \frac{(p^\prime-2)\rho}{p^\prime-1} & \frac{(p^\prime-3)\rho}{p^\prime-1} & \ldots & \frac{\rho}{p^\prime-1} & 0\\
& & & & & \\
\vspace{0.5cm}
\frac{(p^\prime-2)\rho}{p^\prime-1} & 1 & \frac{(p^\prime-2)\rho}{p^\prime-1} & \ldots & \frac{2\rho}{p^\prime-1} & \frac{\rho}{p^\prime-1} \\
\vspace{0.5cm}
\vdots & & \ddots & & & \vdots\\
0  & \frac{\rho}{p^\prime-1} & \frac{2\rho}{p^\prime-1} & \ldots & \frac{(p^\prime-2)\rho}{p^\prime-1} & 1
\end{bmatrix},
\end{equation}
where $p^\prime = p/10$. Throughout, we refer $\rho\in (0, 1)$ as the correlation factor. 

\begin{figure}
\begin{center}
\includegraphics[width=0.45\columnwidth]{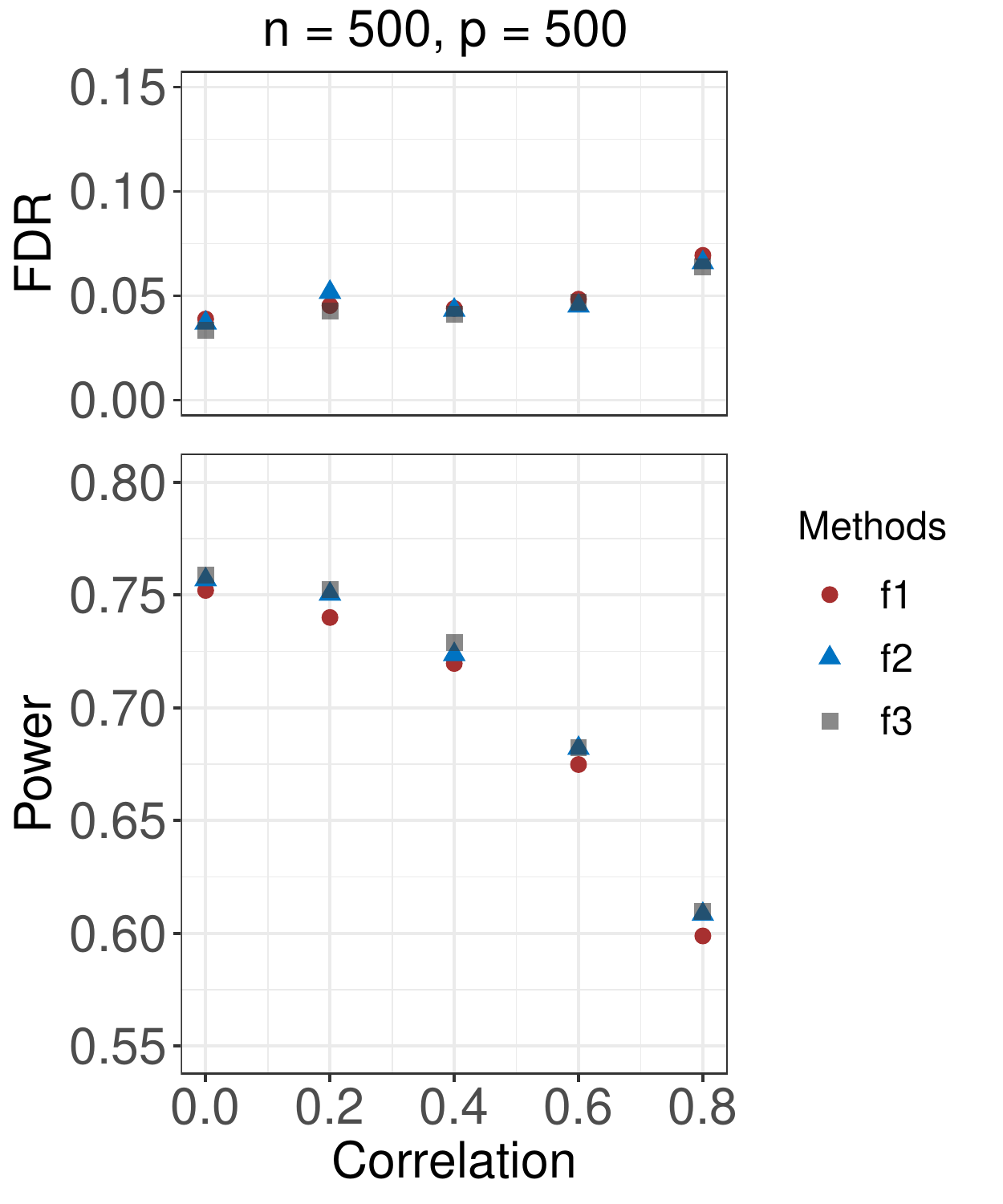}
\includegraphics[width=0.45\columnwidth]{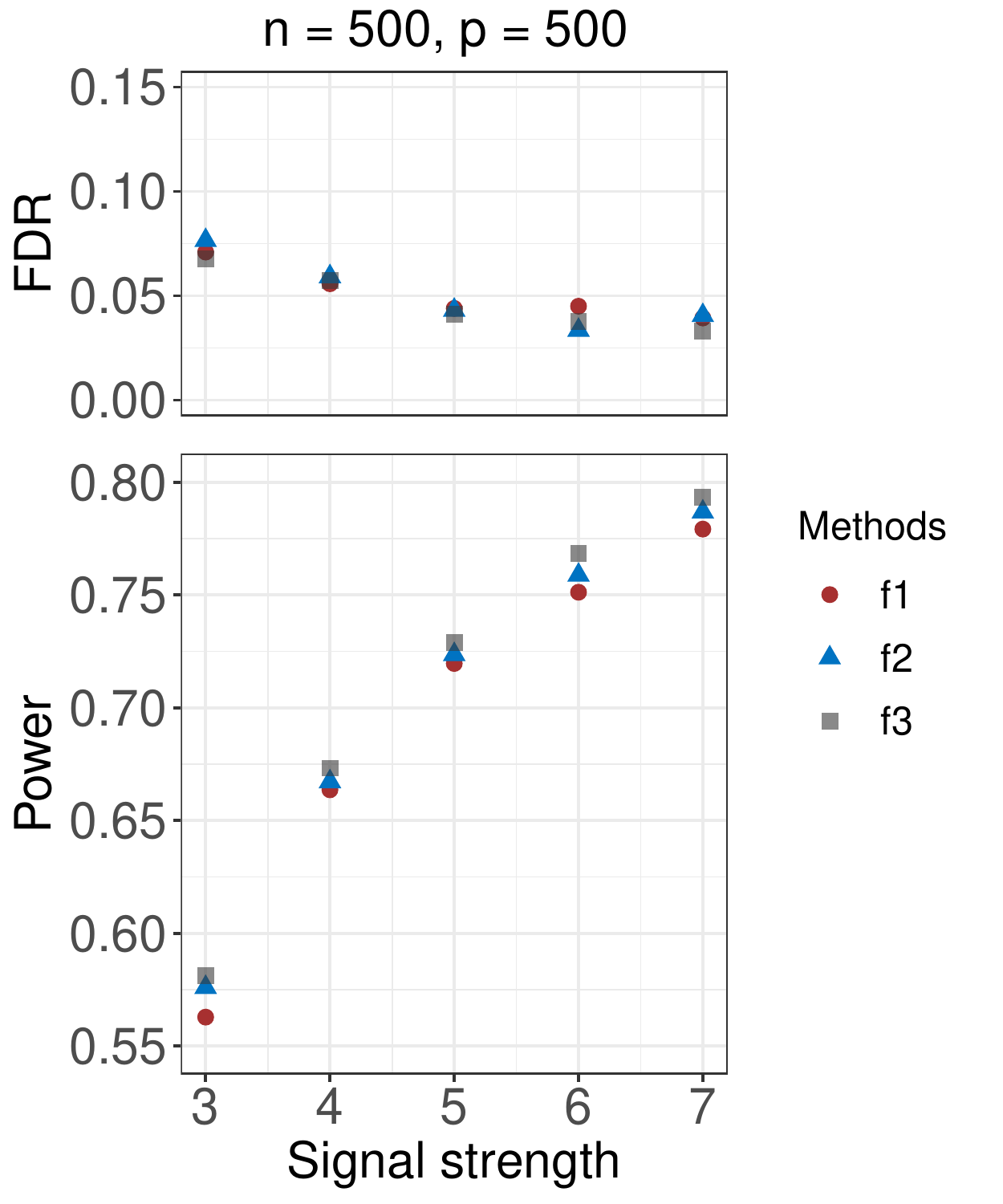}
\end{center}
\caption{Empirical FDRs and powers of MDS using three different mirror statistics constructed with $f_1, f_2, f_3$ specified in  \eqref{eq:contrast_choice}. The algorithmic settings are as per Figure \ref{fig:mirror-statistic-optimality-DS}.}
\label{fig:mirror-statistics-optimality-MDS}
\end{figure}

\begin{figure}
\begin{center}
\includegraphics[width=0.45\columnwidth]{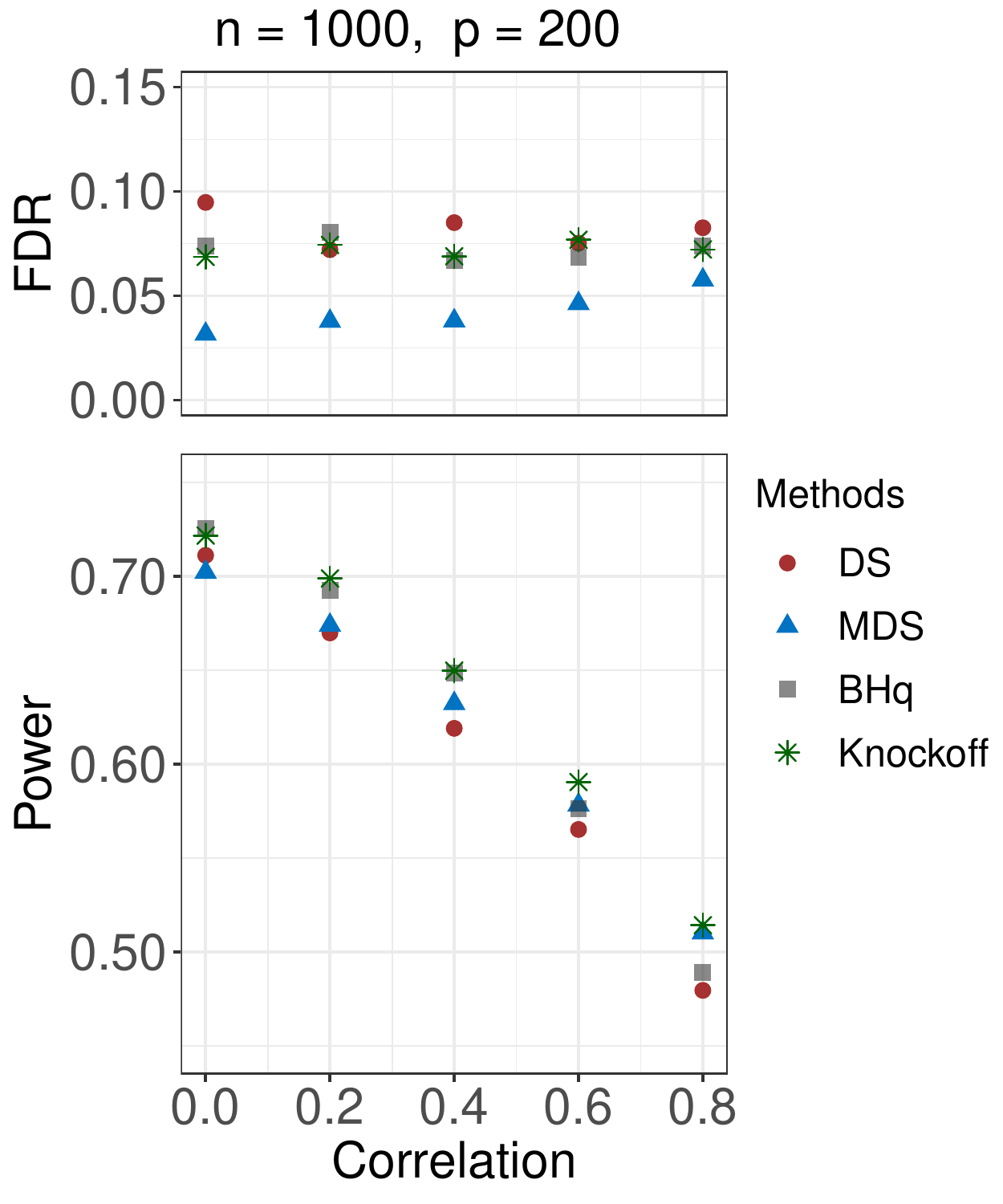}
\includegraphics[width=0.45\columnwidth]{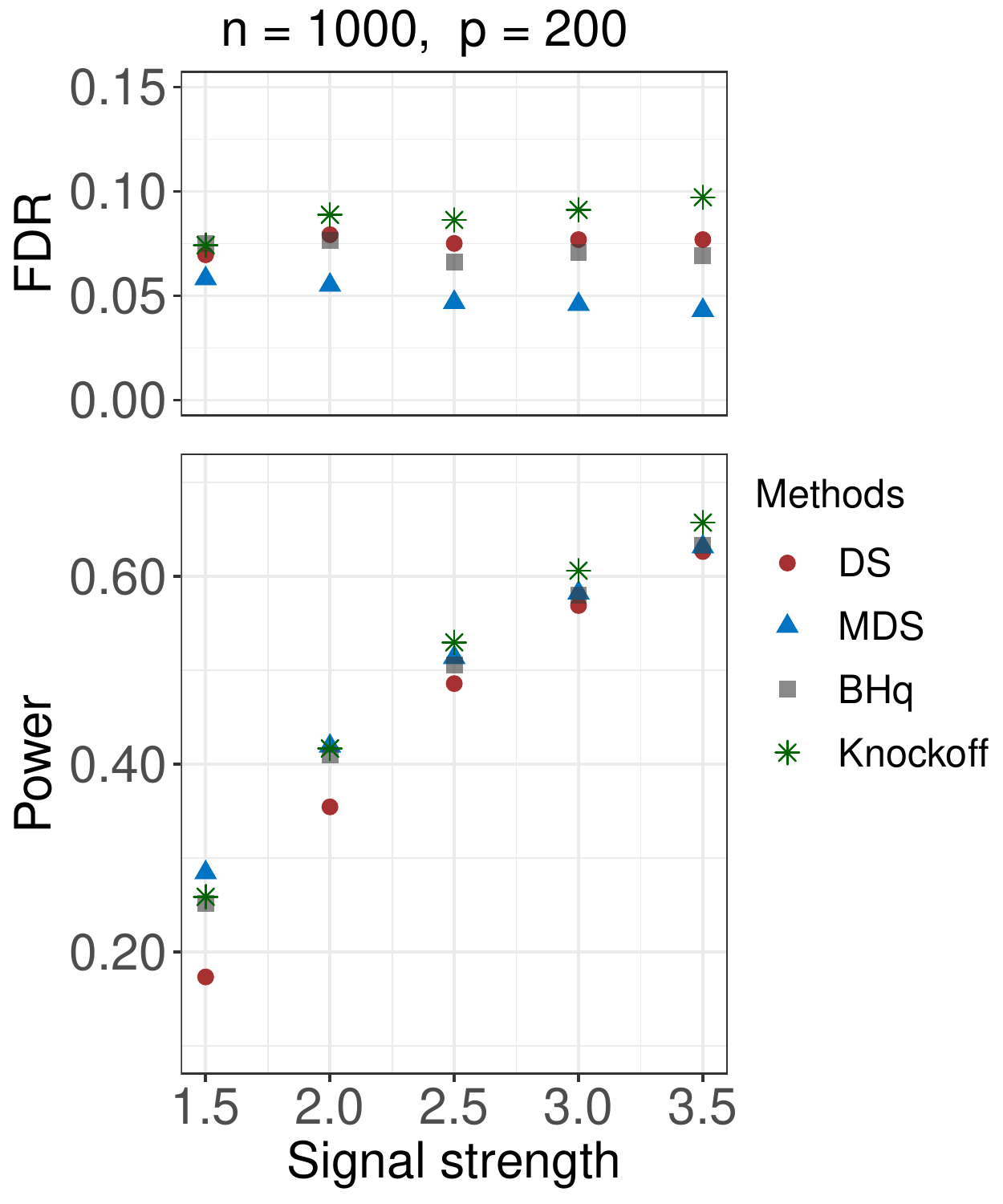}
\end{center}
\caption{Empirical FDRs and powers for low-dimensional linear models. Features are independently drawn from $N(0, \Sigma)$ with $\Sigma$ being a Toeplitz covariance matrix. Knockoff refers to the fixed-design knockoff filter \citep{barber2015controlling}.
In the left panel, we fix  the signal strength at $\delta = 3$ and vary the correlation $\rho$. In the right panels, we fix  the correlation at $\rho = 0.6$ and vary the signal strength $\delta$.
The  designated FDR control level is $q = 0.1$, and the number of relevant features is 50 across all settings.
Each dot in the figure represents the average from 50 independent runs.}
\label{fig:low-d-toeplitz}
\end{figure}

\begin{figure}
\begin{center}
\includegraphics[width=0.49\columnwidth]{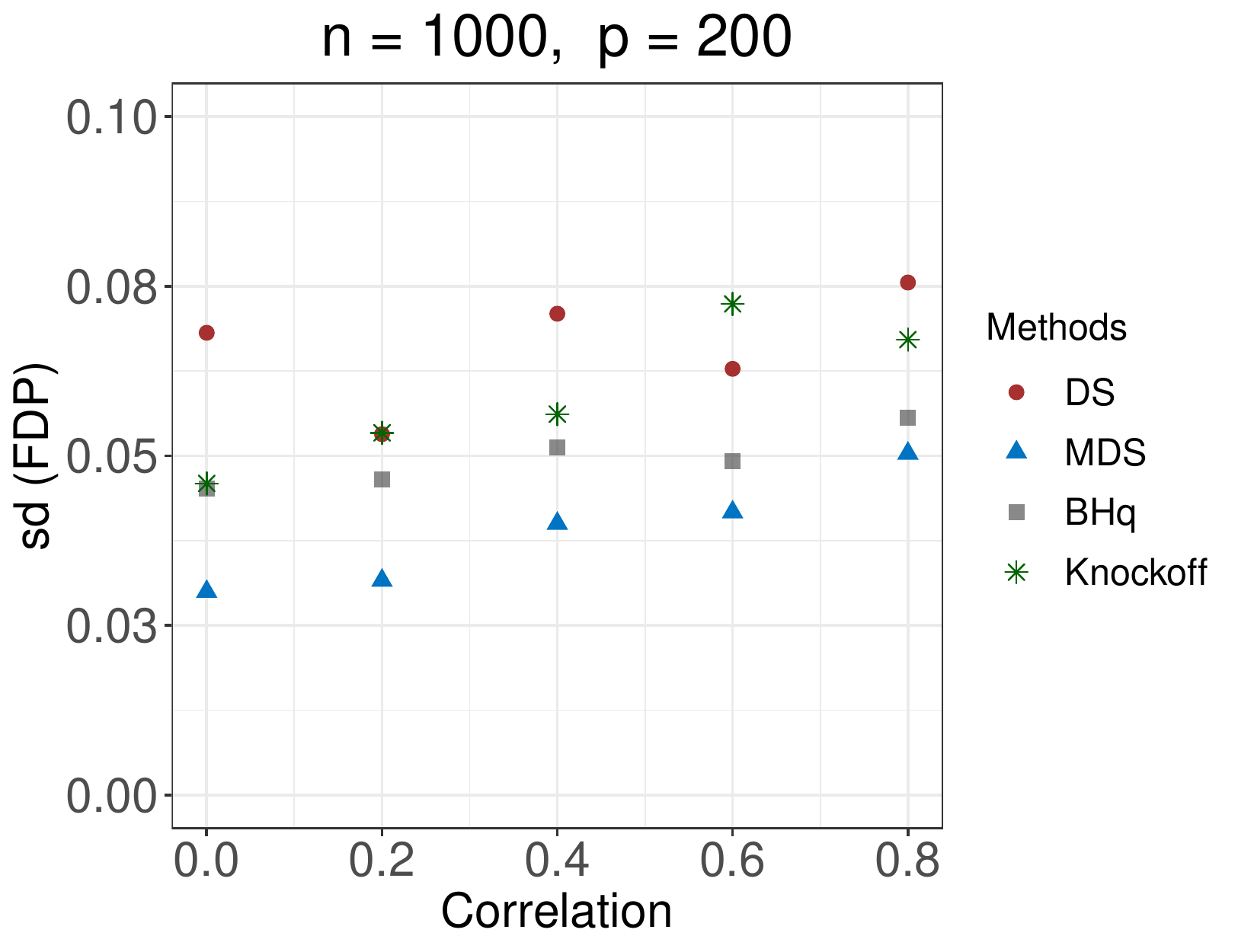}
\includegraphics[width=0.49\columnwidth]{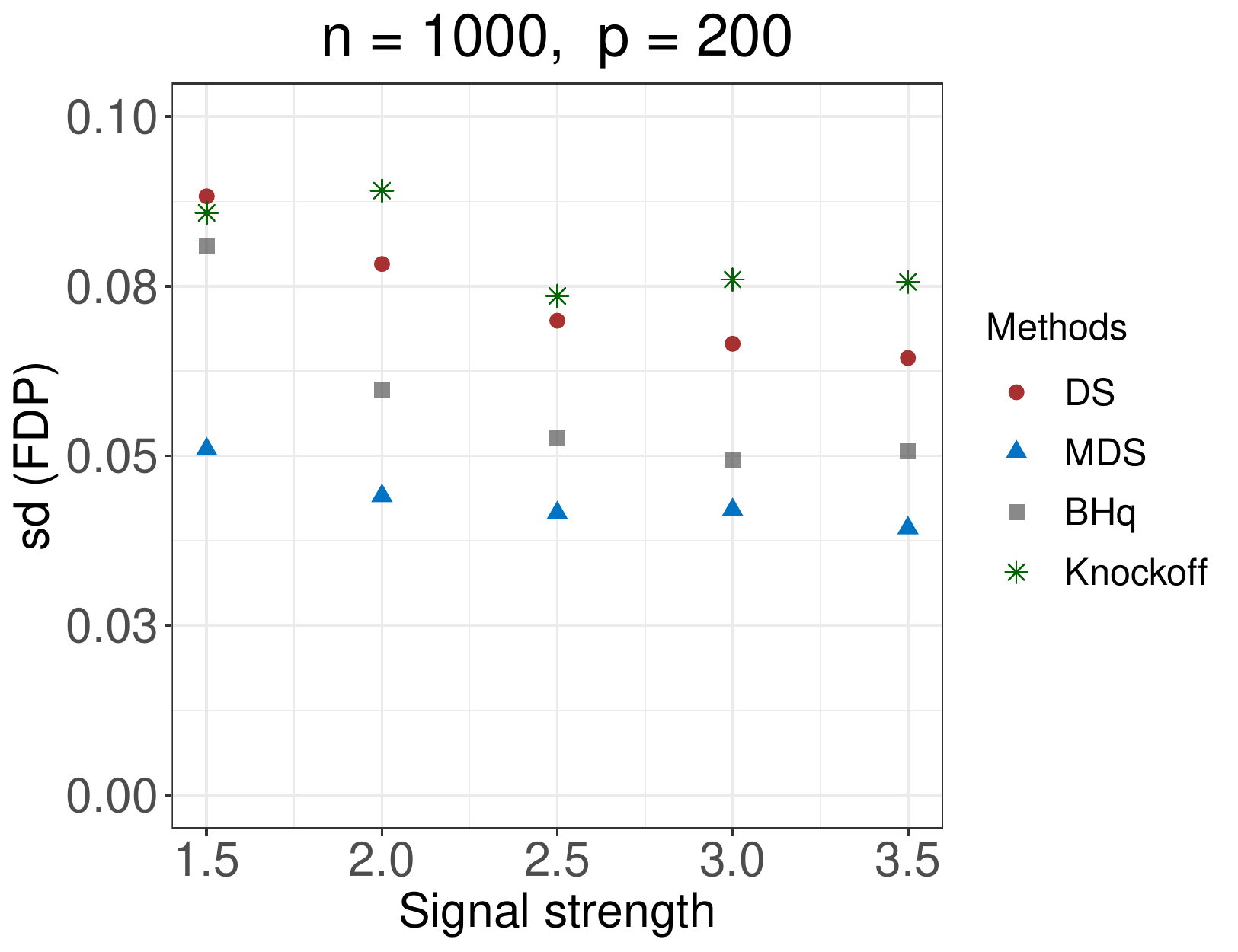}
\end{center}
\caption{
Standard deviations of the FDP for low-dimensional linear models with a Toeplitz correlation structure. The algorithmic settings are as per Figure \ref{fig:low-d-toeplitz}.}
\label{fig:low-d-toeplitz-std}
\end{figure}

\begin{figure}
\begin{center}
\includegraphics[width=0.45\columnwidth]{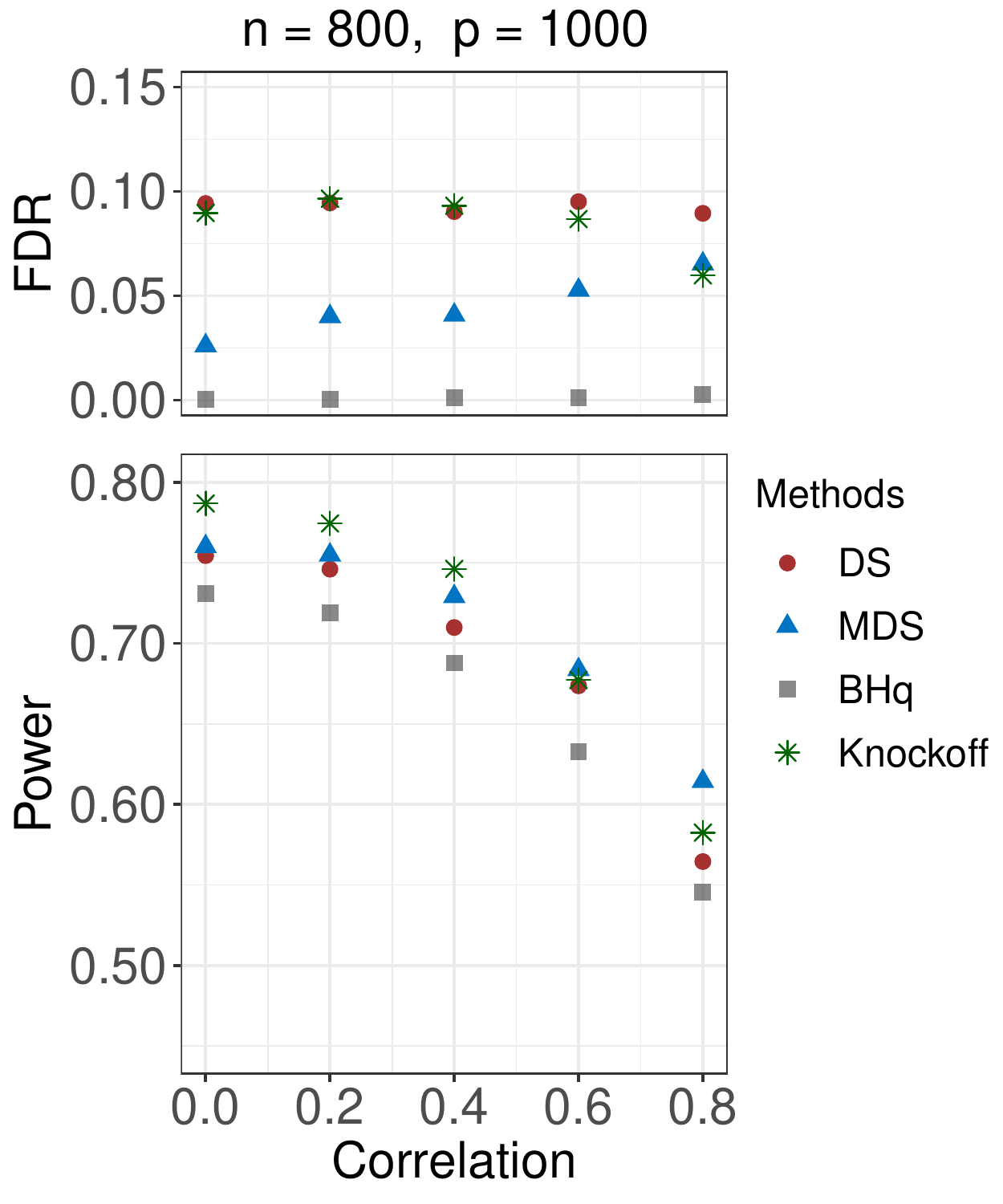}
\includegraphics[width=0.45\columnwidth]{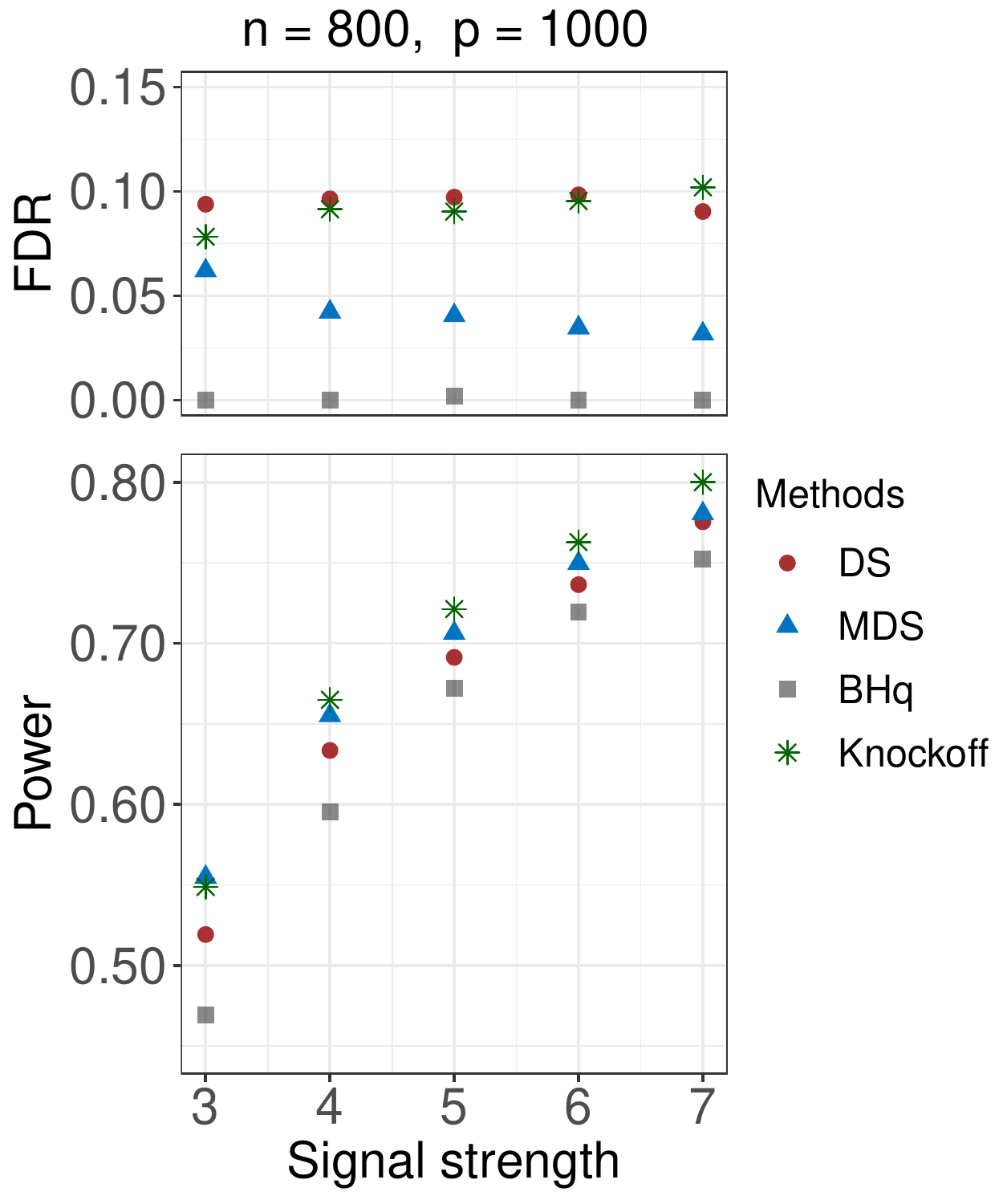}
\end{center}
\caption{Empirical FDRs and powers for linear models. The algorithmic settings are as per Figure \ref{fig:normal-design-toeplitz-correlation-P2000}.}
\label{fig:normal-design-toeplitz-correlation-P1000}
\end{figure}

\begin{figure}
\begin{center}
\includegraphics[width=0.45\columnwidth]{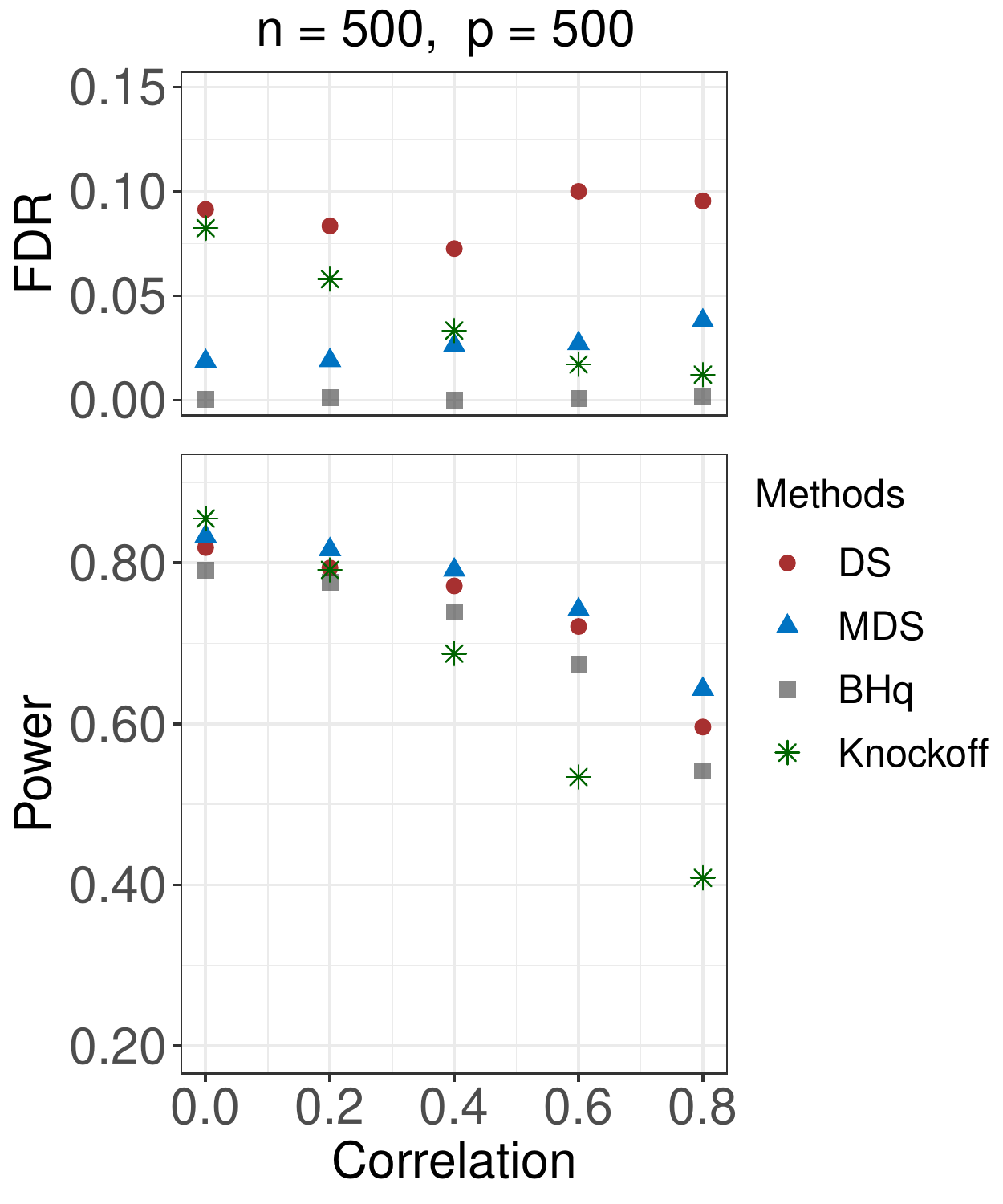}
\includegraphics[width=0.45\columnwidth]{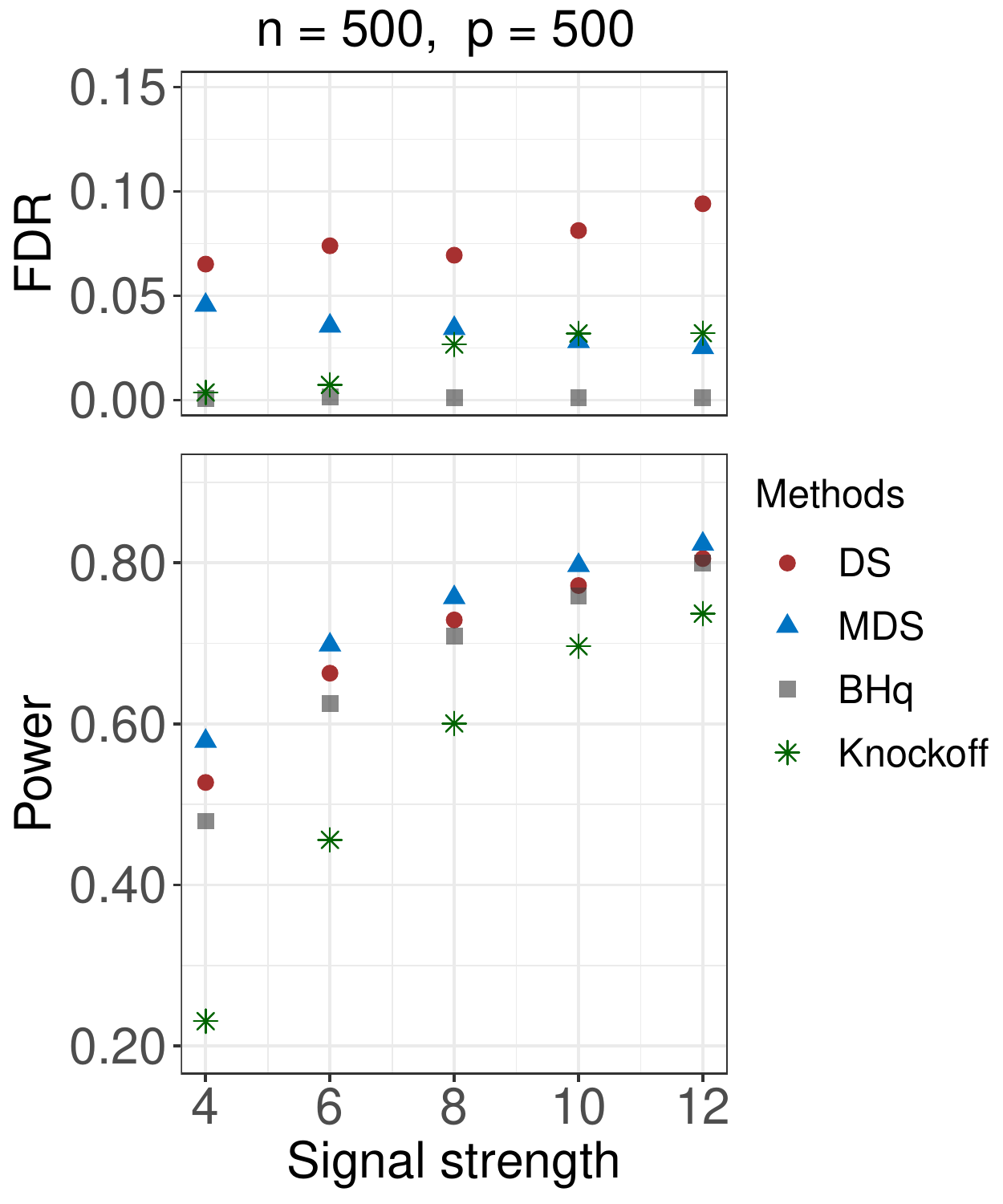}
\includegraphics[width=0.45\columnwidth]{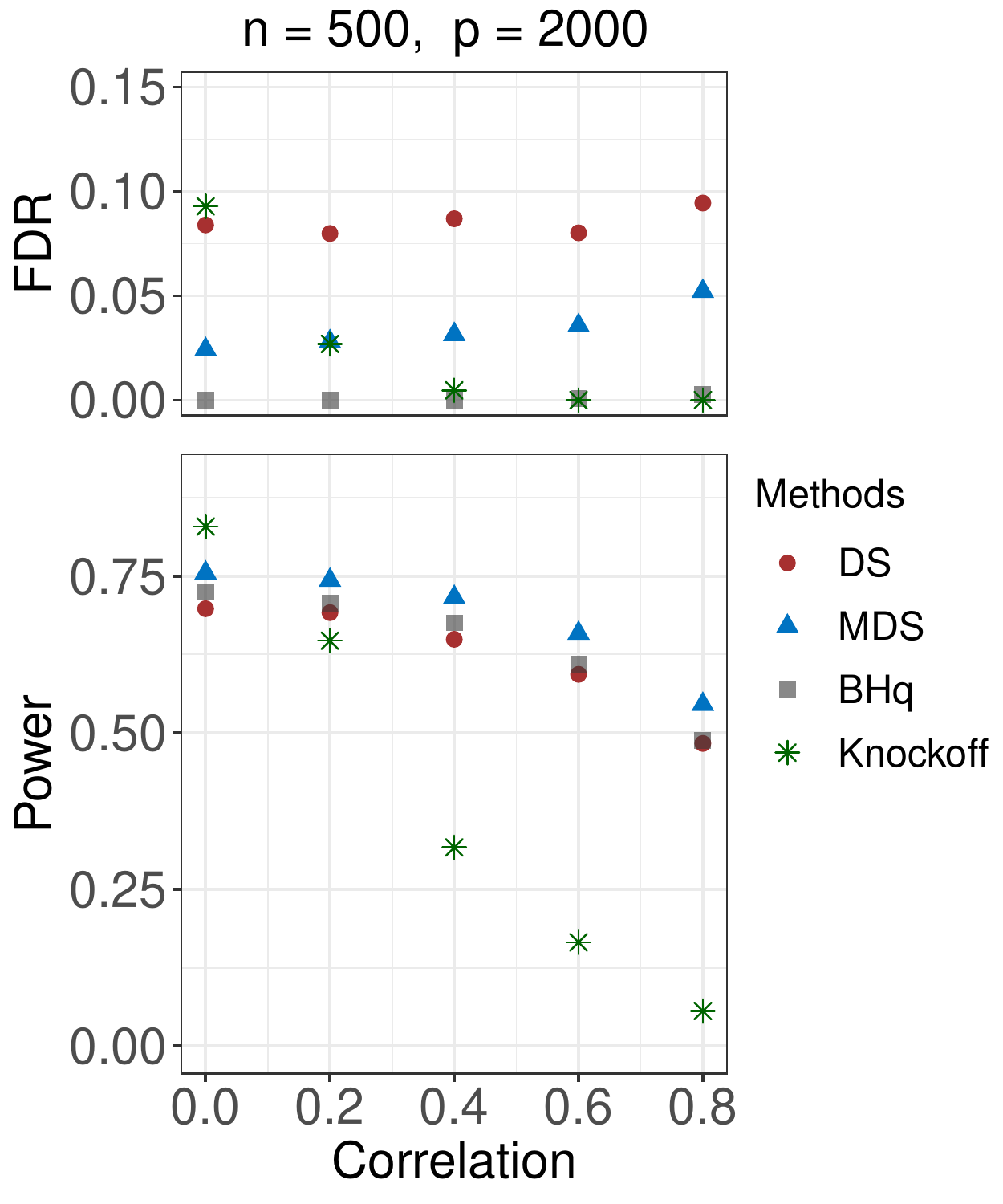}
\includegraphics[width=0.45\columnwidth]{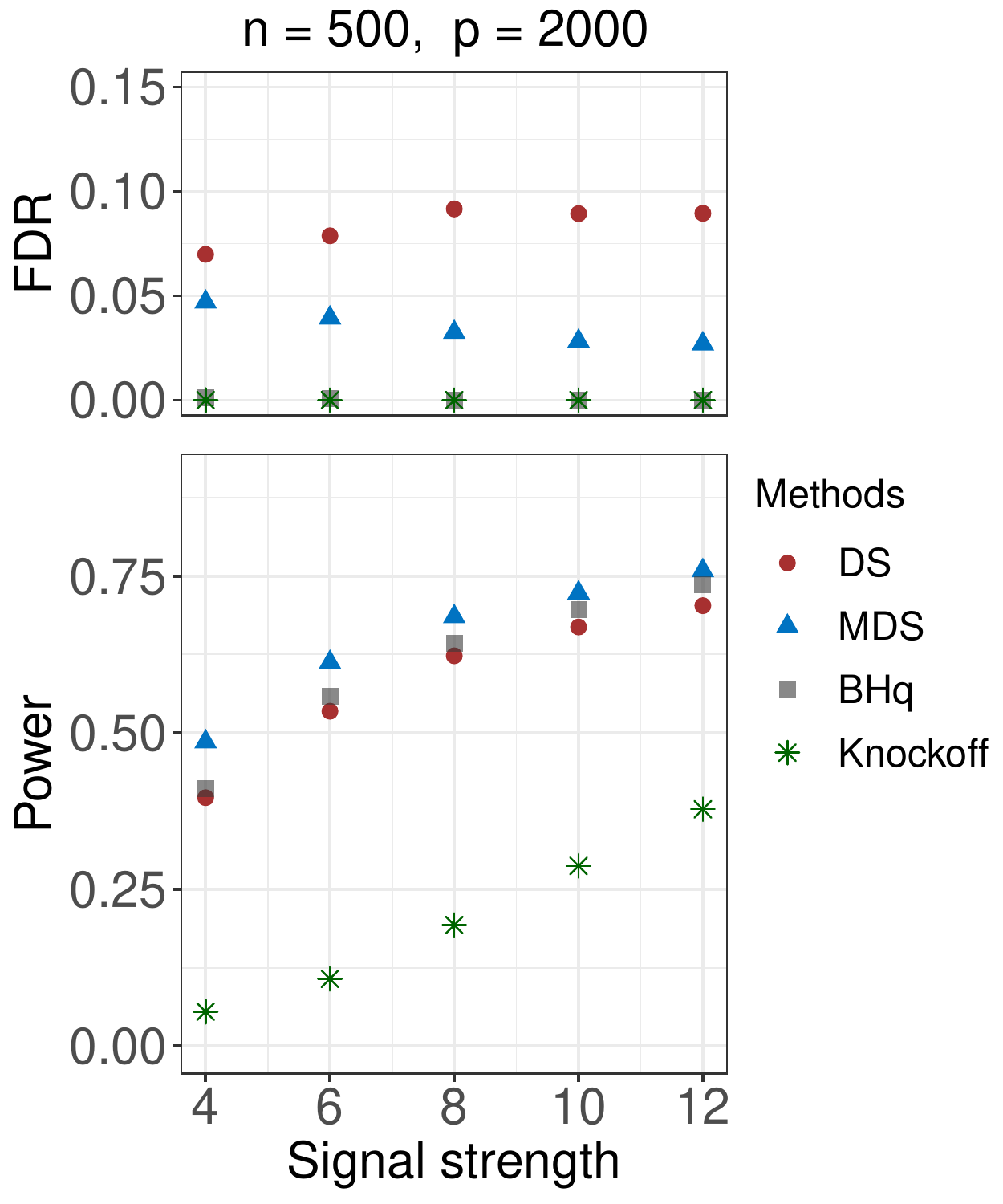}
\end{center}
\caption{Empirical FDRs and powers for linear models. Features are independently drawn from $N(0, \Sigma)$ with $\Sigma_{ij} = \rho^{\mathbbm{1}(i\neq j)}$.
In the left two panels, we fix the signal strength at $\delta = 8$ and vary the pairwise correlation $\rho$.
In the right two panels, we fix the pairwise correlation at $\rho = 0.5$ and vary the signal strength $\delta$.
The number of relevant features is 50 across all settings, and
the designated FDR control level is $q = 0.1$.
Each dot in the figure represents the average from 50 independent runs.}
\label{fig:normal-design-constant-correlation}
\end{figure}

\begin{figure}
\begin{center}
\includegraphics[width=0.49\columnwidth]{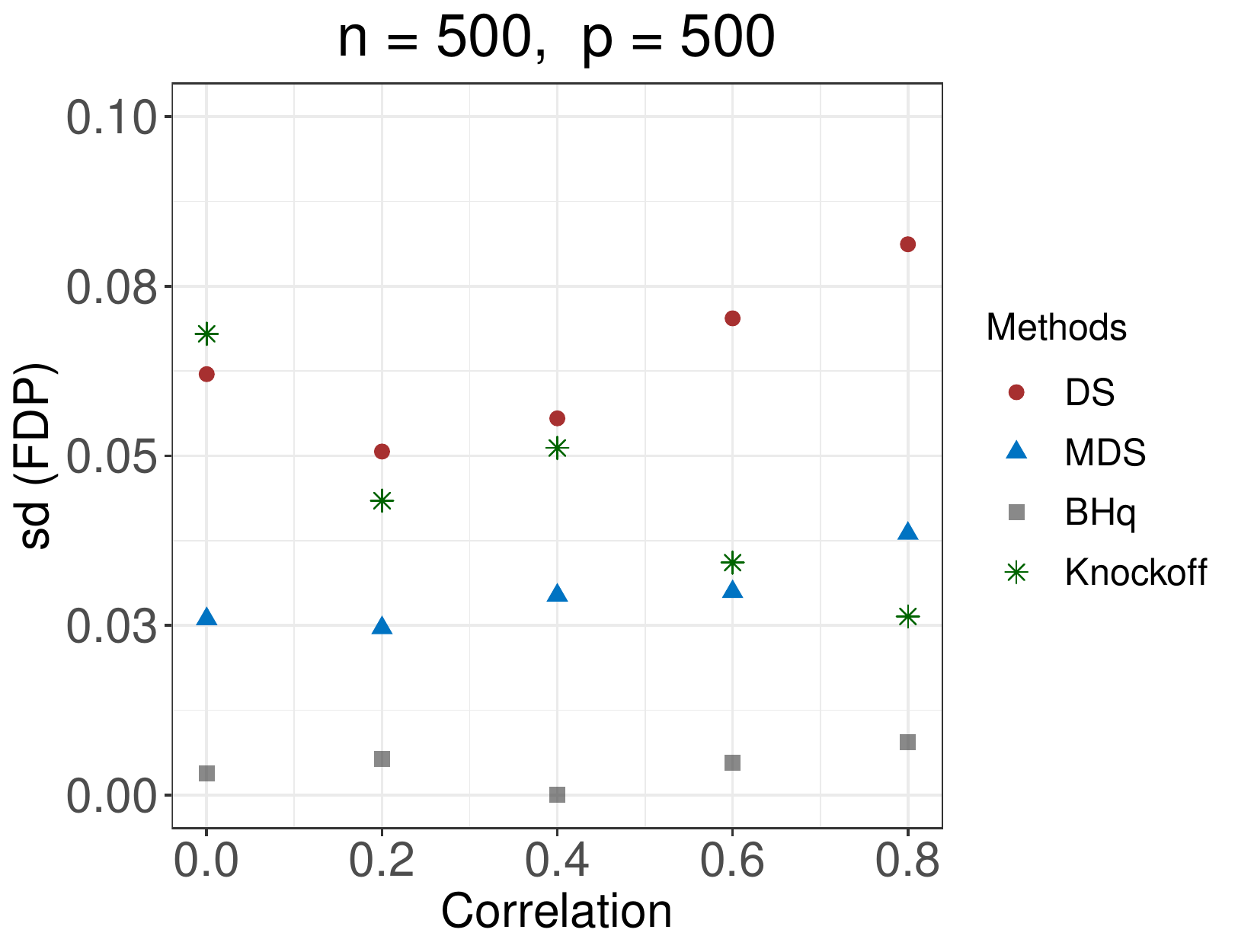}
\includegraphics[width=0.49\columnwidth]{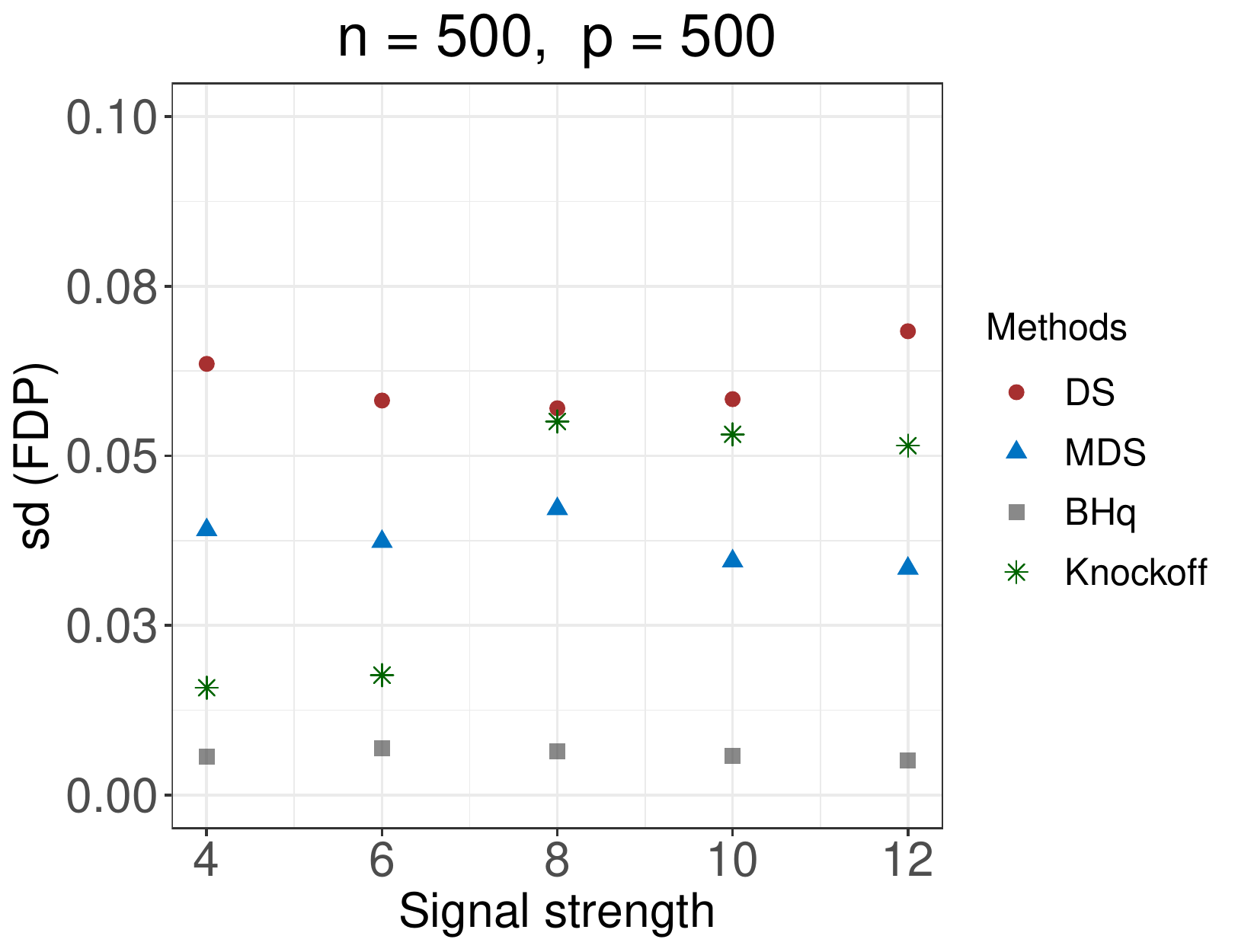}
\includegraphics[width=0.49\columnwidth]{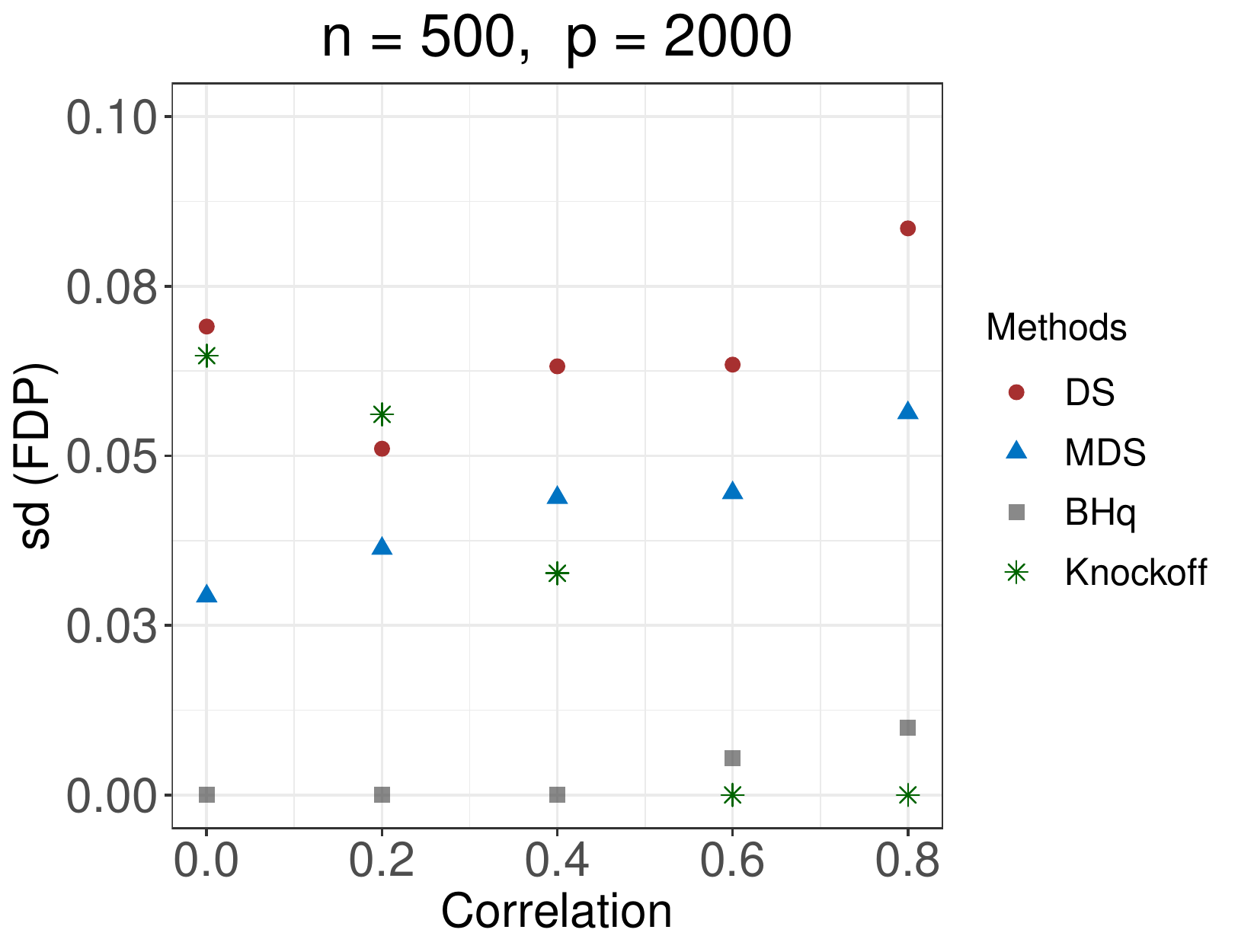}
\includegraphics[width=0.49\columnwidth]{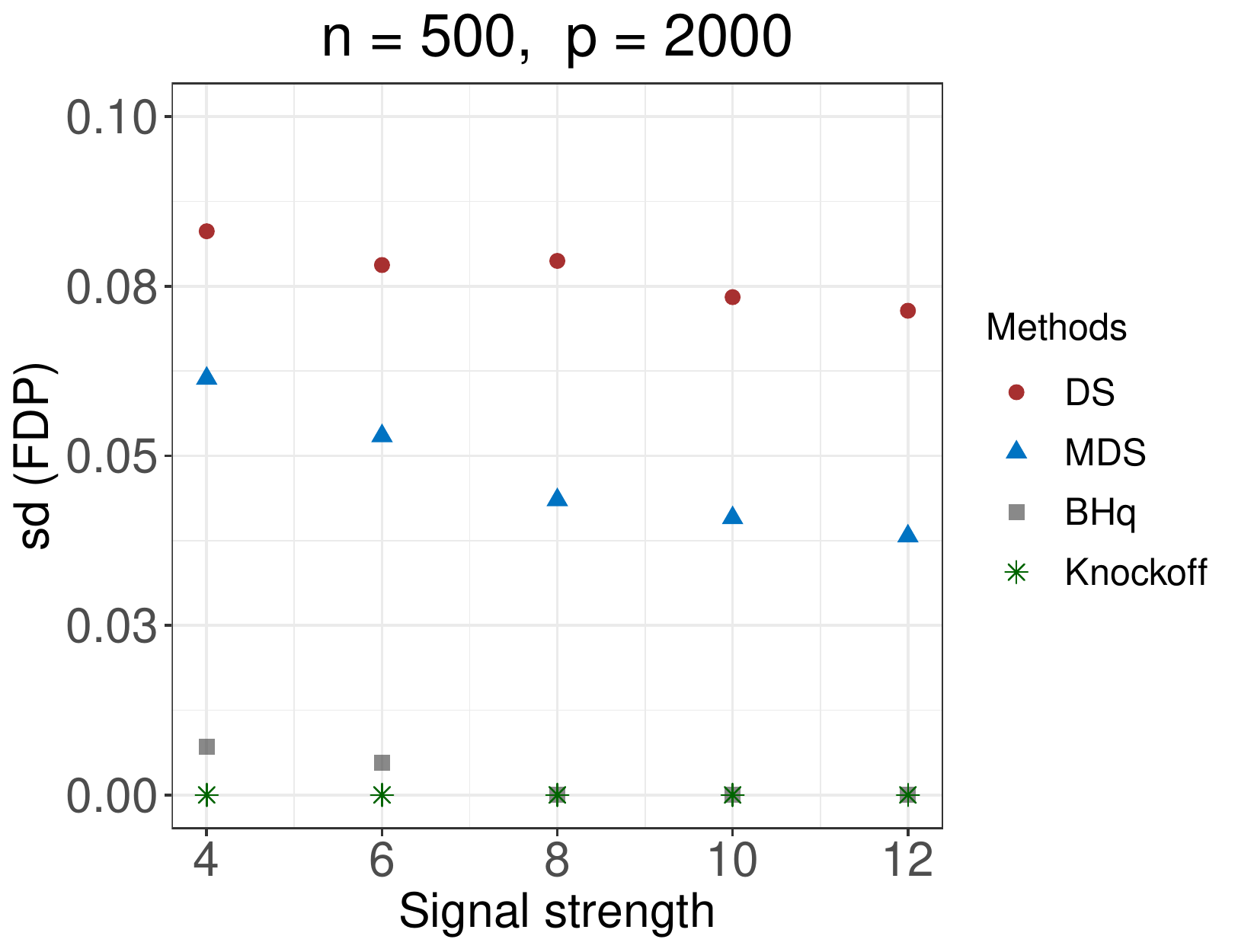}
\end{center}
\caption{Standard deviations of the FDP for linear models with a constant correlation structure. The algorithmic settings are as per Figure \ref{fig:normal-design-constant-correlation}.}
\label{fig:normal-design-constant-correlation-FDP-std}
\end{figure}

\begin{figure}
\begin{center}
\includegraphics[width=0.49\columnwidth]{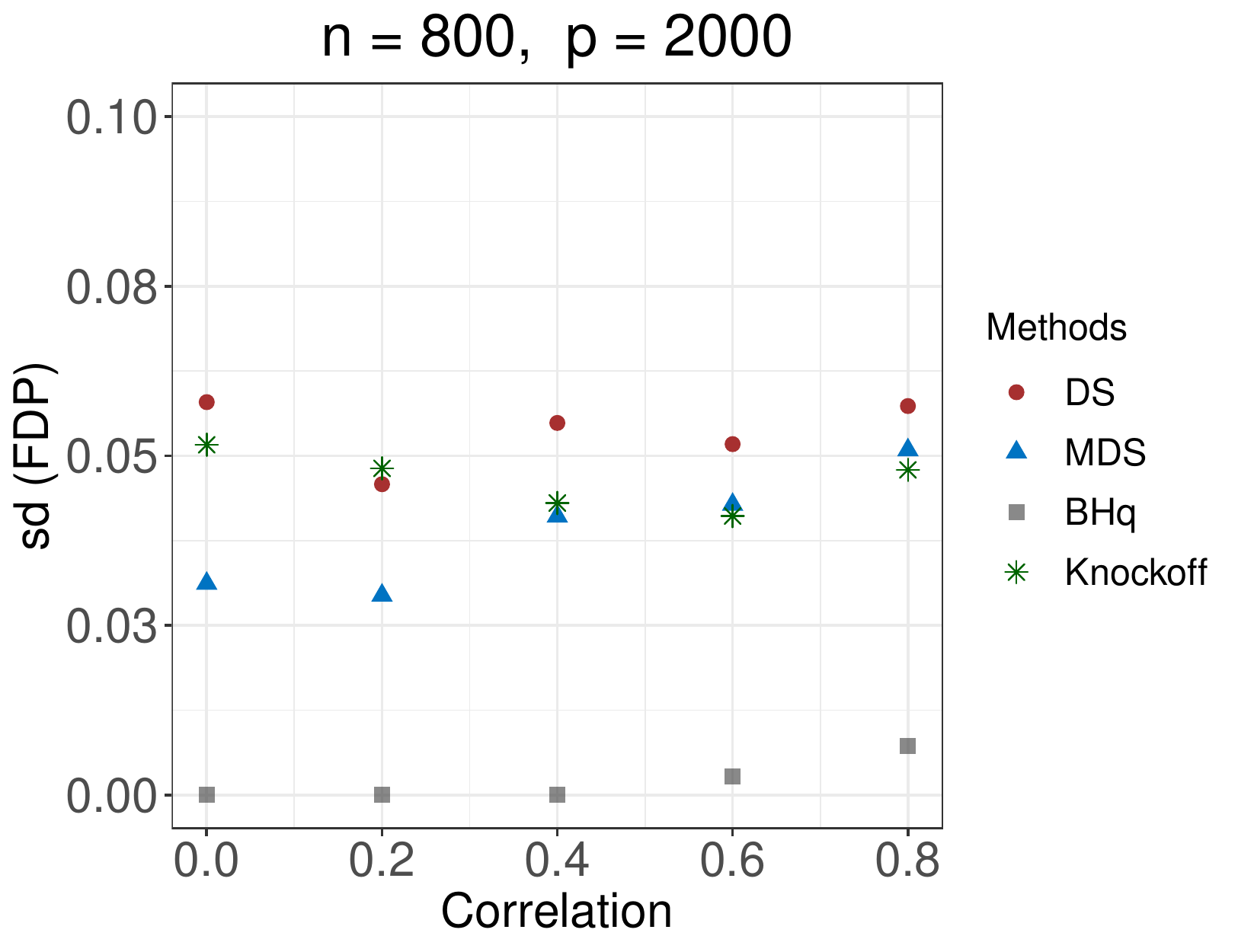}
\includegraphics[width=0.49\columnwidth]{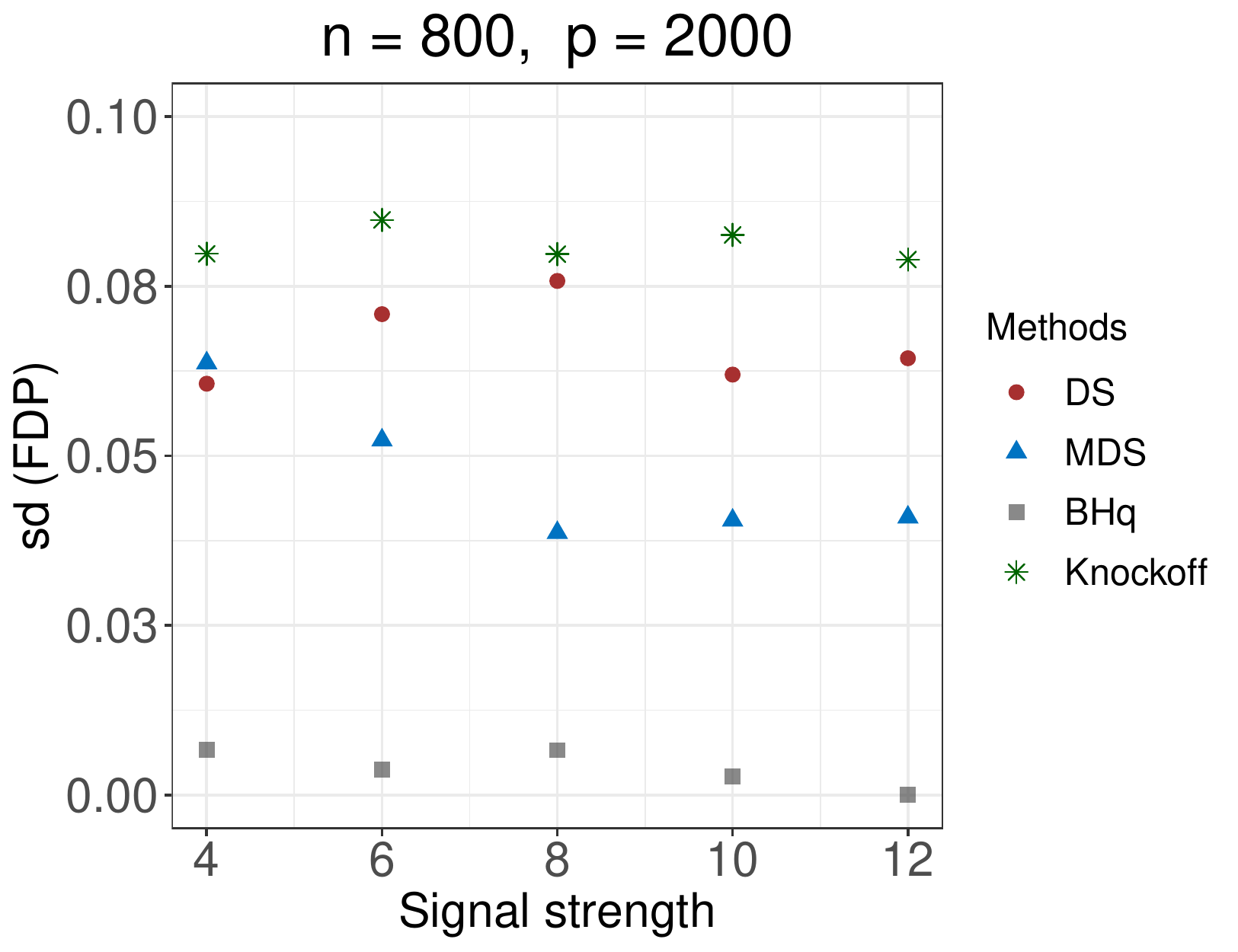}
\end{center}
\caption{Standard deviations of the FDP for linear models with non-normal design matrices. The algorithmic settings are as per Figure \ref{fig:t-design-toeplitz-correlation}.}
\label{fig:t-design-constant-correlation-FDP-std}
\end{figure}

\begin{figure}
\begin{center}
\includegraphics[width=0.49\columnwidth]{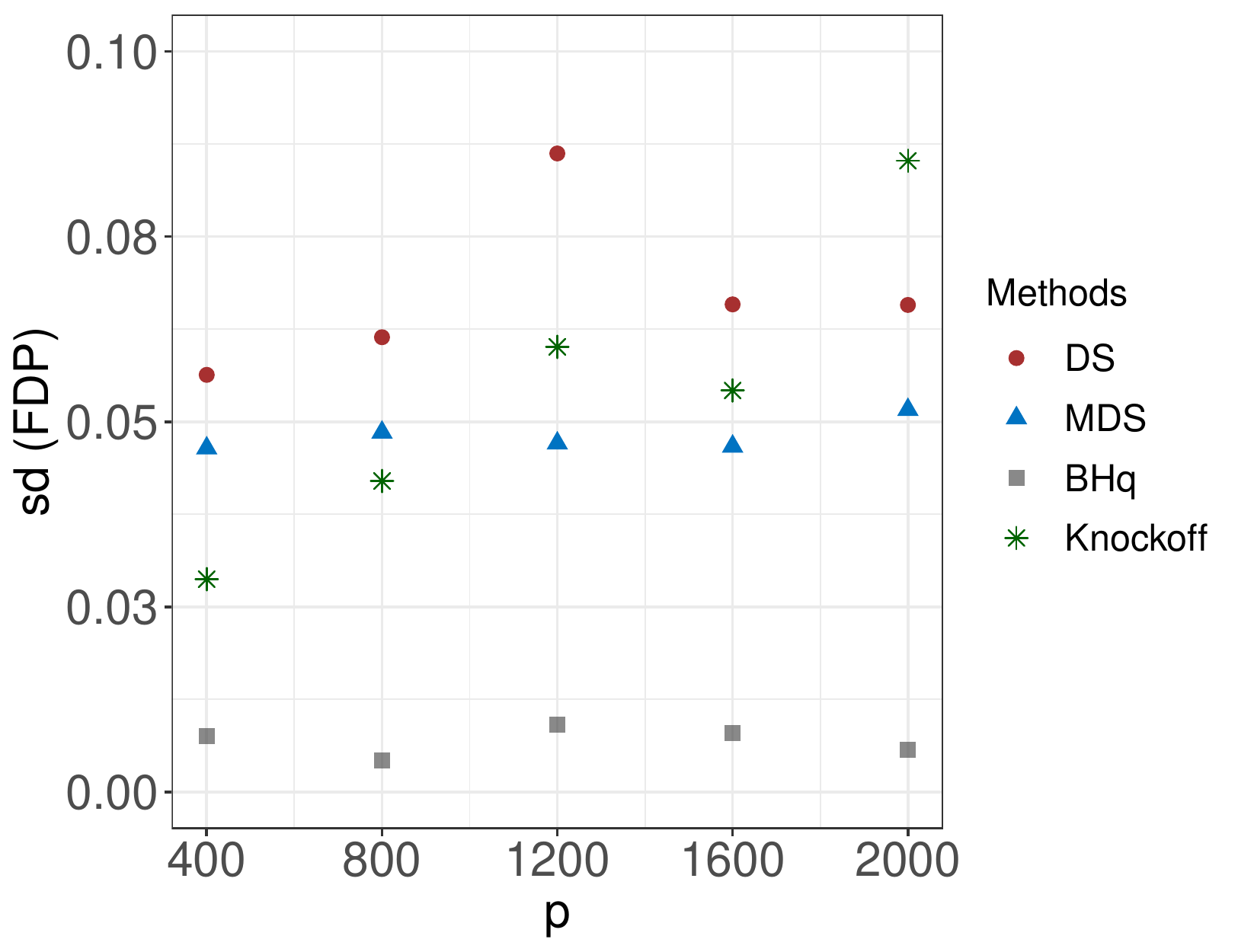}
\includegraphics[width=0.49\columnwidth]{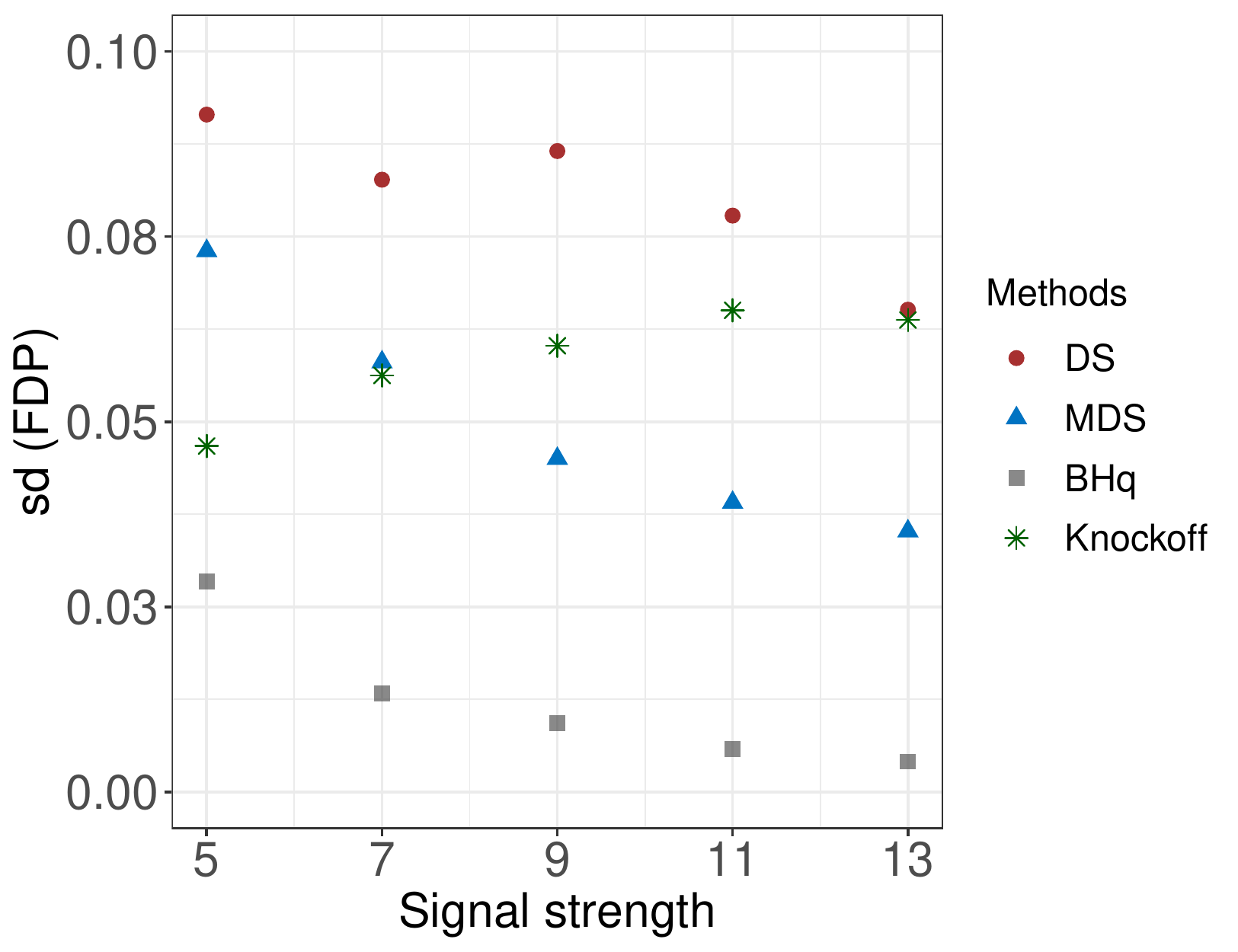}
\end{center}
\caption{Standard deviations of the FDP for linear models with a GWAS design matrix. The algoritmic settings are as per Figure \ref{fig:GWAS-design-toeplitz-correlation}.}
\label{fig:GWAS-design-constant-correlation-FDP-std}
\end{figure}

\end{document}